\documentclass[prd,11pt, %twocolumn,
showpacs, floatfix, nofootinbib
]{revtex4-2}
\usepackage[utf8]{inputenc}
\usepackage{amsmath,amsfonts,mathtools}
\usepackage{amssymb}
\usepackage{epsfig}
\usepackage{graphicx}
\usepackage{color}
\usepackage{hyperref}
\hypersetup{colorlinks=true,linkcolor=blue,citecolor=blue,urlcolor=black}
\usepackage{comment}
\usepackage{cancel}
\usepackage{bbold}
\usepackage{booktabs}
\usepackage{vwcol} % for multiple columnsverbatim
\usepackage[normalem]{ulem}
\usepackage[dvipsnames]{xcolor}
\usepackage{pbox}
\usepackage{multirow,bigdelim} % for braces in tables
\usepackage{caption}
\usepackage{subcaption}
\definecolor{plotBlue}{rgb}{0.25,0.26,0.78}
\definecolor{plotLightBlue}{rgb}{0.29,0.54,0.77}
\definecolor{plotGreen}{rgb}{0.62,0.74,0.35}
\definecolor{plotOrange}{rgb}{0.9,0.49,0.2}
\definecolor{plotBlue2}{rgb}{0,0,1}

\DeclareMathOperator{\sign}{sign}

\newcommand{\GeV}{\, \mathrm{GeV}}

\newcommand{\SU}{\mathrm{SU}}
\newcommand{\SM}{$\mathrm{SU}(3)_c \times \protect\linebreak[0]\mathrm{SU}(2)_L \times \protect\linebreak[0]\mathrm{U}(1)_Y \ $}
\newcommand{\LR}{$\mathrm{SU}(3)_c \times \protect\linebreak[0]\mathrm{SU}(2)_L \times \protect\linebreak[0]\mathrm{SU}(2)_R \times \protect\linebreak[0]\mathrm{U}(1)_{B-L} \ $}

\newcommand{\BLzero}{$\mathrm{SU}(4)_C \times \protect\linebreak[0]\mathrm{SU}(2)_L \times \protect\linebreak[0]\mathrm{U}(1)_R \ $}
\def\EQSPACE{\\[8pt]}
\newcommand{\backspace}{\!\!\!\!\!\!\!\!\!\!\!\!\!}
\newcommand{\forwardspace}{\;\;\;\;\;\;\;}

\newcommand*\xbar[1]{\hspace{0.1em}%
	\hbox{%
		\vbox{%
			\hrule height 0.5pt % The actual bar
			\kern0.4ex%         % Distance between bar and symbol
			\hbox{%
				\kern-0.1em%      % Shortening on the left side
				\ensuremath{#1}%
				\kern-0.1em%      % Shortening on the right side
			}%
		}%
	}\hspace{0.1em}%
}
\def\ZZ{\phantom{0}}

\def\slog{\mathrm{lg}\,}
\newcommand{\TBOX}[1]{\makebox[1.6cm][r]{#1}}

\def\real{\textcolor{blue}}
\def\complex{\textcolor{black}}
\def\conjugate{\textcolor{red}}

\begin{document}

%\numberwithin{equation}{section}
%\tableofcontents
%\newpage
%----------------------------------------------------------------------------------
\title{Quantum nature of the minimal potentially realistic $\mathrm{SO}(10)$ Higgs model}
\preprint{}
%----------------------------------------------------------------------------------
\author{Kate\v{r}ina Jarkovsk\'a}\email{jarkovska@ipnp.mff.cuni.cz}
\affiliation{Institute of Particle and Nuclear Physics,
Faculty of Mathematics and Physics,
Charles University in Prague, V Hole\v{s}ovi\v{c}k\'ach 2,
180 00 Praha 8, Czech Republic}
\author{Michal Malinsk\'{y}}\email{malinsky@ipnp.mff.cuni.cz}
\affiliation{Institute of Particle and Nuclear Physics,
Faculty of Mathematics and Physics,
Charles University in Prague, V Hole\v{s}ovi\v{c}k\'ach 2,
180 00 Praha 8, Czech Republic}
\author{Timon Mede}\email{timon.mede@ijs.si}
\affiliation{%Reactor Engineering Division,
Jo\v zef Stefan Institute, Jamova cesta 39, SI-1000 Ljubljana, Slovenia
}
\author{Vasja Susi\v c}\email{vasja.susic@unibas.ch}
\affiliation{Department of Physics, University of Basel,
Klingelbergstrasse 82, CH-4056 Basel, Switzerland
}
%----------------------------------------------------------------------------------

%%%%%%%%%%%%%%%%%%%%%%%%%%%%%%%%%%%%%%%%%%%%%%%%%%%%%%%
% ABSTRACT
%%%%%%%%%%%%%%%%%%%%%%%%%%%%%%%%%%%%%%%%%%%%%%%%%%%%%%%

\begin{abstract}
We study several aspects of the quantum structure of the minimal potentially realistic renormalizable $\mathrm{SO}(10)$ Higgs model in which the $\mathbf{45}\oplus\mathbf{126}$ scalars spontaneously break the symmetry down to the Standard Model (SM) group $\SU(3)_{c}\times \SU(2)_{L}\times \mathrm{U}(1)_{Y}$. With complete information about the one-loop corrections to the masses of all scalars in the theory and the one-loop beta functions governing the running of all dimensionless scalar self-couplings, the domains of the parameter space where the model can be treated perturbatively are established, along with improved bounds from the requirements of the SM vacuum stability and gauge-coupling unification. We demonstrate that the model is fully consistent and potentially realistic only in very narrow regions of the parameter space corresponding to the breaking chains with well-pronounced  $\SU(4)_{C}\times \SU(2)_L\times \mathrm{U}(1)_R$ and $\SU(3)_{c}\times \SU(2)_L\times \SU(2)_R\times \mathrm{U}(1)_{B-L}$ intermediate symmetries, with a clear preference for the former case. Barring accidental fine-tunings in the scalar sector, this makes it possible to provide a very sharp prediction for the position of the unification scale and the value of the associated gauge coupling, with clear implications for the phenomenology of grand unified models based on this structure.
\vskip 1cm
\end{abstract}
\maketitle
%\tableofcontents
%\newpage

%%%%%%%%%%%%%%%%%%%%%%%%%%%%%%%%%%%%%%%%%%%%%%%%%%%%%%%
\section{Introduction\label{Sec:Introduction}}
%%%%%%%%%%%%%%%%%%%%%%%%%%%%%%%%%%%%%%%%%%%%%%%%%%%%%%%

With the upcoming generation of large-volume experiments aiming to test the potential instability of baryonic matter  
(DUNE~\cite{DUNE:2016hlj,DUNE:2015lol}, Hyper-K~\cite{Abe:2011ts,Hyper-Kamiokande:2018ofw}),
one can expect at least an order-of-magnitude improvement of their sensitivity in most of the relevant nucleon decay channels ($p \to \pi^{0} e^{+}$, $p \to \pi^{+}\bar{\nu}$, $p \to K^{+}\bar{\nu}$, etc.) with respect to the current limits. 

Unfortunately, on the theory side, these efforts are notoriously difficult to meet with good enough estimates that would, at least in principle, make it possible to distinguish among different scenarios. To this end, even the most popular models of baryon number ($B$) violation based on the idea of grand unification, the so called Grand Unified Theories (GUT)~\cite{Georgi:1974sy}, often come short when better than several-orders-of-magnitude predictions are demanded. 
This has to do with various types of obstacles plaguing an accurate determination of some of the critical inputs to such calculations, namely, 
(i) the proximity of the unification scale to the Planck scale which, in general, enhances the effects of higher-dimensional operators inducing out-of-control shifts in, e.g.,~the GUT-scale matching conditions~\cite{Hill:1983xh,Shafi:1983gz,Calmet:2008df}, 
(ii) 
the need to go beyond the leading-order approximation in the high-scale mediator masses in order to deal with the associated theoretical uncertainties in the proton lifetime estimates,
(iii) the need to model the flavour structure of the relevant baryon and lepton number violating (BLNV) currents and, finally, 
(iv) the lack of accurate information about the $B\neq 0$ hadronic matrix elements.

While the last two issues may be alleviated to some degree by, e.g.,~focusing on specific observables with less sensitivity to flavour uncertainties such as branching ratios and/or neutrino production channels (case iii) and, perhaps, investing more resources to accurate lattice QCD modelling (case iv), the first two are difficult in principle. As for point (ii), higher-order calculations in the GUT context are, by definition, complicated by the typically large number of degrees of freedom in the loops, raising questions about the stability of the results obtained at any given order of the perturbative expansion. Concerning (i), there is hardly anything one can do about this issue in general.        

Nevertheless, there are very particular model scenarios in which both (i) and (ii) can be addressed in a relatively satisfactory manner. Among these, a prominent role is played by the minimal renormalizable non-supersymmetric $\mathrm{SO}(10)$ GUT with the adjoint $\mathbf{45}$ triggering the first stage of symmetry breaking (followed by a second stage where the rank of the gauge group is reduced to that of the SM)~\cite{Chang:1984qr,Deshpande:1992au}. Remarkably, the structure of the scalar sector of this theory is such that the most troublesome Planck-scale-associated effective operators are entirely absent~\cite{Nath:2006ut} and, at the same time, the underlying Higgs model is simple enough to admit a comprehensive numerical analysis. Let us note that both these features are not only vital for any sensible physics scrutiny, but they also enable one to overcome~\cite{Bertolini:2009es} a peculiar pathology that the classical version of the model has been known to suffer from for decades~\cite{Buccella:1980qb, Yasue:1980fy, Anastaze:1983zk, Babu:1984mz}, namely, the instability of its SM-like vacua. 
For these reasons, the minimal $\mathrm{SO}(10)$ GUT has attracted a lot of attention in recent years with a number of interesting works touching upon its specific aspects, often within the bigger phenomenological picture; see, e.g.,~\cite{Joshipura:2011nn,Dueck:2013gca,Kolesova:2016ibq,Babu:2016bmy,Deppisch:2017xhv,DiLuzio:2020qio}.

To this end, detailed studies of the minimal renormalizable $\mathrm{SO}(10)$ Higgs model(s) play a central role as precursors to essentially all other activities. To date, these have focused predominantly on the leading quantum corrections to the masses of the $\SU(2)_L$-triplet $(1,3,0)$ and $\SU(3)_c$-octet $(8,1,0)$ pseudo-Goldstone bosons (PGBs)~\cite{Bertolini:2009es}, which were identified long ago as the main culprits behind the tree-level vacuum instability issues~\cite{Bertolini:2009es,Bertolini:2012im,Bertolini:2013vta,Kolesova:2014mfa}. It has recently been noted~\cite{Graf:2016znk} that a third potentially problematic singlet pseudo-Goldstone mode worth detailed scrutiny pops up along the potentially realistic symmetry-breaking chains with the $\mathrm{U}(1)_{B-L}$-breaking (seesaw) scale parametrically smaller than the GUT scale $M_{\text{GUT}}\sim 10^{16}\GeV$. This further complicates matters, since the field in question is a member of a rich SM-singlet family of four scalars, and thus the analysis requires a thorough inspection of a $4\times 4$ mass matrix along with the associated quantum corrections.      

In this paper, we aim to provide the ultimate synthesis of these (and several new) aspects into a decisive and self-contained analysis of the one-loop quantum structure of the minimal potentially realistic renormalizable $\mathrm{SO}(10)$ Higgs model. Besides complementing the previous studies by a refined account of the pseudo-Goldstone sector, including issues related to the previously unnoticed instability also plaguing one of the SM-singlet scalars (which, in certain limits, behaves as a third pseudo-Goldstone boson in the spectrum), we calculate the leading quantum corrections to the masses of all other fields in the scalar sector along with the one-loop beta functions of all the dimensionless scalar-potential couplings. This not only makes it possible to verify the convexity of the local extrema supporting the potentially realistic SM-like vacua, but at the same time opens the door to another important aspect ignored to a large degree so far, namely, that of the perturbative stability of all the results.    

Remarkably enough, such perturbativity requirements turn out to be extremely powerful in eliminating large patches of the formerly allowed parameter space. As we shall demonstrate, the  model entertains a certain level of perturbative stability only in very specific limits corresponding to the breaking chains  with well-pronounced $\SU(4)_C\times \SU(2)_L\times \mathrm{U}(1)_R$ and $\SU(3)_c\times \SU(2)_L\times \SU(2)_R\times \mathrm{U}(1)_{B-L}$ intermediate-level symmetries, with a clear preference for the former. This has to do with an interesting interplay between the three SM-compatible vacuum expectation values (VEVs) available in the scalar sector, which ubiquitously appear in the form of the universal structure
\begin{align}
     \frac{\omega_{BL} \omega_{R} (\omega_{BL} + \omega_{R})}{|\sigma|^2}. \label{eq:VEV-structure}
\end{align}
This structure may give large massive contributions when a hierarchy between the GUT scale (represented by the larger of $\omega_R$ and $\omega_{BL}$) and the seesaw scale (denoted by $\sigma$) is assumed. One possible way to retain perturbativity would be to suppress the structure's dimensionless prefactors, as was often done in previous accounts~\cite{Bertolini:2013vta,Kolesova:2014mfa}. However, due to its ubiquitous appearance also in higher-order loop corrections with different dimensionless prefactors, we would be hard pressed to suppress all the relevant coefficients simultaneously, not least due to the presence of the gauge coupling $g$, whose value is dictated by unification constraints, and as such cannot be suppressed.  We thus conclude that the structure of Eq.~\eqref{eq:VEV-structure} itself needs to be kept under control. The possibility of a small $(\omega_{BL}+\omega_{R})$ turns out to be unviable for phenomenological reasons [unification through an intermediate flipped $\SU(5)$ is unattainable], so the smaller of the two $\omega$ VEVs must therefore be hierarchically smaller than the seesaw scale; i.e.,~it is merely an induced VEV. This implies one of the two mentioned intermediate symmetries must be realized.

The work is organized as follows: In Sec.~\ref{Sec:The Model}, we recapitulate the salient features of the model of interest, specify its field content and scalar potential, as well as recognize the possible breaking patterns. In Sec.~\ref{Sec:analysis}, we discuss at a conceptual level various theoretical constraints bounding the allowed parameter space --- non-tachyonicity of the scalar spectrum, perturbativity, and one-loop gauge-coupling unification. Preliminary analysis of the parameter space based on analytical considerations, the results of our numerical scans and the accompanying discussion are presented in Sec.~\ref{Sec:Results}. In Sec.~\ref{Sec:Conclusions}, we summarize our main conclusions and provide an outlook. All technical details related to the one-loop spectrum computation (including the resulting masses in both symmetry-breaking scenarios of interest), decomposition of the relevant $\mathrm{SO}(10)$ representations under intermediate-scale effective symmetry groups, detailed treatment of the parameter-space constraints, and running of gauge and scalar couplings (including their one-loop beta functions) are relegated to a set of Appendices.

%%%%%%%%%%%%%%%%%%%%%%%%%%%%%%%%%%%%%%%%%%%%%%%%%%%%%%%
\section{The $\mathbf{45}\;\oplus\; \mathbf{126}$ Higgs model\label{Sec:The Model}}
%%%%%%%%%%%%%%%%%%%%%%%%%%%%%%%%%%%%%%%%%%%%%%%%%%%%%%%

The minimal potentially realistic renormalizable Higgs model of our interest features a scalar sector transforming as $\mathbf{45}\oplus \mathbf{126}$ of the $\mathrm{SO}(10)$. In what follows, we  write the $\mathbf{45}$ as a set of real components $\phi_{ij}$, while the $\mathbf{126}$ is parametrized in terms of complex components $\Sigma_{ijklm}$, with latin indices running from $1$ to $10$. Both tensors are completely antisymmetric, and $\Sigma$ is a self-dual tensor; cf.~\cite{Graf:2016znk} for more details. The decompositions of these multiplets into their irreducible components with respect to several subgroups of $\mathrm{SO}(10)$ relevant for our analysis are given in Table~\ref{tab:decompositions} in Appendix~\ref{app:decompositions}. The complex-conjugate representation of $\Sigma$ is denoted by $\Sigma^*$.
The gauge fields (including those of the SM, as well as the extra components with leptoquark/diquark characteristics relevant for proton decay) are accommodated in the $45$-dimensional adjoint representation.

It is perhaps worth noting that a fully realistic symmetry-breaking pattern supporting the observed SM fermion spectrum at the renormalizable level requires at least one more scalar multiplet~\cite{Bajc:2005zf}, typically the $\mathbf{10}$ of $\mathrm{SO}(10)$. The electroweak (EW) VEVs carried by this representation, however, do not impact the high-scale symmetry breaking. Moreover, the one-loop effective-mass contributions coming from the $\mathbf{10}$ are subdominant due to the small dimensionality of the representation. Hence, we mostly ignore such an extra vector representation in the current analysis, since its absence typically makes little difference in the high-scale spectrum and the associated gauge unification constraints. Any possible implications are discussed later as the need arises. Note that the Higgs-doublet mass eigenstate of the SM in an extended scenario must live partly in the $\mathbf{10}$ and partly in the $\mathbf{126}$, which must be consistent with the doublet extended mass matrix, whose new columns and rows contain new scalar-potential parameters introduced by the extension.

%======================================================
\subsection{The classical-level setup\label{sec:generalSetup}}
%======================================================

%......................................................
\subsubsection{The Lagrangian}
%......................................................

Conforming to the notation of~\cite{Graf:2016znk}, the most general form of the Lagrangian in the unbroken phase can be written as $\mathcal{L} = \mathcal{L}_{kin} - V_0$, where the kinetic part is defined as
\begin{align}
\mathcal{L}_{kin} &= \frac{1}{4} \left(F_{\mu\nu}\right)_{ij} \left(F^{\mu\nu}\right)_{ij} 
+ \frac{1}{4}\left(D_\mu \phi_{ij}\right)^\ast
\left(D^\mu \phi_{ij}\right) %+\nonumber\\
%&\quad 
+\frac{1}{5!} \left(D_\mu \Sigma_{ijklm}\right)^\ast
\left(D^\mu \Sigma_{ijklm}\right)
\end{align}
for 
\begin{align}
\left(F^{\mu\nu}\right)_{ij} &= \partial^\mu A^\nu_{ij} - \partial^\nu A^\mu_{ij}-i \, g \left[A^\mu,A^\nu \right]_{ij},\\
D^{\mu}\phi_{ij}&=\partial^{\mu}\phi_{ij}-ig\,[A^\mu,\phi]_{ij},\\
D^{\mu}\Sigma_{ijklm}&=\partial^{\mu}\Sigma_{ijklm}-ig(A^\mu_{in}\Sigma_{njklm}+A^\mu_{jn}\Sigma_{inklm}%+\nonumber\\
%&\quad 
+A^\mu_{kn}\Sigma_{ijnlm}+A^\mu_{ln}\Sigma_{ijknm}+A^\mu_{mn}\Sigma_{ijkln}).
\end{align}
We use the definition $A_{\mu}:= A_{\mu}^{a}\,T^{a}$, where $T^{a}$ denotes the $\mathrm{SO}(10)$ generators in the representation $\mathbf{10}$. The fundamental (latin) indices refer to the real basis of the $\mathrm{SO}(10)$ vector $\mathbf{10}$, and they are always written in the lower position. Summation over repeated indices is implicitly assumed. 

The renormalizable tree-level scalar potential takes the form
\begin{align}
V_0(\phi, \Sigma,\Sigma^*) &= V_{45}(\phi) + V_{126} (\Sigma, \Sigma^*) + V_{mix}(\phi, \Sigma,\Sigma^*), \label{eq:tree-potential}
\end{align}
with
\begin{align}
V_{45} &= -\frac{\mu^2}{4}\left(\phi\phi\right)_0 + \frac{a_0}{4}\left(\phi\phi\right)_0\left(\phi\phi\right)_0 + \frac{a_2}{4} \left(\phi\phi\right)_2\left(\phi\phi\right)_2,\\
\begin{split}
V_{126} &= -\frac{\nu^2}{5!} \left(\Sigma\Sigma^*\right)_0 + \frac{\lambda_0}{(5!)^2} \left(\Sigma\Sigma^*\right)_0\left(\Sigma\Sigma^*\right)_0 + %\nonumber\\
%&\quad + 
\frac{\lambda_2}{(4!)^2}\left(\Sigma\Sigma^*\right)_2 \left(\Sigma \Sigma^*\right)_2 + \frac{\lambda_4}{(3!)^2(2!)^2}\left(\Sigma\Sigma^*\right)_4 \left(\Sigma \Sigma^*\right)_4 + %\nonumber
\\
&\quad + \frac{\lambda^{\prime}_4}{(3!)^2}\left(\Sigma\Sigma^*\right)_{4^{\prime}} \left(\Sigma \Sigma^*\right)_{4^{\prime}}+ \frac{\eta_2}{(4!)^2}\left(\Sigma\Sigma\right)_2 \left(\Sigma \Sigma\right)_2%+ \nonumber\\
%&\quad 
+ \frac{\eta^*_2}{(4!)^2}\left(\Sigma^*\Sigma^*\right)_2 \left(\Sigma^* \Sigma^*\right)_2,
\end{split}
\\
\begin{split}
V_{mix} &= \phantom{+} \frac{i\tau}{4!}\left(\phi\right)_2\left(\Sigma\Sigma^*\right)_2 + \frac{\alpha}{2\cdot 5!}\left(\phi \phi\right)_0\left(\Sigma\Sigma^*\right)_0 + %\nonumber\\
%&\quad + 
\frac{\beta_4}{4\cdot 3!}\left(\phi \phi\right)_4\left(\Sigma\Sigma^*\right)_4 + \frac{\beta^{\prime}_4}{3!}\left(\phi\phi\right)_{4^{\prime}}\left(\Sigma\Sigma^*\right)_{4^{\prime}} + %\nonumber
\\
&\quad + \frac{\gamma_2}{4!}\left(\phi \phi\right)_2\left(\Sigma\Sigma\right)_2 + \frac{\gamma^*_2}{4!}\left(\phi\phi\right)_2\left(\Sigma^*\Sigma^*\right)_2.
\end{split}
\end{align}
As usual, the following abbreviations are used:
\begingroup
\allowdisplaybreaks 
\begin{align}
\begin{split}
\left(\phi\phi\right)_0 &= \phi_{ij}\phi_{ij}, %\nonumber
\\
\left(\phi\phi\right)_2 &= \left(\phi\phi\right)_{jk} = \phi_{ij}\phi_{ik},
%\nonumber
\\
\left(\Sigma\Sigma^*\right)_0 &= \Sigma_{ijklm}\Sigma^*_{ijklm}, %\nonumber
\\
\ \left(\Sigma\Sigma^*\right)_2 &= \left(\Sigma\Sigma^*\right)_{mn} = \Sigma_{ijklm}\Sigma^*_{ijkln}, %\nonumber
\\
\left(\Sigma\Sigma^*\right)_4 &= \left(\Sigma\Sigma^*\right)_{lmno} = \Sigma_{ijklm}\Sigma^*_{ijkno}, %\nonumber
\\
\left(\phi\right)_2\left(\Sigma\Sigma^*\right)_{2} &= \phi_{mn}\left(\Sigma\Sigma^*\right)_{mn},%\nonumber
\\
\left(\phi\phi\right)_2\left(\Sigma\Sigma\right)_{2} &= \left(\phi\phi\right)_{jk}\left(\Sigma\Sigma\right)_{jk}, %\nonumber
\\
\left(\phi\phi\right)_4\left(\Sigma\Sigma^*\right)_{4} &= \phi_{lm}\phi_{no}\left(\Sigma\Sigma^*\right)_{lmno}, %\nonumber
\\
\left(\phi\phi\right)_{4^{\prime}}\left(\Sigma\Sigma^*\right)_{4^{\prime}} &= \phi_{lm}\phi_{no}\left(\Sigma\Sigma^*\right)_{lnmo}, %\nonumber
\\
\left(\Sigma\Sigma^*\right)_2\left(\Sigma\Sigma^*\right)_2 &= \left(\Sigma\Sigma^*\right)_{mn}\left(\Sigma\Sigma^*\right)_{mn}, %\nonumber
\\
\left(\Sigma\Sigma^*\right)_{4}\left(\Sigma\Sigma^*\right)_{4} &= \left(\Sigma\Sigma^*\right)_{lmno}\left(\Sigma\Sigma^*\right)_{lmno}, %\nonumber 
 \\
\left(\Sigma\Sigma^*\right)_{4^{\prime}}\left(\Sigma\Sigma^*\right)_{4^{\prime}} &= \left(\Sigma\Sigma^*\right)_{lmno}\left(\Sigma\Sigma^*\right)_{lnmo}.
\end{split}
\end{align}
\endgroup

The tree-level scalar potential contains $11$ dimensionless parameters: $9$ real couplings $\lbrace a_0$, $\! a_2$, $\!\lambda_0$, $\!\lambda_2$, $\!\lambda_4$, $\!\lambda_4^\prime$, $\!\alpha$, $\!\beta_4$, $\!\beta_4^\prime \rbrace$ and $2$ complex couplings $\lbrace \gamma_2, \eta_2 \rbrace$. Additionally, there are $3$ dimensionful parameters $\lbrace \mu, \nu, \tau \rbrace$ with the numerical coefficients in front of the corresponding terms in $V_0$ chosen such that the expressions $-\mu^2$ and $-\nu^2$ represent the mass squares of scalar fields in $\phi$ and $\Sigma$ in the unbroken phase, respectively.

%......................................................
\subsubsection{Field content }
%......................................................

For later convenience, we gather in Table~\ref{tab:particle-content} a list of all scalar fields in our $\mathbf{45}\oplus\mathbf{126}$ Higgs model in terms of their SM symmetry representations.  Alongside the representation type, we provide the information on whether each representation is real or complex ($\mathbb{R}/\mathbb{C}$), its multiplicity $\#$ in the model, as well as the $\mathrm{SO}(10)$ origins of each instance. 

Note that for each complex representation, we could have equivalently chosen its complex conjugate as the canonical label; our choices are purely conventional in this regard.  
In this paper, we shall denote mass eigenstates by the SM representation labels and add a numbered index when the state has a multiplicity greater than $1$. The value of the index increases with the mass eigenvalue. This labelling scheme will be convenient in our numerical analysis, since the masses can be computed explicitly and ranked for each parameter point.

\def\PG{{\phantom{\dagger}}}
\begin{table}[ht]
\caption{Field content of the $\mathbf{45}\oplus\mathbf{126}$ Higgs model. Each SM representation $R$ has its reality / complexity, multiplicity $\#$, and $\mathrm{SO}(10)$ origins indicated. A dagger $(\dagger)$ indicates the presence of a massless would-be Goldstone mode.
\label{tab:particle-content}}
\begin{tabular}{lcr@{$\quad$}l}
\toprule
\makebox[1.5cm][l]{$R\sim G_{321}$}
&$\mathbb{R}/\mathbb{C}$
&\makebox[0.5cm][r]{$\#$}
&$\subseteq SO(10)$\\
\midrule
$(1,1,0)$			&  $\mathbb{R}$  &  $4^\dagger$ &  $\phi$, $\phi$, $\Sigma$, $\Sigma^{\ast}$ \\
$(1,1,+1)$			&  $\mathbb{C}$  &  $2^\dagger$ &  $\phi$, $\Sigma$ \\
$(1,1,+2)$			&  $\mathbb{C}$  &  $1^\PG$  	&  $\Sigma$ \\
$(1,2,+\tfrac{1}{2})$		&  $\mathbb{C}$  &  $2^\PG$  	&  $\Sigma$, $\Sigma^\ast$ \\
$(1,3,-1)$			&  $\mathbb{C}$  &  $1^\PG$  	&  $\Sigma$ \\
$(1,3,0)$			&  $\mathbb{R}$  &  $1^\PG$  	&  $\phi$ \\
$(3,2,-\tfrac{5}{6})$		&  $\mathbb{C}$  &  $1^\dagger$ &  $\phi$ \\
$(3,2,+\tfrac{1}{6})$		&  $\mathbb{C}$  &  $3^\dagger$ &  $\phi$, $\Sigma$, $\Sigma^\ast$ \\
$(3,2,+\frac{7}{6})$		&  $\mathbb{C}$  &  $2^\PG$  	&  $\Sigma$, $\Sigma^\ast$ \\
$(3,3,-\tfrac{1}{3})$		&  $\mathbb{C}$  &  $1^\PG$  	&  $\Sigma$ \\
$(\bar{3},1,-\frac{2}{3})$	&  $\mathbb{C}$  &  $2^\dagger$ &  $\phi$, $\Sigma$ \\
$(\bar{3},1,+\tfrac{1}{3})$	&  $\mathbb{C}$  &  $3^\PG$  	&  $\Sigma$, $\Sigma$, $\Sigma^\ast$ \\
$(\bar{3},1,+\tfrac{4}{3})$	&  $\mathbb{C}$  &  $1^\PG$  	&  $\Sigma$ \\
$(6,3,+\tfrac{1}{3})$		&  $\mathbb{C}$  &  $1^\PG$  	&  $\Sigma$ \\
$(\bar{6},1,-\tfrac{4}{3})$	&  $\mathbb{C}$  &  $1^\PG$  	&  $\Sigma$ \\
$(\bar{6},1,-\tfrac{1}{3})$	&  $\mathbb{C}$  &  $1^\PG$  	&  $\Sigma$ \\
$(\bar{6},1,+\tfrac{2}{3})$	&  $\mathbb{C}$  &  $1^\PG$  	&  $\Sigma$ \\
$(8,1,0)$			&  $\mathbb{R}$  &  $1^\PG$  	&  $\phi$ \\
$(8,2,+\frac{1}{2})$		&  $\mathbb{C}$  &  $2^\PG$  	&  $\Sigma$, $\Sigma^\ast$ \\
\bottomrule
\end{tabular}
\end{table}
%

%......................................................
\subsubsection{Symmetry breaking and VEVs}
%......................................................

The scalar spectrum contains three SM singlets: two real singlets residing in $\phi$ and one complex in $\Sigma$, for a total of $4$ real SM-singlet degrees of freedom; cf.~Table~\ref{tab:particle-content}. Their vacuum expectation values
are parametrized as
\begin{align}
\begin{split}
	\langle (1,1,1,0)_\phi \rangle &\equiv \sqrt{3} \; \omega_{BL},\\
	\langle (1,1,3,0)_\phi \rangle &\equiv \sqrt{2} \, \omega_{R},\\
	\langle (1,1,3,+2)_\Sigma \rangle &\equiv \sqrt{2} \, \sigma.\\
\end{split} \label{eq:VEV-definitions}
\end{align}
For unambiguous identification of these states, we referred to their \hbox{\LR}  transformation properties. 

The values $\omega_{BL}$ and $\omega_{R}$ are real because $\phi$ is a real representation, while $\sigma$ is, in general, complex. Since the overall phase of $\Sigma$ can be redefined without loss of generality, $\sigma$ can be taken real and positive. Although this freedom is utilized in our scans of the parameter space, we retain a notation consistent with complex $\sigma$ in all analytical expressions. 

Because of phenomenological requirements (gauge-coupling unification and a need for a seesaw scale), the GUT symmetry is assumed to be broken spontaneously in two stages. At the unification scale $M_{\text{GUT}}$, the VEVs in $\phi$ ($\omega_{BL}$ and $\omega_{R}$) break $\mathrm{SO}(10)$ down to one of its subgroups of rank five. The subsequent breaking to the SM, preferably well below $M_{\text{GUT}}$, is then accomplished by the rank-reducing VEV of $\Sigma$ ($\sigma$) which is identified with the seesaw scale. Breaking patterns associated with various VEV directions are summarized in Table~\ref{tab:BreakingSchemes}. 

\begin{table*}[htb]
\caption{
\label{tab:BreakingSchemes}
Residual gauge symmetries (in self-explanatory notation) for various VEV configurations. The $5'\,1_{Z'}$ refers to an intermediate flipped-$\mathrm{SU}(5)$ stage~\cite{Barr:1981qv,Derendinger:1983aj}, while in the last column, the $\mathrm{SU}(5)$ symmetry remains unbroken due to the $\mathrm{SU}(5)$-singlet nature of $\sigma$.}
\begin{tabular}{c@{$\quad$}c@{$\quad$}c@{$\quad$}c@{$\quad$}c@{$\quad$}c} \hline
	& $\omega_{BL } \neq 0, \omega_{R } \neq 0$& $\omega_{BL } = 0, \omega_{R } \neq 0$&$\omega_{R } = 0, \omega_{BL } \neq0$ & $\omega_{BL } = -\omega_{R } \neq 0$&$\omega_{BL } = \omega_{R } \neq 0$\\\hline
	$\sigma = 0$ & $3_c\,2_L\,1_R\,1_{B-L}$&$4_C\,2_L\,1_R$&$3_c\,2_L\,2_R\,1_{B-L}$&$5'\,1_{Z'}$&$5\,1_Z$\\
	$\sigma \neq 0$ & $3_c\,2_L\,1_Y$&$3_c\,2_L\,1_Y$&$3_c\,2_L\,1_Y$&$3_c\,2_L\,1_Y$&$5$
	\\ \hline
\bf\end{tabular}
\end{table*}
%

%......................................................
\subsubsection{The classical vacuum structure and pseudo-Goldstone modes\label{sec:PSG-singlets}}
%......................................................

The three mass parameters $\lbrace \mu, \nu, \tau \rbrace$ are connected to the three VEVs $\{\omega_{R},\omega_{BL},\sigma \}$ by the vacuum stationarity conditions\footnote{Note that for special configurations of VEVs the number of non-trivial relations can be reduced. For instance, only two independent conditions exist in the $\omega_{BL} = \omega_{R}$ case. Thus, one of the $\lbrace \mu, \nu, \tau \rbrace$ parameters remains unspecified [meaning that stable points of the scalar potential with $\mathrm{SU}(5)$ symmetry exist for any possible value of this parameter].}, which take the following form at tree level:
\begin{align}
     \mu^2 & = (12 a_0 + 2 a_2) \omega_{BL}^2 + (8 a_0 + 2 a_2) \omega_{R}^2 
    + 2 a_2 \omega_{BL} \omega_{R} + 4 (\alpha + \beta_4') |\sigma|^2 , \label{eq:mu} \EQSPACE
    \nu^2 & = 3 (\alpha + 4 \beta_4') \omega_{BL}^2 + 2 (\alpha + 3 \beta_4') \omega_{R}^2 
    + 12 \beta_4' \omega_{BL} \omega_{R} + 4 \lambda_0 |\sigma|^2  + a_2 \frac{\omega_{BL} \omega_{R}}{|\sigma|^2} (\omega_{BL} + \omega_{R}) (3 \omega_{BL} + 2 \omega_{R}),
    \label{eq:nu} \EQSPACE
    \tau & = 2 \beta_4' (3 \omega_{BL} + 2 \omega_{R}) + a_2 \frac{\omega_{BL} \omega_{R}}{|\sigma|^2} (\omega_{BL} + \omega_{R}) .\label{eq:tau}
\end{align}
Notice the presence of the VEV structure of Eq.~\eqref{eq:VEV-structure} in both $\nu^2$ and $\tau$. For later convenience, we define $\chi$ as the dimensionless \emph{universal ratio} of VEVs present in that structure:
\begin{align}
    \chi:= \frac{\omega_{BL} \omega_{R}}{|\sigma|^2}. \label{eq:chi-definition}
\end{align}

Given the relations of Eqs.~\eqref{eq:mu}--\eqref{eq:tau}, the VEVs and dimensionless scalar couplings can be taken as independent input parameters that fully determine the (tree-level) scalar and gauge spectra. The key observation made in~\cite{Yasue:1980fy, Anastaze:1983zk} was that the tree-level masses of scalars transforming as $(1,3,0)$ and $(8,1,0)$ under the SM group take the simple form
%
%\begin{align}
%M^2_S(1,3,0) =~& 2 a_2 (\omega_{R} - \omega_{BL})(\omega_{BL} + 2 \, \omega_{R}) , \label{eq:tripletmass} \\
%M^2_S(8,1,0) =~& 2 a_2 (\omega_{BL} - \omega_{R})(\omega_{R} + 2 \, \omega_{BL}) , \label{eq:octetmass}
%\end{align}
%
\begin{eqnarray}
M^2_S(1,3,0) & = 2 a_2 (\omega_{R} - \omega_{BL})(\omega_{BL} + 2 \, \omega_{R}) , \label{eq:tripletmass} \\
M^2_S(8,1,0) & = 2 a_2 (\omega_{BL} - \omega_{R})(\omega_{R} + 2 \, \omega_{BL}) , \label{eq:octetmass}
\end{eqnarray}
and can thus be simultaneously made non-tachyonic if and only if
\begin{align}\label{eq:flippedSU5like}
a_2 >0\quad  \text{and}\quad  -2<\frac{\omega_{BL}}{\omega_{R}}<-\frac{1}{2},
\end{align}
i.e.,~in the vicinity of the intermediate flipped-$\mathrm{SU}(5)$ configuration; cf.~Table~\ref{tab:BreakingSchemes}. This, however, triggers the usual issues with gauge unification and/or baryon number violation --- either one respects the proton lifetime limits and breaks the residual flipped-$\mathrm{SU}(5)$ immediately by lifting $\sigma$ to the vicinity of $M_{\text{GUT}}$ (which corresponds to the problematic one-stage symmetry-breaking pattern), or one postpones the flipped-$\mathrm{SU}(5)$ symmetry breaking and faces light $(3,2,+\tfrac{1}{6})$ gauge leptoquarks in the spectrum with all the implications for matter instability. In either case, the phenomenological constraints are practically impossible to meet. The model has thus been discarded as non-viable and it took almost $30$ years to bring it back from oblivion by invoking radiative effects~\cite{Bertolini:2009es}: for small $a_2$, these may remedy the tree-level tachyonicity of Eqs.~\eqref{eq:tripletmass} and/or~\eqref{eq:octetmass} along the potentially viable breaking chains well outside the (near)flipped-$\mathrm{SU}(5)$ region of Eq.~\eqref{eq:flippedSU5like}.

Remarkably enough, only recently, another tachyonic instability was revealed in the tree-level mass matrix of the SM singlets~\cite{Graf:2016znk} assuming the seesaw-compatible regime $|\sigma| \ll \max[|\omega_{BL }|,|\omega_{R }|]$. 
In the $\sigma\to 0$ limit\footnote{This limit needs to be taken carefully due to the appearance of the $\chi$-structure in the trilinear scalar coupling $\tau$ of Eq.~\eqref{eq:tau}. One assumes $a_2$ or the smaller of the two $\omega$ scales to be taken to zero alongside $\sigma$ in such a way that $\tau$ is kept fixed and sub-Planckian, i.e.,~under perturbative control; cf.~Sec.~\ref{sec:quantum-level}.}, one of the masses of the physical SM-singlet scalars takes the form 
\begin{widetext}
\begin{align}
\begin{split}
M^2_S(1,1,0)_{\text{PGB}} & = 
  4 a_0 \left(3 \omega_{BL}^2 + 2 \omega_{R}^2\right) \left (1-\sqrt{1+\left (\tfrac{a_2}{a_0}\right)\tfrac{3 (3 \omega_{BL}^2 - 2 \omega_{R}^2) (\omega_{BL}^2 - \omega_{R}^2)}{2 \, (3 \omega_{BL}^2 + 2 \omega_{R}^2)^2}+\left(\tfrac{a_2}{a_0}\right )^2\tfrac{9 \, (\omega_{BL}^2 - \omega_{R}^2)^2}{16 \, (3 \omega_{BL}^2 + 2 \omega_{R}^2)^2}} \right )+%\nonumber
  \\
 &\quad + a_2 (\omega_{BL} - \omega_{R})^2 .\label{eq:singlet-PSG}
\end{split}
\end{align}
\end{widetext}
We refer to this state as the pseudo-Goldstone boson singlet. A companion SM singlet to the PGB singlet has the same mass expression as that of Eq.~\eqref{eq:singlet-PSG}, except for changing the sign in front of the square root, while the remaining two SM singlets are true would-be Goldstone boson (WGB) modes\footnote{More precisely, one is a true WGB and one is a $B-L$ breaking Higgs field whose mass is proportional to $|\sigma|^2$ to all orders in the perturbative expansion; see~\cite{Hudec:2019rfi}.} when $\sigma\to 0$. In the well-motivated $|a_2|\ll |a_0|$ limit\footnote{The tree-level triplet and octet PGB masses are proportional to $a_2$, so taking $a_2$ small enables loop corrections to overwhelm them and cure their tree-level tachyonic instability.}, the PGB-singlet mass can be expanded as
\begin{align}
M^2_S(1,1,0)_{\text{PGB}} & \approx a_2\;\Big(-\frac{45 \, \omega_{BL}^4}{3 \, \omega_{BL}^2 + 2 \, \omega_{R}^2} + 13 \, \omega_{BL}^2 %- \nonumber\\
%&\quad 
-2 \, \omega_{BL} \omega_{R} - 2 \, \omega_{R}^2\Big) + {\cal O}(a_2^2/a_0^2). \label{eq:singletmass}    
\end{align}
In the same $|a_2|\ll |a_0|$ limit, the companion singlet has the mass $8 a_0 \left(3 \omega_{BL}^2 + 2 \omega_{R}^2\right) + \mathcal{O}(a_2)$ controlled by the $a_0$ parameter, providing further justification of the limit \emph{a posteriori}. Incidentally, the opposite $|a_2| \gg |a_0|$ regime leads to the masses of the two physical singlets [based on Eq.~\eqref{eq:singlet-PSG}] equal to
the PGB triplet and octet masses of Eqs.~\eqref{eq:tripletmass} and~\eqref{eq:octetmass}, respectively.

The expression of Eq.~\eqref{eq:singletmass} is positive only for
$\omega_{BL} \simeq -\omega_R$ assuming $a_2>0$.
Hence, one reveals again that the tachyonic instabilities are absent from the tree-level mass spectrum only in the vicinity of the phenomenologically problematic flipped-$\mathrm{SU}(5)$-breaking direction.

%======================================================
\subsection{The quantum-level situation \label{sec:quantum-level}}
%======================================================

As argued in Sec.~\ref{sec:generalSetup}, potentially viable scenarios require 
dealing with \emph{three} rather than the two previously identified instabilities in the scalar spectrum. Since the SM-singlet PGB mass is buried within a $4\times 4$ mass matrix, a far more elaborate account of radiative corrections to the scalar spectrum of the model is required than that available in the existing literature~\cite{Graf:2016znk}. Hence, for the purposes of this study, we have developed a numerical code that calculates the one-loop quantum corrections to \emph{all} scalars of the model, i.e.,~including the modes that should not suffer from any issues inherent to the pseudo-Goldstone nature of the three culprits of Eqs.~\eqref{eq:tripletmass}, \eqref{eq:octetmass}, and~\eqref{eq:singlet-PSG}. Even though the quantum effects should not significantly affect the heavy non-tachyonic part of the tree-level spectrum, this additional information enables a comprehensive analysis of perturbativity, a feature that is seldom addressed in the existing GUT literature.

The first perturbativity issue to be addressed is the potentially large terms with the universal ratio $\chi$ defined in Eq.~\eqref{eq:chi-definition}. Remarkably, this structure pops up not only in the tree-level vacuum conditions of Eqs.~\eqref{eq:mu}--\eqref{eq:tau} (and by extension in the tree-level spectrum), but independently also at the level of quantum corrections.

At tree level, its tendency to diverge in various limits can be compensated by taking the accompanying $a_2$ parameter appropriately small (which in addition helps to keep the tachyonic instabilities of PGB states under control). All symmetry-breaking patterns identified in Table~\ref{tab:BreakingSchemes} can therefore be, at least in principle, consistently attained.

The situation changes dramatically at the loop level where the same $\chi$-structure appears in the (polynomial part of the) one-loop stationarity conditions~\cite{Graf:2016znk} but with parameters other than $a_2$ present in the prefactors. It is difficult to keep all these contributions simultaneously under control merely by suppressing the value of relevant scalar parameters due to the presence of the (relatively large) gauge coupling $g$ among them. The only way to retain control over such $\chi$-terms is by (i) sticking to small fine-tuned patches of the parameter space where the scalar couplings just cancel the effects of $g$ and/or (ii) keeping the universal ratio $\chi$ itself under control by pushing the VEVs into several ``prophylactic'' corners of the parameter space.

%......................................................
\subsubsection{Landau poles in scalar couplings}
%......................................................

To this end, case (i) is generally difficult to achieve because cancellations of the gauge coupling effects necessarily invoke relatively large scalar couplings. This, unfortunately, brings in another aspect of the overall perturbativity issue, namely, the potential proximity of the scalar-sector Landau pole(s) to $M_{\text{GUT}}$. In order to address this, we have derived one-loop beta functions for all scalar couplings at play and used them to look for and inspect the regions where the scalar couplings are stable enough to support such a regime. Remarkably, these constraints turn out to be extremely powerful in excluding large patches of the parameter space that were formerly thought to be viable; cf.~Sec.~\ref{sec:Perturbativity}.

%......................................................
\subsubsection{Perturbative VEV configurations}
%......................................................

Consequently, one can expect that at the quantum level the omnipresent factor $\chi$ will have to be dealt with along the lines of option (ii) above. Hence, in what follows, we shall require
\begin{align}
 \frac{|\omega_{BL} \omega_{R} (\omega_{BL} + \omega_{R})|}{|\sigma|^2} & \lesssim \max\left[|\omega_{BL}|,|\omega_{R}|\right] , \label{eq:VEVConstraint}
\end{align}
which confines the viable VEV configurations to four distinct classes corresponding to four different breaking patterns in Table~\ref{tab:BreakingSchemes}:
\begin{enumerate}
	\item $|\sigma| \approx \max[|\omega_{BL }|,|\omega_{R }|]$ corresponding to approximate single-stage spontaneous symmetry breaking $\mathrm{SO}(10)\to$ \SM
	\label{class1}
	\item $\omega_{BL } \approx - \omega_{R }$ with a flipped-$\mathrm{SU}(5)$ intermediate-symmetry stage
	\label{class2}
	\item$|\omega_{BL}| \ll |\sigma| \ll |\omega_{R }|$ with \BLzero intermediate-symmetry stage 
	\label{class3}
	\item $|\omega_{R}| \ll |\sigma| \ll |\omega_{BL}|$ with \LR intermediate-symmetry stage
	\label{class4}
\end{enumerate}
As already mentioned, the first two options are strongly phenomenologically disfavoured either by gauge unification constraints or by proton longevity. Hence, we shall focus predominantly on the latter two scenarios and present the results obtained in the corresponding $\omega_{BL}\to 0$ and $\omega_{R}\to 0$ limits. From them, it will become evident that the $\omega_{R}\to 0$ case is disfavoured in several aspects. Therefore, a clear quantum-level preference for the symmetry-breaking chains passing through a well-pronounced intermediate \BLzero symmetry stage can be identified.

%%%%%%%%%%%%%%%%%%%%%%%%%%%%%%%%%%%%%%%%%%%%%%%%%%%%%%%
\section{Analysis of constraints\label{Sec:analysis}}
%%%%%%%%%%%%%%%%%%%%%%%%%%%%%%%%%%%%%%%%%%%%%%%%%%%%%%%

In this section, we provide a systematic account of the different types of constraints that a consistent GUT theory amenable to a perturbative expansion must satisfy. We start with those for which rigorous criteria can be implemented more easily (non-tachyonicity of the scalar spectrum and gauge-coupling unification) and follow with those requiring a subjective choice of the used criterion (perturbativity). 

The constraints of this section can be considered for any given parameter point of the theory. The goal ultimately is to identify the parameter-space region(s) which pass all the viability criteria. This analysis is carried out later in Sec.~\ref{Sec:Results}.
To facilitate the readability of the paper and streamline the main text to reach our numerical results quicker, we discuss in this section the constraints used later only at a conceptual level. The interested reader is kindly referred to Appendices~\ref{App:one-loop masses} and~\ref{App:viability-considerations} for technical details of the implementation of the criteria presented in this section.

%======================================================
\subsection{Non-tachyonicity of the scalar spectrum\label{sec:Tachyonicity}}
%======================================================

A consistent broken-phase perturbative expansion is developed around the true vacuum, i.e.,~around a (possibly local) minimum of the scalar potential at which all physical scalar masses are non-negative. Hence, a parameter point is not considered viable if some of the masses are found to be tachyonic.

To this end, we provide in Appendix~\ref{App:one-loop masses} a detailed description of the procedure used to calculate the one-loop scalar spectrum of the model in any given parameter point. Two conceptual considerations are important to note here:
\begin{enumerate}
    \item The masses of the fields associated with the $\mathrm{U}(1)_{B-L}$-breaking scale are proportional to the $|\sigma|$ VEV; i.e.,~they naturally live at the intermediate (seesaw) scale rather than in the vicinity of the GUT scale.\footnote{Scalar fields with $|\sigma|$-proportional masses belong to the same \BLzero (for $\omega_{BL} \to 0$) or \LR (for $\omega_{R} \to 0$) representation as the SM-singlet Higgs field which breaks the $\mathrm{U}(1)_{B-L}$ symmetry.} Because of symmetry reasons, not only their tree-level masses but even the corresponding loop corrections are $|\sigma|$-proportional~\cite{Hudec:2019rfi}. Since, conceptually, these should be computed in the effective field theory at the intermediate scale --- either in the \BLzero model (for $\omega_{BL} \to 0$) or in the \LR theory (for $\omega_R \to 0$), where both of these setups contain a significantly smaller number of degrees of freedom than the full $\mathrm{SO}(10)$ Higgs model --- the seesaw-scale fields are expected to receive rather small loop corrections for any values of couplings in the perturbative regime.  
    We thus simplify the analysis by taking only the tree-level expressions for the $|\sigma|$-proportional part of the spectrum. %
    \item In the minimal realistic scenario, the scalar SM multiplets $(\xbar{3},1,+\tfrac{1}{3})$ and $(1,2,+\tfrac{1}{2})$ are eventually mixed with their counterparts from additional scalar $\mathbf{10}$'s of $\mathrm{SO}(10)$. Although one can neglect its impact in almost all aspects of our analysis,\footnote{Note that adding a $\mathbf{10}$ also introduces new terms in the scalar potential, which generate further one-loop corrections to the original $\mathbf{45}\oplus\mathbf{126}$ states. However, the number of additional fields appearing in loops is small and their loop contributions can be neglected.} the triplet and doublet mass matrices in the model without the $\mathbf{10}$ should be treated only as subparts of larger structures. Nevertheless, it is still possible to formulate a necessary condition for the non-tachyonicity of these states even with just partial information of the complete mass matrices in this sector.
    
    According to Sylvester's criterion~\cite{PrussingSylvester}, a Hermitian matrix $\mathcal{M}^2$ is positive definite (it has positive eigenvalues) if and only if all its \emph{leading principal minors} are positive, i.e.,~if the determinants of all upper-left submatrices of $\mathcal{M}^2$ (its upper-left $1\times 1$, $2 \times 2$, $3\times 3$,~$\ldots$ blocks up to $\mathcal{M}^2$ itself) are positive.\footnote{An analogue of Sylvester's criterion for positive semi-definite matrices requires all \emph{principal minors} of $\mathcal{M}^2$ to be non-negative. The $(1,2,+\tfrac{1}{2})$ mass matrix in the full theory could be regarded as positive semi-definite, since the electroweak Higgs mass eigenvalue therein can be considered as effectively zero. This, however, is the only vanishing eigenvalue, and it can be imposed by performing a fine-tuning (outside the Higgs-model block) of one of the new scalar couplings associated with the additional $\mathbf{10}$. The here considered doublet block is thus strictly positive definite, and hence, the same positive definiteness criterion is used as for the triplet block.} Since the $\mathbf{10}$ introduces no new SM-singlet VEV, the mass matrices of $(\xbar{3},1,+\tfrac{1}{3})$ and $(1,2,+\tfrac{1}{2})$ used here represent upper-left blocks of the corresponding full mass matrices in the extended models. The positivity of all their leading principal minors (and thus of the eigenvalues presented in Table~\ref{tab:tree-level-scalar-masses}) forms a necessary condition for the non-tachyonicity of the $(\xbar{3},1,+\tfrac{1}{3})$ and $(1,2,+\tfrac{1}{2})$ SM multiplets within.
\end{enumerate}
%

%======================================================
\subsection{Gauge unification constraints\label{sec:unification-constraints}}
%======================================================

Consistency of a GUT model demands that the SM gauge couplings unify at some high scale, starting with their measured EW-scale values. Since our $\mathrm{SO}(10)$ Higgs model is envisioned to be employed as a subsector of such a realistic GUT model, we include gauge unification as a viability criterion for our points.

We perform the unification test top down, i.e.,~starting with the value of the unified gauge coupling $g$. We use the renormalization group equation (RGE) to compute the EW-scale values of the SM gauge couplings and accept only points whose coupling values 
match the experimental ones.

Technically, the computation is done using standard techniques; cf.~Appendix~\ref{app:unification}.
Although a reasonable account of the relevant BLNV phenomenology (such as proton lifetime calculations) in potentially realistic models requires a detailed two-loop gauge running analysis (such as~\cite{Kolesova:2014mfa}),
a one-loop approximation is sufficient for the purposes of this Higgs-model study.

Despite considering a $2$-stage breaking $\mathrm{SO}(10)\to G \to \mathrm{SU}(3)_c \times \mathrm{SU}(2)_L \times \mathrm{U}(1)_Y$, with the $2$ scenarios of interest having \BLzero and \LR intermediate symmetries for $G$, the one-loop analysis can be performed
in the SM effective theory, provided we know the spectrum.

We therefore need to consider how best to mimic the spectrum in a realistic model, e.g.,~an extension with an extra $\mathbf{10}$ scalar representation. The only non-trivial issue arises for SM representation types introduced by the extension: the doublets $(1,2,+\tfrac{1}{2})$ and triplets $(\bar{3},1,+\tfrac{1}{3})$. Recall that in the current Higgs-model setting we have access only to incomplete doublet and triplet mass matrices of the realistic theory. The relevant masses (as inputs to the RGE analysis) thus cannot be fully determined, yet we can still make use of the (positive) eigenvalues $M_{S}^{2}(1,2,+\tfrac{1}{2})_{1,2}$  and $M^{2}_{S}(\xbar{3},1,+\tfrac{1}{3})_{1,2,3}$ as computed in the Higgs model; cf.~Table~\ref{tab:tree-level-scalar-masses} in the Appendix. The best approximation to the realistic case involves the following $2$ considerations:

\begin{enumerate}
    \item As one of the doublets plays the role of the light SM Higgs doublet, it should be removed from the heavy RGE-contributing scalar spectrum. However, there is no point in imposing the corresponding fine-tuning on either of the $M_{S}^{2}(1,2,+\tfrac{1}{2})_{1,2}$ eigenvalues of the \emph{incomplete} doublet mass matrix. What we instead do is to model their effect in the full setting by taking into account only one copy of a doublet (not two) and assign it a mass corresponding to the geometric mean of the two $M_{S}^{2}(1,2,+\tfrac{1}{2})_{1,2}$.
    As explained above, the other doublet is taken at the EW scale.
    \item There exist some ambiguities related to the possible admixture of additional doublet and triplet fields from the extra $\mathbf{10}$'s into the physical mass eigenstates in models with a fully realistic Yukawa sector.
    If these additional multiplets came exactly degenerate in mass at around  $M_{\text{GUT}}$ [i.e.,~as complete $\mathrm{SO}(10)$ multiplets], they would inflict no change at all to the position of the unification scale and only a very small (practically irrelevant) shift to the value of $g$. In the realistic case, the new doublets and triplets from the $\mathbf{10}$'s mix with the old ones; hence, they are not exactly degenerate, and even a shift in the doublet and triplet Higgs-model eigenvalues is induced. However, the net effect on $M_{\text{GUT}}$ and $g$ is still expected to be small for at least two reasons. First, the 
    beta-function contributions of these new scalar states (both of them in the vector representations of their associate gauge factors) are minute, and thus the corresponding changes to the  renormalization group (RG) running are generically subleading. Second, as all the heavy doublets and triplets are clustered around the GUT scale, the interval of scales between which the running is non-trivial (corresponding to the mass differences between the heavy doublets and triplets) is very short. Hence, in most cases the associated uncertainties in the RG evolution are negligible, and we shall not consider the effects of the extra $\mathbf{10}$'s here. To summarize, we use the computed spectrum of the triplets from the Higgs model, while the treatment of doublets was described in the previous point.
\end{enumerate}
%

%======================================================
\subsection{Perturbativity aspects\label{sec:Perturbativity}}
%======================================================

Since all calculations in the model rely on perturbative methods, some type of perturbativity test needs to be performed to check for their self-consistency.  
The loss of order-by-order robustness in a perturbative calculation can manifest itself in different ways, so we consider a number of different perturbativity constraints. Needless to say, their definitions are typically subject to some arbitrariness, so we shall often test them at different levels of strictness (producing different datasets).

We conceptually discuss the considered perturbativity criteria one at a time in the numbered subsections below. The technical details of their implementation are found in Appendix~\ref{app:perturbativity}.

%......................................................
\subsubsection{The global-mass-perturbativity test\label{sec:massPerturbativity}}
%......................................................

The first obvious restriction that we impose concerns the relative size of the one-loop shifts to the tree-level scalar masses. As simple as it sounds, it is not necessarily trivial in practice for at least two reasons:
\begin{itemize}
    \item There are accidentally light pseudo-Goldstone modes in the tree-level scalar spectrum for which a large one-loop shift is not only admissible but, in most cases, even mandatory; cf.~Sec.~\ref{sec:generalSetup}. Thus, we should exclude the relative shifts to these fields' masses from the assessment. In practice, the one-loop scalar-mass correction largest in magnitude is compared to the average of the \emph{heavy} tree-level masses.\footnote{The \emph{heavy} (tree-level) masses are those scalar masses that are not $|\sigma|$-proportional and do not belong among the would-be Goldstone bosons or the pseudo-Goldstone bosons.}
    \item The relatively simple effective potential methods that we use for the computation of the leading quantum corrections to the scalar masses do not, in fact, provide the fully physical one-loop masses but rather their counterparts calculated in one of the unphysical schemes such as $\overline{\text{MS}}$. These, however, suffer from several drawbacks such as sensitivity to potentially large IR or UV logs and residual renormalization-scale dependence. As for the former, we work with the regularized one-loop effective-mass spectrum [see Appendix~\ref{App:one-loop masses}, Eq.~\eqref{eq:RegMassDef}] which, in the current situation, is perhaps the closest attainable approximation to the actual physical spectrum. However, even in such a case there is a residual renormalization-scale dependence that should be kept under control.
\end{itemize}
Considering the above, we define the quantity $\overline{\Delta}$, which represents an overall measure 
of mass shifts:
\begin{align}
    \overline{\Delta} &:= \frac{\max_{i,j \in \text{\emph{heavy} fields}}[|M^2_{ij, \text{one-loop}}-M^2_{ij,\text{tree}}|]}{\overline{M^2}_{\rm heavy}}. \label{eq:definition-delta}
\end{align}
    This quantity effectively compares the largest one-loop correction in the \emph{heavy} fields' masses to their average; see Appendix~\ref{app:perturbativity} for further details. A necessary condition for perturbativity can be imposed by only accepting parameter points with $\overline{\Delta}$ below a chosen threshold. 

%......................................................
\subsubsection{Renormalization-scale dependence and stability under the RG running\label{sec:couplingsStability}}
%......................................................

Since the earlier ``global-mass-perturbativity'' test is not entirely renormalization-scale independent, we need to ensure that the computed one-loop scalar masses are kept under control under a change of renormalization scale. Note that this issue can be rather severe in the busy environment of grand unified models with typically many degrees of freedom ``flying around'' the loops. Technically, such pathologies exhibit themselves as Landau-pole instabilities in the RG flows which, at the given level of perturbative expansion, can be studied in terms of the corresponding beta functions. To this end, the complete system of the one-loop beta functions for dimensionless scalar couplings has been derived (see Appendix~\ref{App:BetaFunctions}) and used as a basis for the study of RGE stability of the scalar-mass spectrum. 

In this context, we label the initial renormalization scale by $\mu_{R}$, the upper and lower scales where the RGE system blows up by $\mu_{R+}$ and $\mu_{R-}$, respectively, and define the useful perturbativity measures
\begin{align}
    t_\pm &:= \log_{10} \frac{\mu_{R\pm}}{\mu_R},\label{eq:definition-tplusminus}\\
     \overline{t}&:= \sqrt{t_- t_+}. \label{eq:definition-tbar}
\end{align}
The quantity $t_{+}$ ($t_{-}$) tells us how many orders of magnitude above (below) the initial scale $\mu_{R}$ the theory can be run in its full form (i.e.,~with no degrees of freedom integrated out), while $\bar{t}$ as the geometric mean tells us the average amount of allowed running up or down. Further details are given in Appendix~\ref{app:perturbativity}.

Note that checking RG stability above the Planck scale is physically not required, and a switch to an effective theory should be performed below the scale where most of the spectrum lies. Nevertheless, persistence of perturbativity under RGE demonstrates numerical robustness of the calculation. RG stability can be checked by imposing a minimum threshold of $t_{\pm}$ or $\bar{t}$ for viable points.

%......................................................
\subsubsection{Vacuum position stability\label{sec:vacuumStability}}
%......................................................

Another aspect of perturbativity, though perhaps even more arbitrary than the two discussed so far, concerns the stability of the location of the broken-phase-theory vacuum in the VEV space. On one hand, it deals with quantities which do not have a clear physical interpretation unlike masses or couplings\footnote{Moreover, the criterion even depends on the rescaling of unphysical parameters $\lbrace \mu, \nu, \tau \rbrace$, which have a ``natural'' normalization chosen so that $-\mu^2$ and $-\nu^2$ are exactly the masses of the scalar fields in the unbroken phase.} but, on the other hand, it is still quite intuitive and can be seen as a one-point complement to the two- and four-point Green's functions' constraints above.  
Since the vacuum position is used in the computation of one-loop scalar masses, a big shift in vacuum typically causes also a large numerical shift in the masses.
Technically, the requirement that the  position of the one-loop vacuum in the VEV space should not be ``too far'' from the tree-level one is also one of the easiest conditions to test in practice (the vacuum position is determined by one-loop stationarity condition expressions that represent only a  subset of those that enter the mass corrections) and, as such, it can be used as a fast first perturbativity check. The reader is referred to Appendix~\ref{app:perturbativity} for details of implementation.

%......................................................
\subsubsection{Iterative pseudo-Goldstone masses}
%......................................................

The last perturbativity constraint arises purely from the technical aspects of the one-loop mass calculations (see Appendix~\ref{App:one-loop masses}), in particular, whether the regularized effective mass of Eq.~\eqref{eq:RegMassDef} in Appendix~\ref{app:regularized-mass} is a good approximation of the physical mass. Essentially, the main concern arises from diagrams with pseudo-Goldstone bosons in both the outer legs and the loop, which lead to one-loop mass corrections of PGBs proportional to logs of masses of those same PGBs.
The regularized-effective-mass approach breaks down for points overly sensitive to these contributions; i.e.,~our method is unreliable at those points, and hence, we do not accept them as valid. We check for stability by iterative computation, initially feeding the unreliable tree-level PGB masses into the logs: The first and final converged iteration should not be ``too far separated'' from each other.

%%%%%%%%%%%%%%%%%%%%%%%%%%%%%%%%%%%%%%%%%%%%%%%%%%%%%%%
\section{Results\label{Sec:Results}}
%%%%%%%%%%%%%%%%%%%%%%%%%%%%%%%%%%%%%%%%%%%%%%%%%%%%%%%

Having established the Higgs model in Sec.~\ref{Sec:The Model}
and presented the vital considerations required for its analysis in Sec.~\ref{Sec:analysis}, we now turn to the results.

In Sec.~\ref{sec:analytical}, we first discuss the results of a simplified non-tachyonicity analysis 
based on analytic considerations, which help to build the initial intuitive picture. This analysis includes one-loop contributions to PGB masses in a simplified $a_2\to 0$ regime, while allowing for $|\gamma_{2}|\neq 0$, which was inaccessible in previous works~\cite{Graf:2016znk}. This already introduces an important novel result.

From Sec.~\ref{sec:numerical-scan} onward, we proceed with a discussion of the full numerical analysis and its results, implementing all viability criteria from Sec.~\ref{Sec:analysis}. In Secs.~\ref{sec:numerical-scan}, \ref{sec:parameter-space},
\ref{sec:spectrum}, and~\ref{sec:unification}, respectively, the used datasets of points, the viable parts of parameter space, the predictions for the masses, and the analysis of gauge-coupling unification are described.

%======================================================
\subsection{Analytical aspects of non-tachyonicity\label{sec:analytical}}
%======================================================

%
\begin{figure*}[thb!]
    \centering
    \mbox{
    \includegraphics[width=9cm]{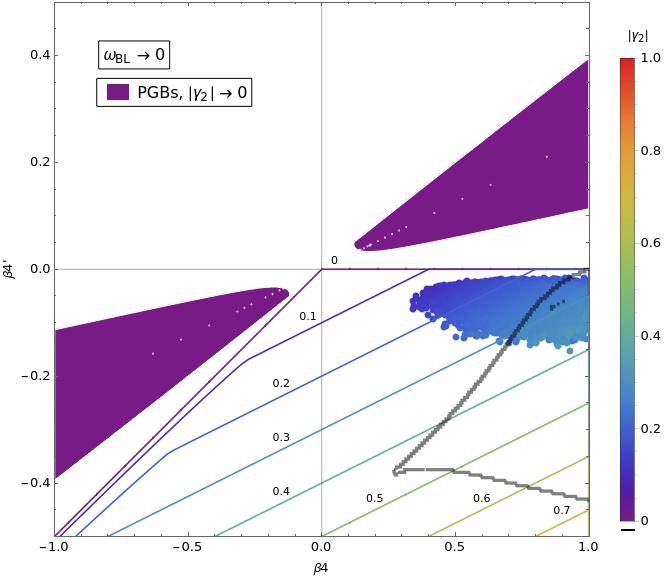}
    \includegraphics[width=9cm]{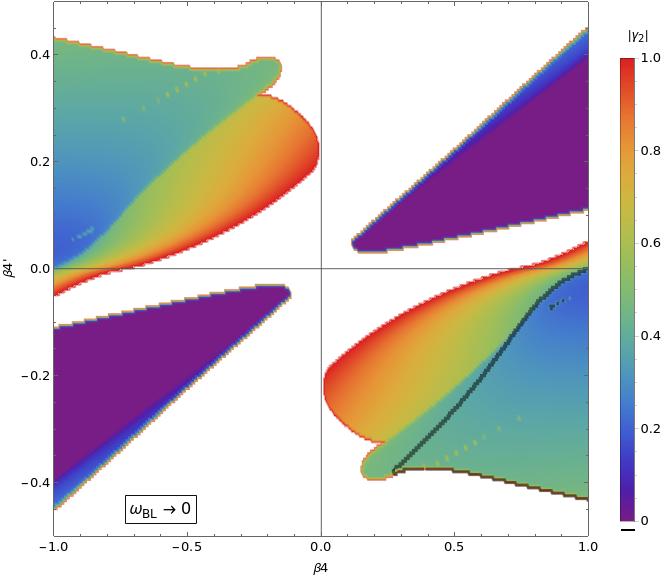}
    }
    \caption{Regions in the $\beta_4$--$\beta_4'$ plane where PGB masses (solid shapes) and non-PGB masses (areas enclosed by thin contours) are non-tachyonic in the $a_2\to 0$, $\sigma\to 0$, and $\omega_{BL}\to 0$ approximation for different values of $|\gamma_2|$ (indicated by different colors). In the left panel, the solid purple region depicts the $\beta_4$--$\beta_4'$ domain where PGB masses are non-tachyonic for $\gamma_2=0$, and one can notice no overlap with the corresponding area supporting the non-tachyonic non-PGB spectrum (stretching down and right from the thin purple line). Retreating colored contours labelled by the corresponding $|\gamma_{2}|$ values display areas supporting non-tachyonic non-PGBs. In the right panel, the color code denotes the minimal $|\gamma_2|$ required for all PGB masses to be non-tachyonic for each $\beta_4$ and $\beta_4'$. The black contour in both panels encloses the area in which a $|\gamma_{2}|$ value exists so that both PGBs and non-PGBs are non-tachyonic. For comparison, we present the results of the full numerical scans described in Secs.~\ref{sec:numerical-scan}--\ref{sec:unification} (discrete color-coded points in the left panel). The white dots along the $\beta'_{4}=\tfrac{1}{4}\beta_{4}$ line in the solid purple region are numerical artifacts related to additional zero-mass scalars
    inflicting large logs.
    }
   \label{fig:AnalyticalBL}
\end{figure*}
\begin{figure*}[thb!]
    \centering
    \mbox{
    \includegraphics[width=9cm]{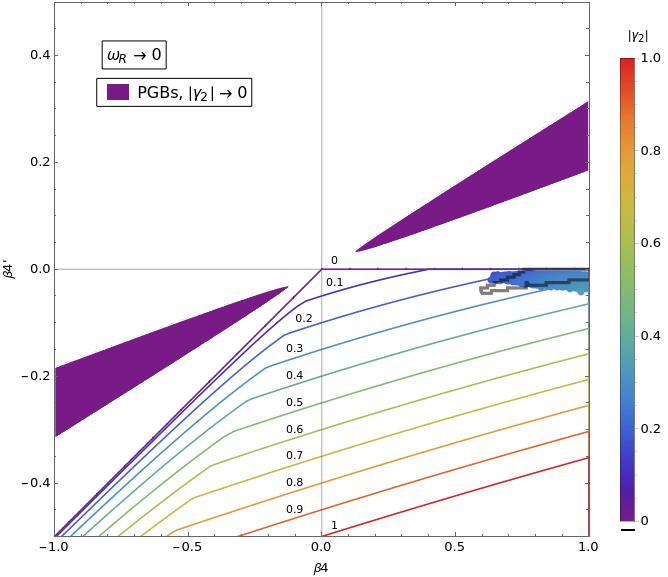}
    \includegraphics[width=9cm]{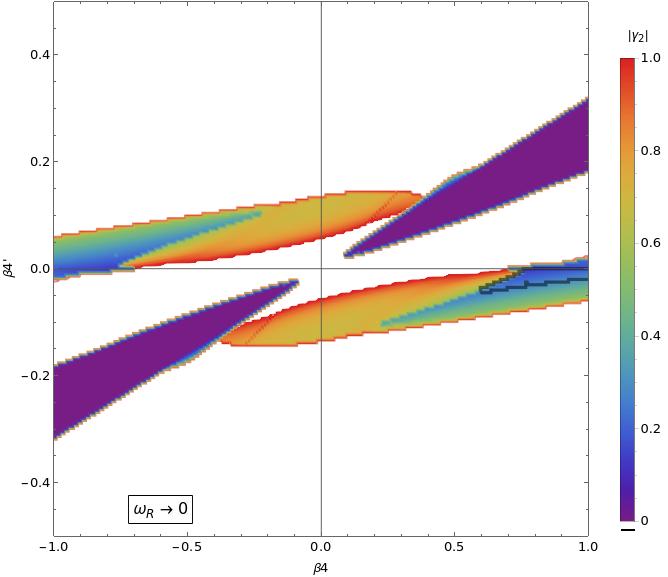}
    }
    \caption{
    The same panels of the $\beta_{4}$--$\beta'_{4}$ plane as in Fig.~\ref{fig:AnalyticalBL}, but for the case $\omega_{R} \to 0$.
    }
   \label{fig:AnalyticalR}
\end{figure*}
Remarkably enough, in the  $\sigma\ll M_{\text{GUT}}$ regime of the two relevant scenarios ($\omega_{BL}\to 0$ and $\omega_{R}\to 0$), one can get good insight into the full numerical results of subsequent sections by means of a semi-analytic account of the non-tachyonicity criterion. 

Assuming perturbativity, the scalar masses of the non-PGBs are expected to be approximated well by their tree-level contributions; see Table~\ref{tab:tree-level-scalar-masses}. Some of these states actually have $|\sigma|$-proportional masses, namely, the $(\xbar{6},1,-\tfrac{4}{3})$ and $(1,1,0)$ for $\omega_{BL}\to 0$, and the $(1,1,+2)$ and $(1,1,0)$ for $\omega_R\to 0$. These vanish identically with $\sigma\to 0$.

We further assume $a_2\ll 1$ in order to suppress the potentially large tachyonic contributions to the tree-level masses of the PGBs $(8,1,0)$, $(1,3,0)$, and/or $(1,1,0)$. These then become dominated by the one-loop effects given in Table~\ref{tab:one-loop-scalar-masses}. Note that in the full numerical scans of Secs.~\ref{sec:numerical-scan}--\ref{sec:unification}, this assumption on $a_2$ is discarded, but the obtained results nevertheless still strongly prefer small values of $a_2$.
The $a_2\to 0$ limit in the semi-analytic approximation is thus justified \emph{a posteriori}. 
Furthermore, the gauge coupling at the GUT scale is fixed to $g=0.5$, which is consistent with the numeric results of Sec.~\ref{sec:unification}. 

With all this in hand, there are only $4$ relevant parameters driving the shape of the entire scalar spectrum, namely, $a_0$, $\beta_4$, $\beta'_4$ (all real), and $\gamma_2$ (complex). Among these, $a_0$ appears solely in the mass of the heaviest singlet [the $\mathrm{SO}(10)$-breaking Higgs field], which can thus always be made positive by a suitable choice of $a_0$. The rest of the scalar spectrum is then in this limit determined by only three parameters: $\beta_4$, $\beta'_4$, and $|\gamma_2|$. We consider two $\gamma_2$ regimes separately. 

%......................................................
\subsubsection{The $\gamma_2\to 0$ regime}
%......................................................

In this case, everything depends predominantly on $\beta_4$ and $\beta'_4$, and complete analytical results are available not only for the tree-level masses of non-PGBs (Table~\ref{tab:tree-level-scalar-masses}), but also for the one-loop masses of PGBs (Table~\ref{tab:one-loop-scalar-masses}).
\begin{itemize}
\item The (tree-level) non-PGB masses are all non-tachyonic in the $\beta_4'< 0$ and $\beta_4'< \frac{1}{2} \, \beta_4$ area depicted in the left panels of  Figs.~\ref{fig:AnalyticalBL} and~\ref{fig:AnalyticalR} --- the allowed region stretches down and right from the thin purple contour. 
\item The PGB masses (which for $a_2\to 0$ do not obtain any tree-level contribution) are all non-tachyonic in the solid purple region in the first and third quadrants in the same plots. In both scenarios ($\omega_{BL}\to 0$ and $\omega_{R}\to 0$), the viable regions are typically bounded from above by the non-tachyonicity of the PGB singlet and from below by the PGB triplet. Note also that the tips of the purple triangular shapes do not extend all the way to the origin of the  $\beta_4$--$\beta_4'$ plane. The reason is that gauge loop contributions to the triplet and octet PGB masses cannot be made simultaneously non-negative, and for small $\beta_4$ and $\beta'_{4}$ cannot be overcome in the $a_{2}\to 0$ limit.
\end{itemize}
The main lesson to be learned here is that the two listed regions do not overlap at all, and there is thus no way to make the entire scalar spectrum non-tachyonic in the $\gamma_2\to 0$ limit.

%......................................................
\subsubsection{The $\gamma_2\neq 0$ regime}
%......................................................

For $|\gamma_{2}|\neq 0$, analytic formulae for the tree-level non-PGB masses retain their relatively simple form, in which the complex phase of $\gamma_{2}$ plays no role. The one-loop PGB masses, however, have to be calculated numerically. The situation then changes as follows:
\begin{itemize}
\item The non-tachyonicity regions for the (tree-level) non-PGB masses are given by the inequalities
\begin{align}
\begin{split}
\omega_{BL}\to 0: &\quad \beta_4'<0, 
\quad \beta_4'<\tfrac{1}{4}\beta_4-|\gamma_2|, \quad a_0 > 0,%\\ 
%& 
\quad (\beta_4'-\tfrac{1}{2}\beta_4) (\beta_4'-\tfrac{1}{18}\beta_4)>\tfrac{4}{9}|\gamma_2|^2,  \label{eq:nontachyonicity-analytic-BL}\\
\end{split}\EQSPACE
\begin{split}
\omega_{R}\to 0: &\quad \beta_4'<0, 
\quad \beta_4'<\tfrac{7}{18}\beta_4-\tfrac{2}{9}|\gamma_2|,
\quad a_0 > 0, %\\ 
%&
\quad 
(\beta_4'-\tfrac{1}{2}\beta_4) (\beta_4'-\tfrac{1}{50}\beta_4)>\tfrac{4}{25}|\gamma_2|^2,\\
& 
\quad
(\beta_4'-\tfrac{1}{4}\beta_4) (\beta_4'-\tfrac{1}{16}\beta_4)>\tfrac{1}{4}|\gamma_2|^2.
\label{eq:nontachyonicity-analytic-R}\\
\end{split}
\end{align}

They are derived by applying the $a_2\to 0$, $\sigma \to 0$ limit to the masses listed in Table~\ref{tab:tree-level-scalar-masses}.
The above conditions introduce the boundaries depicted by a set of $|\gamma_2|$-labelled colored contours in the left panels of Figs.~\ref{fig:AnalyticalBL} and~\ref{fig:AnalyticalR} (as before, the viable regions stretch down and right of these contours). Interestingly, the non-tachyonic region recedes toward the lower-right corner of the $\beta_4$--$\beta_4'$ plane with increasing $|\gamma_2|$. 
\item The rather complicated non-tachyonic regions for numerically calculated PGB masses are displayed in the right panels of Figs.~\ref{fig:AnalyticalBL} and~\ref{fig:AnalyticalR} 
for various $|\gamma_2|$ values. It is perhaps worth noting that they retain their $(\beta_4, \beta_4')\to (-\beta_4, -\beta_4')$ symmetry because the relevant radiative corrections are still quadratic in both $\beta$ couplings  (i.e.,~there are no $\beta_{4}\,\gamma_2$ or $\beta'_{4}\,\gamma_{2}$ mixed terms).
In the right panels of Figs.~\ref{fig:AnalyticalBL} and~\ref{fig:AnalyticalR}, we plot for a given $\beta_{4}$ and $\beta'_{4}$ the value of \emph{minimal} $|\gamma_2|$ for which 
a non-tachyonic PGB spectrum is attainable. It is particularly interesting that for $|\gamma_2|\gtrsim 0.2$, one can find such points even in the 4th quadrant of the $\beta_4$--$\beta_4'$ plane into which also the non-tachyonic region for non-PGBs retreats. 
\item The last observation provides a clear hint where to look for a fully non-tachyonic scalar spectrum.
The black contours depict the overlap of the regions corresponding to the non-tachyonic non-PGB spectrum (the receding polygons in the left panels of Figs.~\ref{fig:AnalyticalBL} and~\ref{fig:AnalyticalR}) with these newly emerging $\beta_4$--$\beta_4'$ areas supporting non-tachyonic PGB masses (the colored shapes in the right panels). Note that in doing so, we need to look for the overlap of the corresponding viable regions for each value of $|\gamma_2|$ separately; only then can these be superimposed and projected onto the $\beta_4$--$\beta_4'$ plane. Remarkably, in the $\omega_{BL}\to 0$ case, a fully consistent region exists for
$0.19\lesssim|\gamma_2|\lesssim 0.47$ within a relatively wide $\beta_4$--$\beta_4'$ range, while for $\omega_R\to 0$, a valid region is obtained for
$0.14\lesssim |\gamma_2|\lesssim 0.29$ in only a very narrow sliver in the $\beta_4$--$\beta_4'$ plane corresponding to small $\beta_4'$ and relatively large $\beta_4$. This indicates that the $\omega_{R}\to 0$ scenario is far more restrictive, and it is correctly anticipated that this remains so even in the full-fledged numerical scans performed later.
\end{itemize} 

To demonstrate the relevance of the simplified picture we have just outlined, we add into the left panels of Figs.~\ref{fig:AnalyticalBL} and~\ref{fig:AnalyticalR} the results of the full numerical scans of Secs.~\ref{sec:numerical-scan}--\ref{sec:unification} (where the entire spectrum has been treated numerically at one loop). One can see that the viable points are essentially where they are expected to be based on the black contours (i.e.,~in the fourth quadrant of the $\beta_4$--$\beta_4'$ plane with a clear affinity toward the larger $\beta_4$ and the smaller $\beta_4'$ values). The slight discrepancy between the results of the simplified semi-analytic account given here and the data from scans can be attributed to a non-zero $a_2$ value admitted in the latter case.  
It is interesting that for $\omega_R\to 0$ a non-zero $a_2$ is actually enforced (cf.~Sec.~\ref{sec:doublets-and-triplets}), yet the overlap of the results of the two methods is almost perfect.

%======================================================
\subsection{Data from numerical scans \label{sec:numerical-scan}}
%======================================================

We now turn to the full numerical analysis and its results. We explore the space of parameters defined by the dimensionless couplings
\begin{align}
\label{eq:dimless_couplings}
    & a_2,\; a_0,\; \lambda_0,\; \lambda_2,\; \lambda_4,\; \lambda_4^\prime,\; \alpha,\; \beta_4,\; \beta_4^\prime,\; \gamma_2,\; \eta_2,\; g,
\end{align}
which are all assumed to be within the $\mathcal{O}(1)$ domain, and the dimensionful VEVs
\begin{align}
    & \omega_{BL},\; \omega_R,\; \sigma,
\end{align}
whose values are restricted by the perturbativity constraint of Eq.~\eqref{eq:VEVConstraint} and unification. 

We evaluate the suitability of a parameter point by its viability with respect to non-tachyonicity, gauge-coupling unification, and perturbativity, as discussed in Sec.~\ref{Sec:analysis} (the technical procedure is described in all detail in Appendix~\ref{app:numerical}). The suitability criteria are numerically implemented as a penalization function, which gives zero when all criteria are satisfied. Furthermore, the penalization function rises monotonically with the quantitative size of the violation of any suitability criterion. We use a stochastic version\footnote{In particular, we use version ``DE/rand/1'' with a random choice $F\in (0.5,2)$ for each candidate point; cf.,~e.g.,~\cite{diffevolalg}.} of the differential evolution algorithm to find and explore viable regions of the parameter space. 

\begin{table*}[htb]
    \centering
    \caption{The datasets obtained by full numerical scans that are analyzed in Secs.~\ref{sec:parameter-space}--\ref{sec:unification}; for a detailed description, the reader is referred to the main text.\label{tab:datasets}}
    \vskip 0.1cm
    \begin{tabular}{l@{$\quad$}r@{$\qquad$}r@{$\qquad$}r@{$\quad$}r@{$\qquad$}l}
    \hline
    Dataset&VEV regime&RG range&Bias&$\#$ of points&Comment\\
    \hline
    $B_{+}$&$\omega_{BL}\to 0$&$t_+ > 0.5$&&30000&Main dataset\\
    $B_{1}$&$\omega_{BL}\to 0$&$\bar{t} > 1.0$&&20000&\\
    $B_{2}$&$\omega_{BL}\to 0$&$\bar{t} > 2.0$&&20000&\\
    $B_{3}$&$\omega_{BL}\to 0$&$\bar{t} > 3.0$&&20000&\\
    $B'_{+}$&$\omega_{BL}\to 0$&$t_+ > 0.5$&&30000&No Sylvester's criterion\\
    $B_{\text{RG}}$&$\omega_{BL}\to 0$&$t_+ > 0.5$&$\bar{t}$&10000& RG perturbativity\\
    $B_{\text{GM}}$&$\omega_{BL}\to 0$&$t_+ > 0.5$&$\overline{\Delta}$&8000&Global mass perturbativity\\[6pt]
    $R_{+}$&$\omega_{R}\to 0$&$t_+ > 0.5$&&30000&Main dataset\\
    $R_{1}$&$\omega_{R}\to 0$&$\bar{t} >1.0$&&20000&\\
    $R'_{+}$&$\omega_{R}\to 0$&$t_+ > 0.5$&&30000&No Sylvester's criterion\\
    $R_{\text{RG}}$&$\omega_{R}\to 0$&$t_+> 0.5$&$\bar{t}$&10000&RG perturbativity\\
    $R_{\text{GM}}$&$\omega_{R}\to 0$&$t_+ > 0.5$&$\overline{\Delta}$&8000&Global mass perturbativity\\[2pt]
    \hline
    \end{tabular}
\end{table*}

Since the threshold values in the perturbativity criteria are to some degree arbitrary, we performed a number of numerical scans with varying degrees of strictness. We consider two main perturbativity measures: 
\begin{enumerate}
\item The persistence of perturbativity at different RG scales, referred to as \emph{RG perturbativity}, is encoded in the quantity $\bar{t}$; cf.~Eq.~\eqref{eq:definition-tbar}. Intuitively, it tells us how many orders of magnitude (in powers of $10$) a point can be run either up or down via RGEs before at least one of the couplings blows up. A similar measure is also $t_{+}$ [cf.~Eq.~\eqref{eq:definition-tplusminus}], which considers only RG running upward in scale.
\item The ratio of the largest one-loop correction to the average of the \emph{heavy} masses is denoted by $\overline{\Delta}$; cf.~Eq.~\eqref{eq:definition-delta}. This measures \emph{global mass (GM) perturbativity}.
\end{enumerate}

With these definitions, a bigger $\bar{t}$ (or $t_{+}$) and smaller $\overline{\Delta}$ imply a better perturbativity of the point. We impose $t_{+}>0.5$ and $\overline{\Delta}<1$ in all our datasets, which are conveniently listed in Table~\ref{tab:datasets}. The main datasets $B_{+}$ and $R_{+}$ do not have any additional constraints, while $B_{1,2,3}$ and $R_{1}$ have a stricter RG-perturbativity criterion imposed in the form of an acceptance threshold for $\bar{t}$. There are no datasets $R_{2}$ and $R_{3}$ because no points with $\bar{t}>2$ were found in the  $\omega_{R}\to 0$ case. Note that all datasets consist only of viable points, i.e.,~those passing all criteria from Sec.~\ref{Sec:analysis}.

For some datasets, we used an additional penalization of how well a perturbativity criterion is satisfied, so as to push the parameter scan to be biased with respect to this quantity; i.e.,~new points are accepted only when they are at least as good as the old ones with respect to that criterion. In such cases, we refer to the scans as \emph{biased}. The biased datasets searching for the best values of $\bar{t}$ and $\overline{\Delta}$ are labelled as RG and GM, respectively; cf.~Table~\ref{tab:datasets}. For each dataset in that table, we denote its label, the VEV regime explored (either $\omega_{BL}\to 0$ or $\omega_{R}\to 0$), the RG range in terms of $t_{+}$ or $\bar{t}$, the bias criterion used for optimization (if any), and the number of points in the dataset. 

All numerical results are based on the datasets from Table~\ref{tab:datasets} and are presented in the form of figures. A list of figures, alongside the used datasets for each figure and a brief description, are gathered in Table~\ref{tab:figures}.
For readability, we separate the results into three sections: viable regions for input parameters are identified in Sec.~\ref{sec:parameter-space}, results for the observables (masses) are collected in Sec.~\ref{sec:spectrum}, and sample patterns of the unification of gauge couplings for selected points are presented in Sec.~\ref{sec:unification}.

\begin{table*}[htb]
    \centering
    \caption{Table of figures in Secs.~\ref{sec:parameter-space}--\ref{sec:unification} and the datasets used for each of them. Semicolons separate datasets of either different figures or different panels of the same figure, while commas separate datasets whose information is presented separately in the same figure.  \label{tab:figures}}
    \begin{tabular}{c@{$\qquad$}l@{$\qquad$}l}
        \hline
        Figure &Datasets&Brief description\\
        \hline
        \ref{fig:ScatterPlotsVSBL}\ ;\ \ref{fig:ScatterPlotsVSR}&$B_{+} \cup B_{\text{RG}}\;;\; R_{+}\cup R_{\text{RG}}$&
            Parameter correlation plots, $\bar{t}$ hot spots\\
        \ref{fig:ScatterPlotsTPBL}\ ;\ \ref{fig:ScatterPlotsTPR}&$B_{+}\cup B_{\text{GM}}\;;\; R_{+} \cup R_{\text{GM}}$&
            Parameter correlation plots, $\overline{\Delta}$ hot spots\\
        \ref{fig:ScalarParameters}&$B_{+}, B_{1}, B_{2}, B_{3}\;;\;R_{+}, R_{1}$&
        Likelihood $\sigma$-ranges for scalar parameters\\
        \ref{fig:Scales}&$B_{+}, B_{1}, B_{2}, B_{3}\;;\;R_{+}, R_{1}$& Comparison of scales (dimensionful parameters)\\
        \ref{fig:Sylvester}&$B'_{+},B_{+}\;;\; R'_{+},R_{+}$&Effect of non-tachyonicity of doublets and triplets\\
        \ref{fig:PGBFields}&$B_{+}, B_{1}, B_{2}, B_{3}\;;\;R_{+}, R_{1}$&
            Likelihood $\sigma$-ranges for PGB particles\\
        \ref{fig:HeavyFields}&$B_{+}, B_{1}, B_{2}, B_{3}\;;\;R_{+}, R_{1}$&
            Likelihood $\sigma$-ranges for \emph{heavy} particles\\
        \ref{fig:LightFields}&$B_{+}, B_{1}, B_{2}, B_{3}\;;\;R_{+}, R_{1}$&
            Likelihood $\sigma$-ranges for intermediate-scale particles\\
        \ref{fig:Unification}&Points in Table~\ref{tab:points-for-unification}&Gauge-coupling unification\\[2pt]
        \hline
    \end{tabular}
\end{table*}
%

%======================================================
\subsection{Viable regions of the parameter space \label{sec:parameter-space}}
%======================================================

In this subsection, we present the viable regions of the parameter space for both $\omega_{BL}\to 0$ and $\omega_{R}\to 0$ scenarios. As we are limited to $2$-dimensional projections, the information contained in the plots can never be complete. In what follows, we thus provide two complementary perspectives: planar \emph{correlation plots} for chosen pairs of parameters in Sec.~\ref{sec:parameter-correlations} and \emph{likelihood $\sigma$-ranges} for each individual parameter in Sec.~\ref{sec:parameter-ranges}.

%......................................................
\subsubsection{Correlation plots for different pairs of scalar parameters \label{sec:parameter-correlations}}
%......................................................

Altogether, there are $11$ real dimensionless scalar parameters of interest:
\begin{align}
a_0,\; a_2,\; \lambda_0,\; \lambda_2,\; \lambda_4,\; \lambda'_4,\; \alpha,\; \beta_4,\; \beta'_4,\; |\gamma_2|,\; |\eta_2|.
\end{align}
We hence choose $6$ correlation pairs (with $\beta_{4}$ --- one of the two main parameters of interest [cf.~Sec.~\ref{sec:analytical}] --- included twice) in a way that best demonstrates the salient features of our results. Since $\gamma_{2}$ and $\eta_{2}$ are complex, they also carry phases $\delta_{\gamma_2}$ and $\delta_{\eta_2}$. However, we omit these phases from the plots, as it turns out that the distributions of both are practically uniform on the entire $[0,2\pi)$ interval. Interesting patterns correlating the two phases appear only by employing stricter RG- or GM-perturbativity constraints in the $\omega_R \to 0$ limit, in which case the parameter space strongly prefers a $2 \delta_{\gamma_2} = \delta_{\eta_2}$ relation that prevents both phases from changing under one-loop RG running; see Eqs.~\eqref{eq:beta-gamma_2}--\eqref{eq:beta-eta_2}.

\begin{figure*}[htb]
    \centering
    \mbox{
    \includegraphics[width=6cm]{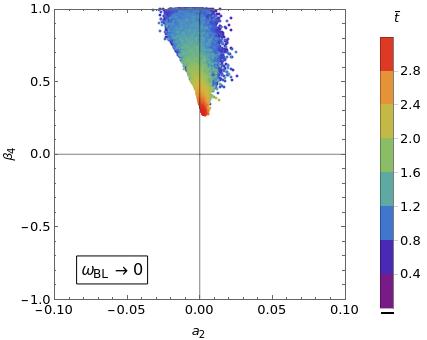}
    \includegraphics[width=6cm]{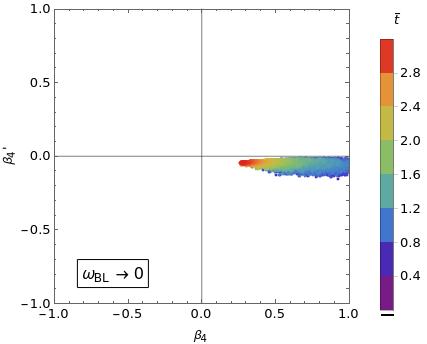}
    \includegraphics[width=6cm]{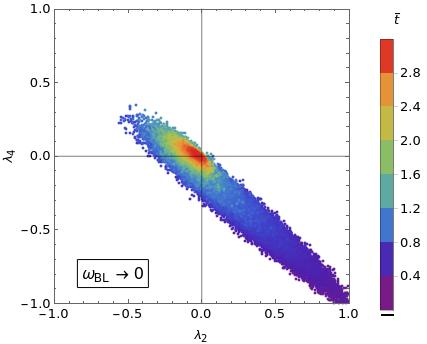}
    }
    \vskip 3mm
    \mbox{
    \includegraphics[width=6cm]{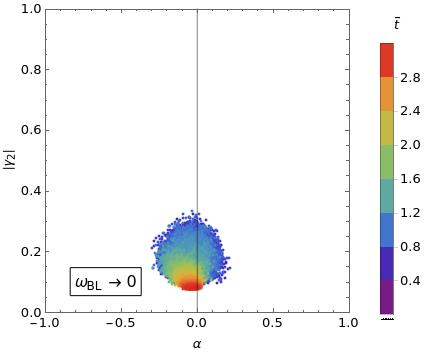}
    \includegraphics[width=6cm]{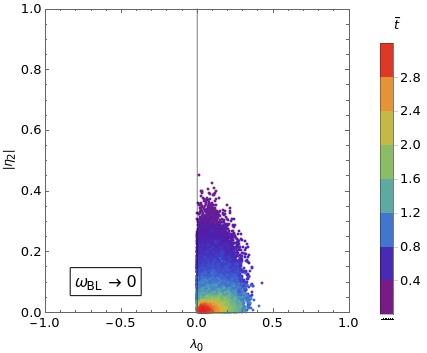}
    \includegraphics[width=6cm]{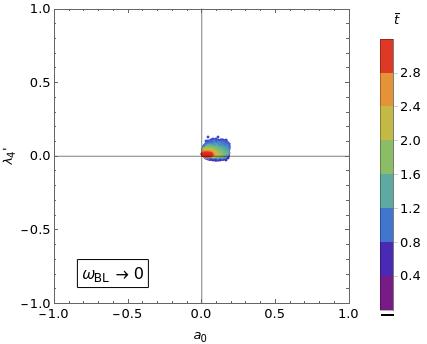}
    }
    \caption{Correlation plots showing viable points projected onto pairs of scalar parameters. Plots consider the $\omega_{BL} \to 0$ scenario, and points are color coded according to the  RG-perturbativity measure $\bar{t}$ defined in Appendix~\ref{app:perturbativity}.    \label{fig:ScatterPlotsVSBL}}
\end{figure*}
\begin{figure*}[hbt]
    \centering
    \mbox{
    \includegraphics[width=6cm]{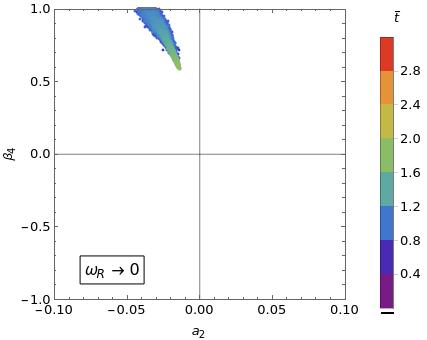}
    \includegraphics[width=6cm]{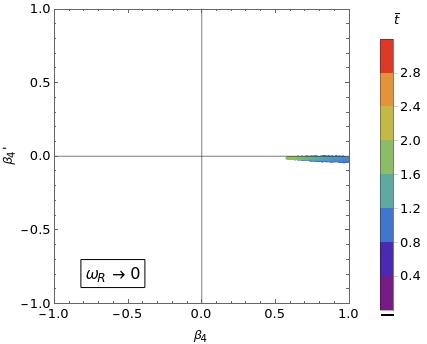}
    \includegraphics[width=6cm]{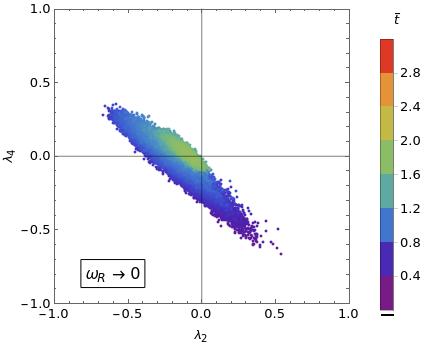}
    }
    \vskip 3mm
    \mbox{
    \includegraphics[width=6cm]{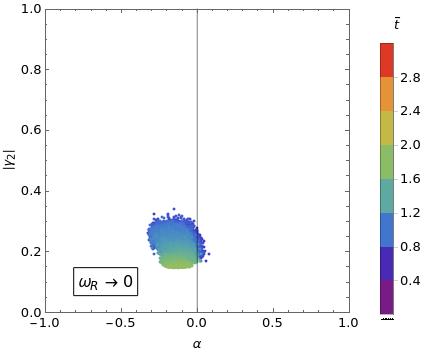}
    \includegraphics[width=6cm]{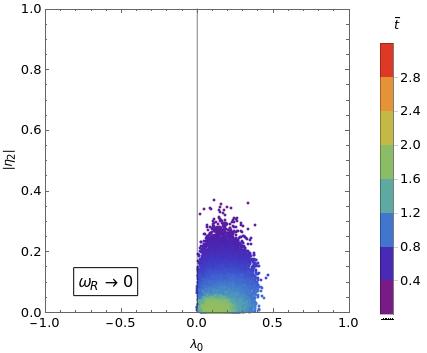}
    \includegraphics[width=6cm]{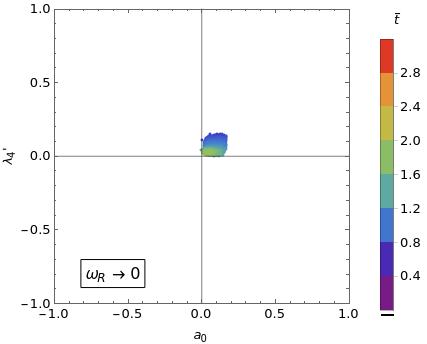}
    }
    \caption{Correlation plots showing viable points projected onto pairs of scalar parameters. Plots consider the $\omega_{R} \to 0$ scenario, and points are color coded according to the  RG-perturbativity measure $\bar{t}$ defined in Appendix~\ref{app:perturbativity}. 
    \label{fig:ScatterPlotsVSR}}
\end{figure*}
\begin{figure*}[htb]
    \centering
    \mbox{
    \includegraphics[width=6cm]{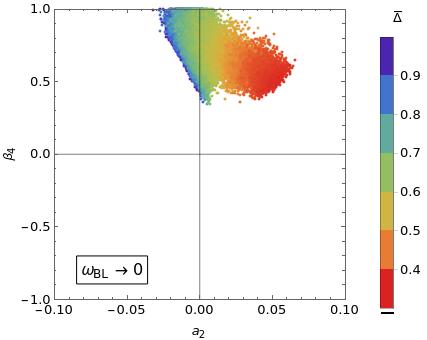}
    \includegraphics[width=6cm]{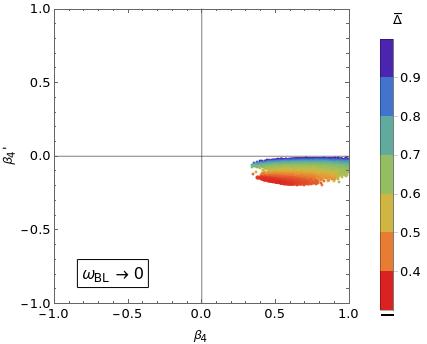}
    \includegraphics[width=6cm]{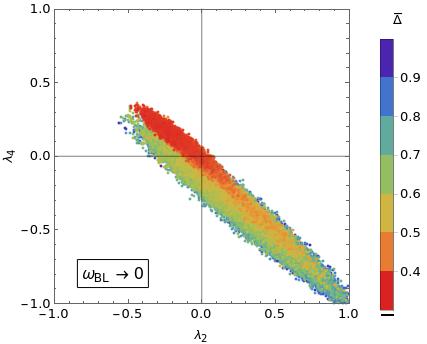}
    }
    \vskip 3mm
    \mbox{
    \includegraphics[width=6cm]{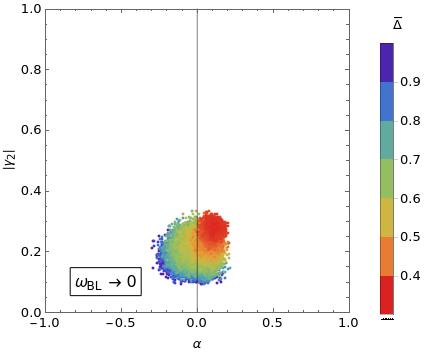}
    \includegraphics[width=6cm]{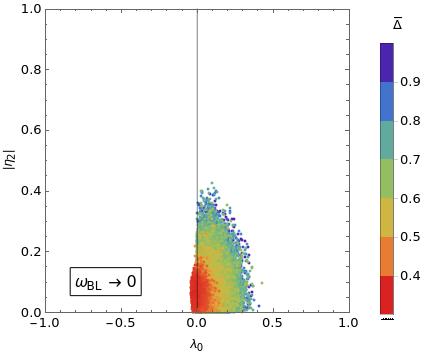}
    \includegraphics[width=6cm]{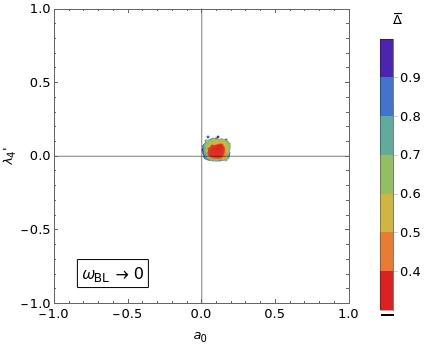}
    }
    \caption{
    Correlation plots showing viable points projected onto pairs of scalar parameters. Plots consider the $\omega_{BL} \to 0$ scenario, and points are color coded according to the mass-perturbativity measure $\overline{\Delta}$ defined in Appendix~\ref{app:perturbativity}.
    \label{fig:ScatterPlotsTPBL}}
\end{figure*}
\begin{figure*}[hbt]
    \centering
    \mbox{
    \includegraphics[width=6cm]{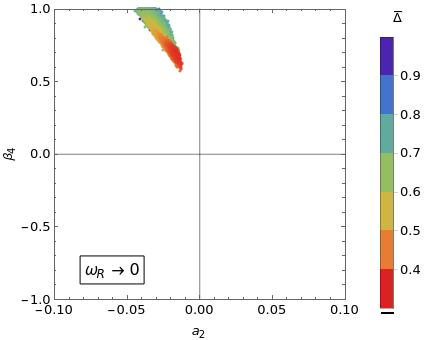}
    \includegraphics[width=6cm]{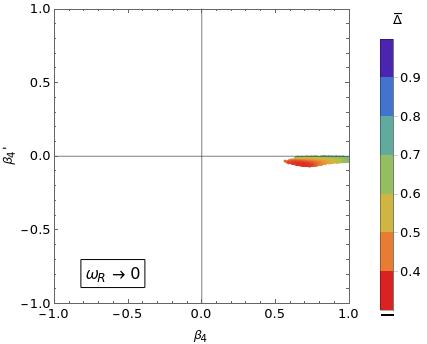}
    \includegraphics[width=6cm]{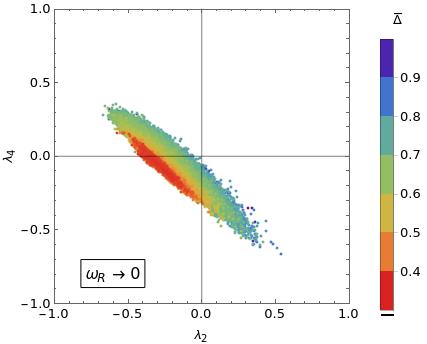}
    }
    \vskip 3mm
    \mbox{
    \includegraphics[width=6cm]{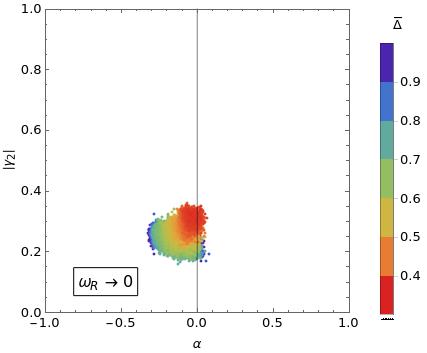}
    \includegraphics[width=6cm]{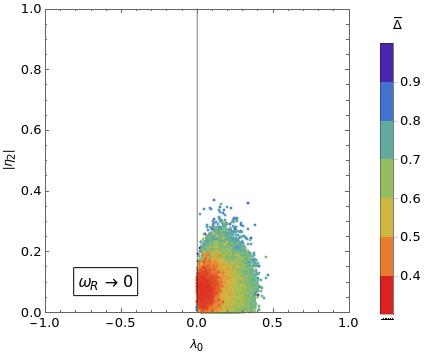}
    \includegraphics[width=6cm]{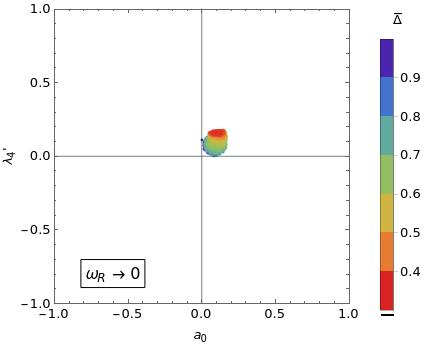}
    }
    \caption{Correlation plots showing viable points projected onto pairs of scalar parameters. Plots consider the $\omega_{R} \to 0$ scenario, and points are color coded according to the mass-perturbativity measure $\overline{\Delta}$ defined in Appendix~\ref{app:perturbativity}.
    \label{fig:ScatterPlotsTPR}}
\end{figure*}

The correlation plots for the $\omega_{BL}\to 0$ hierarchy are given in Figs.~\ref{fig:ScatterPlotsVSBL} and~\ref{fig:ScatterPlotsTPBL} and those for the $\omega_R\to 0$ case are collected in  Figs.~\ref{fig:ScatterPlotsVSR} and~\ref{fig:ScatterPlotsTPR}. Two different color-coding schemes are employed in these plots indicating various levels of perturbativity with respect to two associated measures discussed in Sec.~\ref{sec:numerical-scan}: the $\bar{t}$ quantity corresponding to the RG stability of individual points (Figs.~\ref{fig:ScatterPlotsVSBL} and~\ref{fig:ScatterPlotsVSR}; higher $\bar{t}$ is better) and $\overline{\Delta}$, which quantifies the relative size of loop corrections to masses (Figs.~\ref{fig:ScatterPlotsTPBL} and~\ref{fig:ScatterPlotsTPR}; lower $\overline{\Delta}$ is better). The plots are produced by merging the main datasets denoted by $+$, which consist of unbiasedly sampled viable points, with the RG or GM biased datasets; cf.~Tables~\ref{tab:datasets} and~\ref{tab:figures}. 
When points are overlapping, those considered better with respect to the relevant perturbativity measure are drawn in front. This allows for identification of \emph{hot spot} regions, where the best points (those colored toward red) were found. 

We make the following observations for the correlation plots:
\begin{itemize}
		\item \underline{$a_0,\;\lambda_0 \gtrsim 0$:} 
		The positivity of $a_0$ can be understood by investigating the mass of the $(1,1,0)_4$ \emph{heavy} non-PGB SM singlet [i.e.,~the $\mathrm{SO}(10)$-breaking Higgs field]. In the $|a_2|\ll |a_0|$ regime, its tree-level mass-square value is approximately $8 \, a_0 \left(3 \, \omega_{BL}^2 + 2 \, \omega_{R}^2\right)$; cf.~Sec.~\ref{sec:PSG-singlets}. Hence, it is non-tachyonic only if $a_0\gtrsim 0$. 
		\par
		Note that $a_0$ does not appear in any tree-level mass apart from the $(1,1,0)_{4}$ and $(1,1,0)_{2}$, with the latter being the $\mathrm{U}(1)_{B-L}$-breaking SM-singlet-Higgs field. The mass of this field is $|\sigma|$-proportional and only its tree-level value is relevant; see Sec.~\ref{sec:Tachyonicity}. It is effectively governed by the $\lambda_0$ parameter:
		For small $a_2$ that is needed to change the tachyonic character of PGBs by loop corrections,
		non-tachyonicity requires $\lambda_0 \gtrsim \frac{(\alpha+\beta_4')^2}{4 a_0}$ in both scenarios; cf.~Table~\ref{tab:tree-level-scalar-masses}. Then, $a_0 > 0$ implies $\lambda_0>0$.
		\item \underline{$\beta_4 > 0$, $\beta_4' \lesssim 0$}: The overall negativity of $\beta_4'$
		is required for non-tachyonicity of the \emph{heavy} tree-level spectrum. The domain $\beta_4>0$, $\beta_4'\sim 0$ then corresponds to the overlap region with non-tachyonic PGBs; see Sec.~\ref{sec:analytical}.
		\item \underline{$|a_2|\ll 1$:} As expected, $a_2$ 
		is small since it controls the PGB tree-level masses (note the different scaling of the associated axes in the relevant panels). While $a_2$ can be of either sign in the $\omega_{BL}\to 0$ case and can even vanish, it turns out to be strictly negative in the $\omega_{R}\to 0$ case. We explicitly confirmed this by an unsuccessful dedicated search for viable points in the $a_2>0$ region of the $\omega_{R}\to 0$ case. The main obstruction turns out to be the non-tachyonicity of the doublets and triplets; cf.~Sec.~\ref{sec:doublets-and-triplets}. Incidentally,  the \hbox{$a_2\in (-0.05,-0.01)$} in the $\omega_{R}\to 0$ case implies that the triplet PGB is always non-tachyonic because  $M^2_S(1,3,0)\approx -2 a_2 \omega_{BL}^2$ at tree level. 
        Interestingly, for $\omega_{BL}\to 0$ the points with larger RG-perturbativity ranges prefer the $a_2\approx 0$ region (cf.~Fig.~\ref{fig:ScatterPlotsVSBL}), while global mass perturbativity prefers pushing $a_2$ toward $0.05$ (cf.~Fig.~\ref{fig:ScatterPlotsTPBL}), generating a slight tension if the scans are biased simultaneously toward both these criteria.
		\item \underline{$0.1 \lesssim |\gamma_2| \lesssim 0.4$:} As discussed in Sec.~\ref{sec:analytical}, a compact range for $|\gamma_2|$ with a lower bound of around $0.1$ is expected for a non-tachyonic scalar spectrum. While RG perturbativity strongly prefers smaller values of $|\gamma_2|$ near this bound (see lower-left panels in Figs.~\ref{fig:ScatterPlotsVSBL} and~\ref{fig:ScatterPlotsVSR}), the GM-perturbativity criterion is optimized in the higher $|\gamma_2|$ region. 
		\item \underline{$\lambda_{4}\sim-\lambda_{2}$:}
	    This pair of quantities exhibits the strongest visible linear correlation among all parameter combinations. Its appearance is mostly due to the shape of the intermediate-scale ($|\sigma|$-proportional) scalar masses. 
		\item \underline{General remarks on scalar parameters' domains:}
		Except for $\lambda_2$, $\lambda_4$, and $\beta_4$, the allowed ranges of the scalar parameters are typically much smaller than the standard $[-1,1]$ domain.\footnote{Note that this expectation depends on the actual definition of the scalar parameters (cf.~Sec.~\ref{Sec:The Model}). They were chosen in our case so that all trivial combinatorial factors just cancel.} On the other hand, the region where all scalar couplings almost vanish is not viable. The main reason is the need to compensate for the
		large gauge coupling contributions in their beta functions (cf.~Appendix~\ref{App:BetaFunctions}) that would otherwise lead to a rapid breakdown of their RG perturbativity.\footnote{To this end, it is perhaps worth noting that we observe a very clear correlation between the locations of the (approximate) fixed points of the scalar couplings' RG flow and the regions of the parameter space in Figs.~\ref{fig:ScatterPlotsVSBL} and~\ref{fig:ScatterPlotsVSR} in which viable points cluster.} Moreover, the tachyonicity issues when $\beta_4$ and $\beta'_4$ simultaneously vanish in the $|\gamma_2|\to 0$ regime have been discussed in Sec.~\ref{sec:analytical}. 
		
		Nevertheless, smaller-coupling regions are still preferred from the point of view of RG perturbativity, as seen from higher $\bar{t}$ values on the color scale in Figs.~\ref{fig:ScatterPlotsVSBL} and~\ref{fig:ScatterPlotsVSR}. Global mass perturbativity in Figs.~\ref{fig:ScatterPlotsTPBL} and~\ref{fig:ScatterPlotsTPR}, on the other hand, prefers some parameters (e.g.,~$|\gamma_2|$ or $\beta_{4}'$) to be on the larger side of their allowed ranges, indicating a complicated interplay between the tree-level and one-loop contributions to scalar masses. This makes the numerical analysis presented here not only technically necessary but also highly non-trivial.
\end{itemize}
Note that the lack of red points with high $\bar{t}$ in Fig.~\ref{fig:ScatterPlotsVSR} (compared with  Fig.~\ref{fig:ScatterPlotsVSBL}) makes the $\omega_{R}\to 0$ case significantly less favourable than the $\omega_{BL}\to 0$ scenario from the perturbativity point of view. Remarkably enough, this is indeed consistent with the results of the highly simplified semi-analytic account of Sec.~\ref{sec:analytical}.
%

%......................................................
\subsubsection{Ranges for individual scalar parameters \label{sec:parameter-ranges}}
%......................................................

\begin{figure*}[htb]
    \centering
    \mbox{
    \includegraphics[width=9cm]{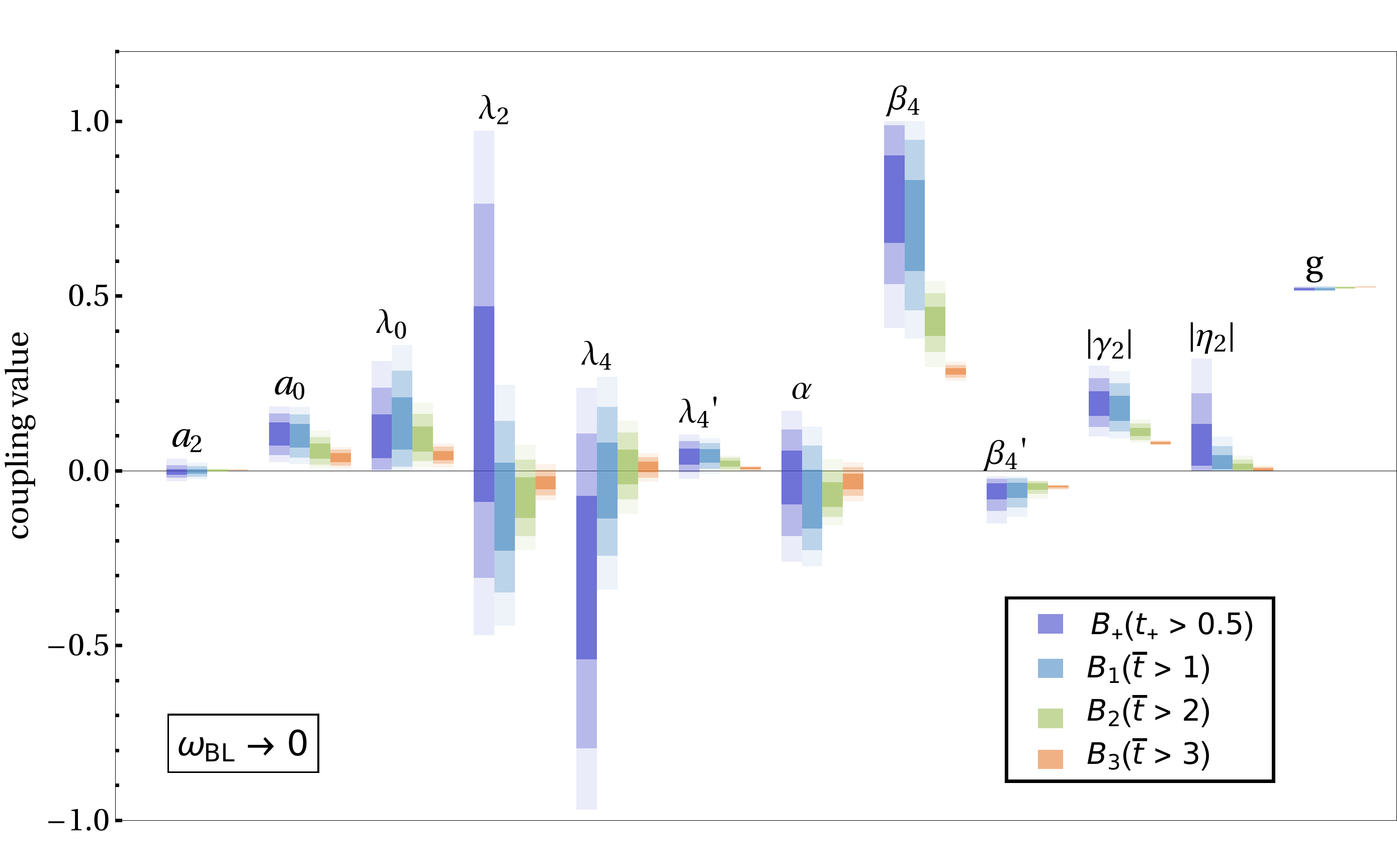}
    \includegraphics[width=9cm]{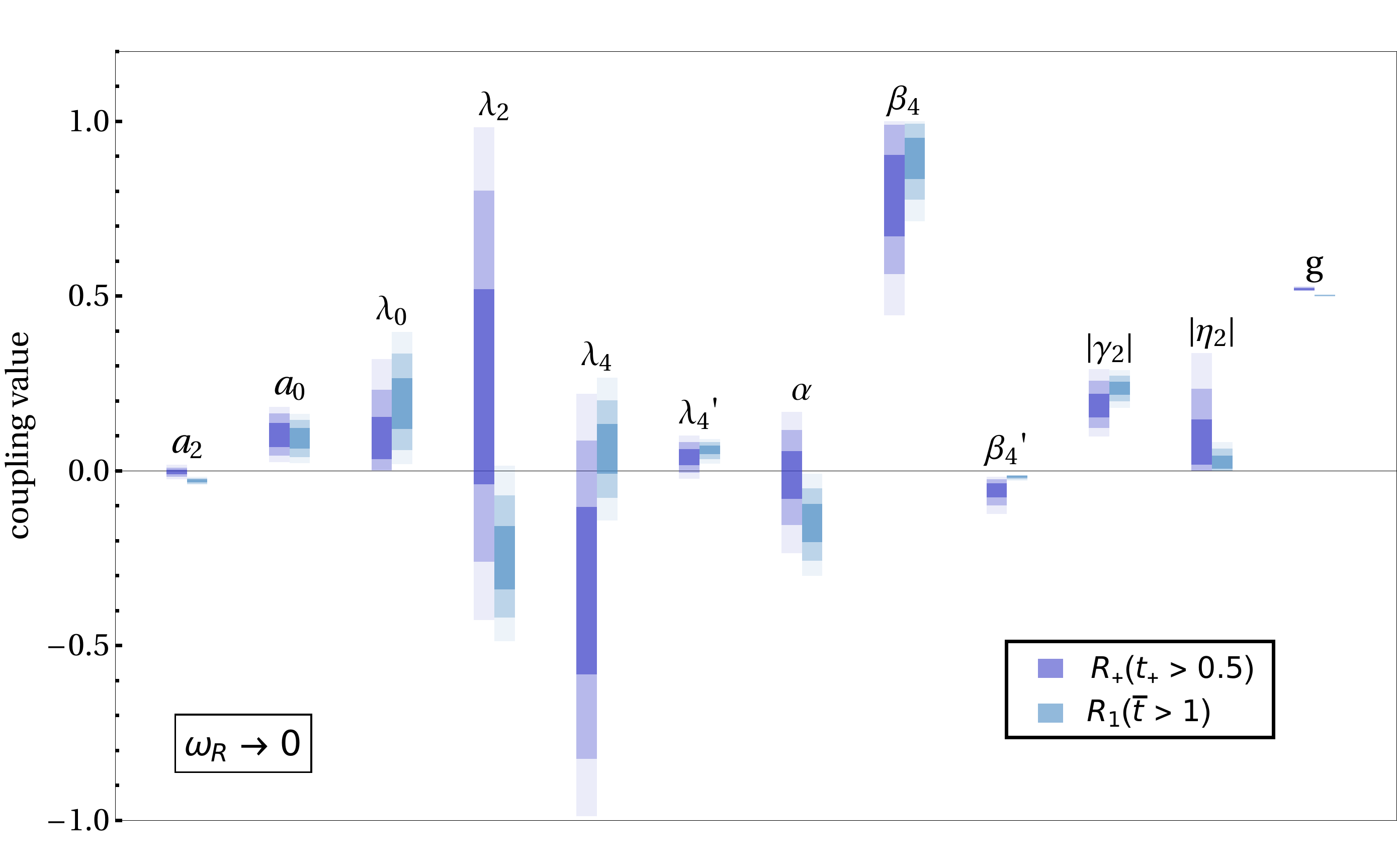}
    }
    \caption{Allowed $1$-, $2$- and $3\sigma$ ranges (with decreasing opacity) of scalar parameters in the $\omega_{BL} \to 0$ and $\omega_R \to 0$ case. Colors encode different strictness of the RG-perturbativity measure defined in Appendix~\ref{app:perturbativity}; 
    see legend and Table~\ref{tab:datasets}.
    }
   \label{fig:ScalarParameters}
\end{figure*}

An alternative way of presenting the viable regions for the scalar parameters at hand is to show the individual range each of them can take. We present these results in Fig.~\ref{fig:ScalarParameters} for both the $\omega_{BL}\to 0$ (left panel) and $\omega_{R}\to 0$ (right panel) cases.

Let us note that the datasets  $B_{+}$ and $R_{+}$  used therein (see Table~\ref{tab:datasets}) essentially correspond to uniform sampling of points from the viable subregion of the parameter space (due to the stochastic nature of the differential evolution sampler). 
Projecting such a dataset to one parameter thus represents an approximation of a  marginal probability distribution in the Bayesian interpretation, effectively providing information about the volume of viable parameter space associated with a particular parameter attaining values close to a certain point. Borrowing tools from Bayesian statistics, we thus present the ranges of each parameter in terms of their \emph{highest density intervals} (HDI): The vertical extent of the bars of decreasing opacity and same horizontal position represents the $1\sigma$, $2\sigma$, and $3\sigma$ HDIs.

Furthermore, the plots include the information obtained from multiple datasets (cf.~Table~\ref{tab:figures}), which is encoded by different colors. We make use of our main datasets with $t_{+}>0.5$ (in \textbf{\textcolor{plotBlue}{blue}}), as well as those where the viability criterion with stricter threshold values for the RG-perturbativity measure was imposed: $\bar{t}>1$, $\bar{t}>2$, and $\bar{t}>3$ colored, respectively, by \textbf{\textcolor{plotLightBlue}{light blue}}, \textbf{\textcolor{plotGreen}{green}}, and \textbf{\textcolor{plotOrange}{orange}}. 

Note that the best points in the $\omega_{R}\to 0$ case have $\bar{t}\approx 1.86$, so there are no $\bar{t}>2$ datasets $R_{2,3}$; i.e.,~no points can be run up and down by $2$ orders of magnitude on average in the renormalization scale without blowing up. 
As expected, increasing the strictness with respect to the RG-perturbativity measure shrinks the allowed parameter ranges, as can be consistently seen in the narrowing of the vertical bars in Fig.~\ref{fig:ScalarParameters} for more constrained datasets. The complex quantum-level interplay between different parameters generates severe constraints even for couplings of seemingly little impact on the observables of our main interest if highest-level RG perturbativity is required. For instance, $|\eta_2|$ is pushed to $0$ for $\bar{t}>3$ (in both scenarios), despite appearing only in one-loop corrections to the \emph{heavy} fields' masses and the scalar-sector beta functions.

The final observation concerns the fact that the allowed ranges of certain parameters within a stricter dataset may be in an unlikely region from the point of view of less strict datasets; i.e.,~the HDIs of a strict dataset may not overlap with even the $3\sigma$ HDI of a less strict one. This implies that enhancing RG perturbativity sometimes requires a push toward a very particular corner of the allowed parameter space. Note that this was already indicated by the positions of the hot spots appearing at the very edges of the clouds of viable points for some parameters in Figs.~\ref{fig:ScatterPlotsVSBL} and~\ref{fig:ScatterPlotsVSR}. A prominent example of this effect is $\beta_{4}$ in the $\omega_{BL}\to 0$ case.

%......................................................
\subsubsection{The VEVs and the renormalization scale \label{sec:VEVs-and-muR}}
%......................................................

We now turn our attention to dimensionful input parameters. The tree-level potential in Eq.~\eqref{eq:tree-potential} contains three dimensionful parameters $\mu$, $\nu$, and $\tau$, which we compute via one-loop stationarity conditions from the three VEVs 
$\omega_{BL}$, $\omega_{R}$, and $\sigma$ of Eq.~\eqref{eq:VEV-definitions}. Together with the renormalization scale $\mu_{R}$, one has four dimensionful parameters in total.  

\vspace{0.2cm}
\paragraph{The VEVs:} \mbox{}\\
The complex VEV $\sigma$ can be made positive and real by a phase redefinition of the $\mathbf{126}$ tensor of $\mathrm{SO}(10)$, while the bigger of the two real VEVs $\omega_{BL}$ and $\omega_{R}$ can be made positive by a sign redefinition of the (real) adjoint $\mathbf{45}$. Since we are interested only in the $\omega_{BL}\to 0$ or $\omega_{R}\to 0$ regimes (see Sec.~\ref{sec:quantum-level}), the bigger of the $\omega$'s sets the GUT scale, and $\sigma$ plays the role of the intermediate $\mathrm{U}(1)_{B-L}$-breaking (seesaw) scale. The smaller of the $\omega$'s must then be small enough to keep the universal VEV ratio $\chi$ defined in Eq.~\eqref{eq:chi-definition} under perturbative control, i.e.,~ensure that~$|\chi|\lesssim 1$. The subdominant $\omega$ thus plays the role of an induced VEV. Since it is far smaller than the other two VEVs and it is not associated with any distinct physical scale either, we shall not pay much attention to it in what follows.

\begin{figure*}[htb]
    \centering
    \mbox{
    \includegraphics[width=9cm]{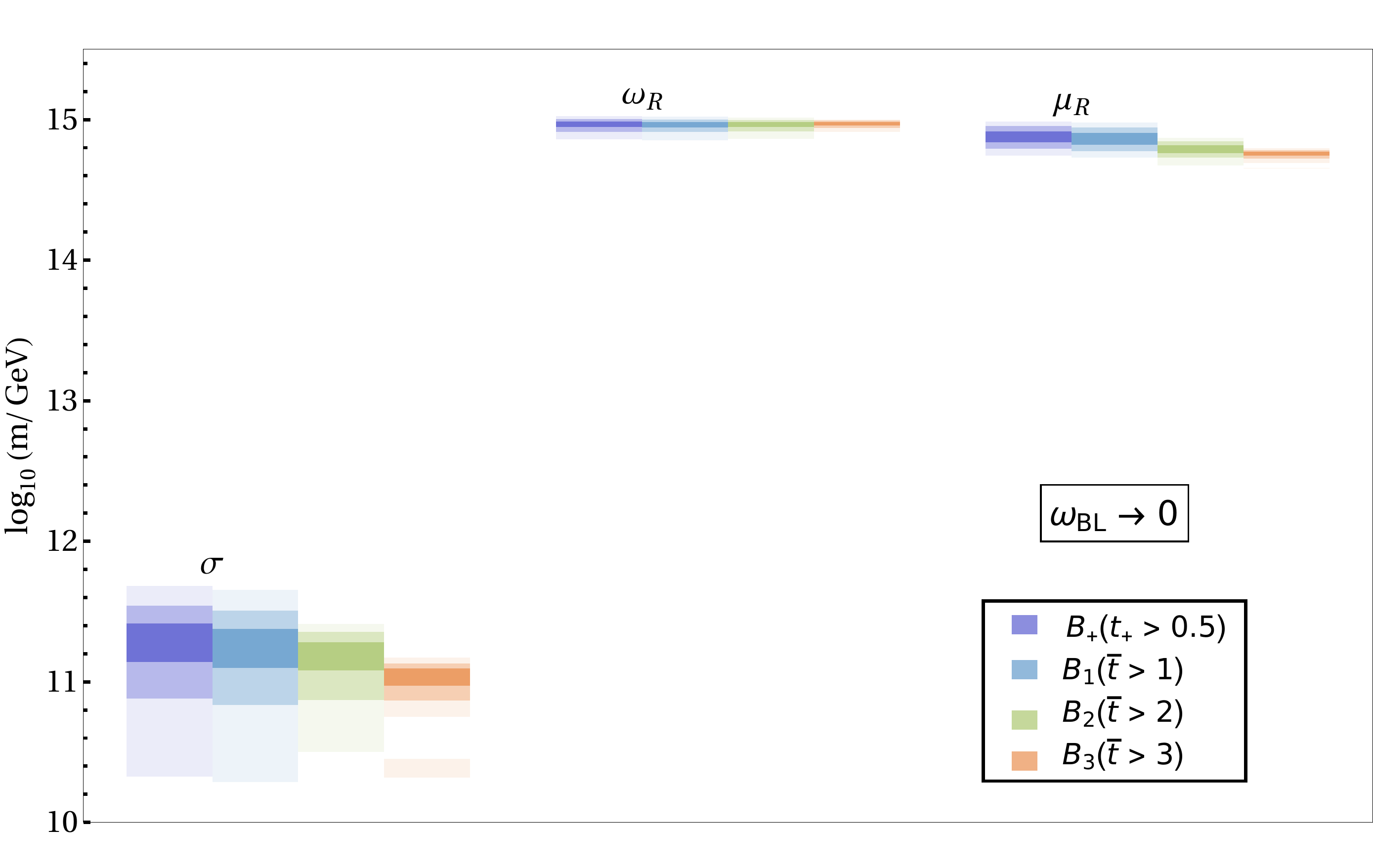}
    \includegraphics[width=9cm]{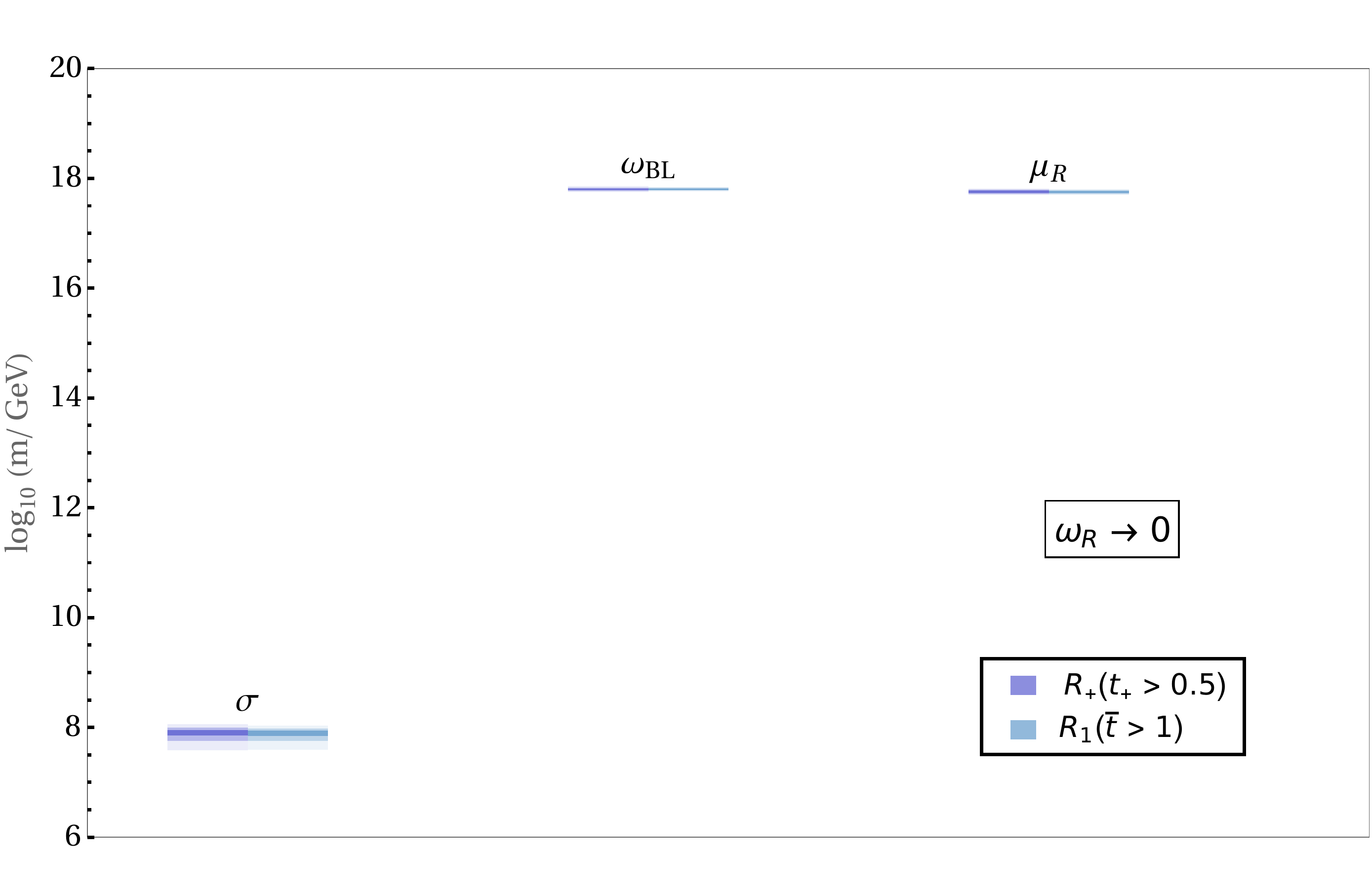}
    }
    \caption{
    $1$-, $2$-, and $3\sigma$ HDIs for the dominant $\omega$ VEV, the $\sigma$ VEV, and the relevant renormalization scale $\mu_{R}$ in the cases $\omega_{BL} \to 0$ (left) and $\omega_R \to 0$ (right). Colors encode different strictness of the RG-perturbativity measure defined in Appendix~\ref{app:perturbativity}; see the legend and Table~\ref{tab:datasets}.
   \label{fig:Scales}
   }
\end{figure*}

The allowed ranges for the relevant VEVs, i.e.,~$\max(|\omega_{BL}|,|\omega_R|)$ and $\sigma$ corresponding to the viable points in both the $\omega_{BL}\to 0$ (left panel) and $\omega_{R}\to 0$ (right panel) limits, are given in Fig.~\ref{fig:Scales}. As before, the data 
corresponding to $1$-, $2$-, and $3\sigma$ HDIs are represented by decreasing opacity, while the colors code different levels of strictness imposed on the RG-perturbativity side; cf.~Sec.~\ref{sec:parameter-ranges}.

From the perturbativity and tachyonicity perspective, the absolute sizes of $\max[|\omega_{BL}|,|\omega_{R}|]$ and $\sigma$  play no role, as nothing changes if these parameters were freely rescaled by a common factor. Thus, the main constraint here comes from the gauge-coupling unification (cf.~Sec.~\ref{Sec:analysis}) in which the $\max[|\omega_{BL}|,|\omega_R|]$ plays the role of the GUT scale, while $\sigma$ sets the seesaw scale. 
To this end, one can expect that the freedom of choosing these two scales together with the value of the unified gauge coupling should, in principle, always admit good fits to the low-energy data given in Appendix~\ref{app:procedure}.

The results in Fig.~\ref{fig:Scales} show that the two scenarios of interest are rather different from this perspective. The $\omega_{R}\to 0$ case [corresponding to the \LR intermediate symmetry] requires the GUT scale to be almost as high as the Planck scale and a very low (yet more constrained) seesaw scale. Consequently, the GUT-to-seesaw-scale hierarchy ratio is rather large. In the opposite case [i.e.,~for $\omega_{BL}\to 0$ with \BLzero as the intermediate symmetry], this hierarchy is generally milder, and the GUT scale of $\sim 10^{15}\GeV$ is rather close to the lower bound implied by proton lifetime limits~\cite{Super-Kamiokande:2012zik,Super-Kamiokande:2014otb,Super-Kamiokande:2016exg,Mine:2016mxy,Super-Kamiokande:2020wjk}. These results agree very well with previous estimates~\cite{Gipson:1984aj,Chang:1984qr,Deshpande:1992au,Deshpande:1992em,Bertolini:2009qj} based on the minimal survival hypothesis\footnote{
It is the presence of lighter gauge bosons in the RGE that crucially contributes to gauge-coupling unification. The scalars
that are accidentally light then mostly just shift the seesaw and GUT scales; see, e.g.,~\cite{Bertolini:2013vta,Kolesova:2014mfa}.}~\cite{delAguila:1980qag,Mohapatra:1982aq}. A more detailed account of the unification constraints is provided in Sec.~\ref{sec:unification}.

\vspace{0.2cm}
\paragraph{The renormalization scale $\mu_{R}$:} \mbox{} \\
\begin{figure*}[htb]
    \centering
    \mbox{
    \includegraphics[width=9cm,height=6cm]{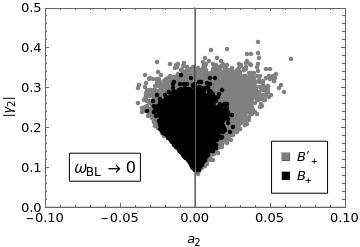}
    \includegraphics[width=9cm,height=6cm]{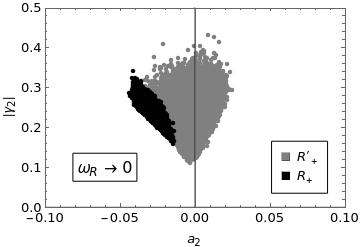}
    }
    \caption{
    Viable parameter-space points projected onto the $a_2$--$|\gamma_2|$ plane in the cases with the non-tachyonicity conditions on the $(\xbar{3},1,+\tfrac{1}{3})$ and $(1,2,+\tfrac{1}{2})$ mass matrices either imposed (in \textbf{\textcolor{black}{black}}) or not imposed (in \textbf{\textcolor{gray}{gray}}). 
 The relevant datasets are detailed in Table~\ref{tab:datasets}.
   \label{fig:Sylvester}   
   }
\end{figure*}
For each point, the quantum-level scalar spectrum computation is performed at a specific renormalization scale $\mu_{R}$ which, in order to tame potentially large logs, we choose to be the square root of the average of all \emph{heavy} scalar tree-level masses-squared (weighed by the numbers of the corresponding real degrees of freedom; for technical details of the procedure,\footnote{Note that $\mu_{R}$ is subject to iterative changes throughout the procedure because the overall scale of all the \emph{heavy} spectrum must be re-adjusted to attain gauge unification, and as such, it cannot be anticipated in advance.} see~Appendix~\ref{app:procedure}). All couplings then depend on the selection of such a $\mu_{R}$ for any particular point. 

Different parameter-space points can be directly compared only when taken at the same $\mu_{R}$ which, in principle, requires RG evolution from their specific renormalization scale(s) to the universal one (using, among other things, the beta functions given in Appendix~\ref{App:BetaFunctions}).
This procedure would be further complicated if some of the points began diverging before they reached the common $\mu_{R}$ or ceased satisfying some of the other viability criteria, some of which are not RG invariant. 

As it turns out, this is more of an academic interest rather than a real hurdle to our analysis because the range of $\mu_R$'s corresponding to fully viable points does not exceed half-an-order of magnitude in either of the two limits; see Fig.~\ref{fig:Scales}. Thus, different points can be compared right away as they are calculated at nearly identical scales. Moreover, the choice of our RG-perturbativity requirements ensures that the viable parameter points could, if desired, all be run safely to a common scale without blowing up.

Numerically, the resulting ranges of $\mu_{R}$ are close to those of the largest VEV for both the $\omega_{BL}\to 0$ and $\omega_{R}\to 0$ limits. 

%......................................................
\subsubsection{Effects of non-tachyonicity conditions applied to the  $(1,2,+\tfrac{1}{2})$ and $(\xbar{3},1,+\tfrac{1}{3})$ multiplets \label{sec:doublets-and-triplets}}
%......................................................

Finally, let us discuss the effect of imposing the non-tachyonicity condition on the SM multiplets  $(1,2,+\tfrac{1}{2})$ and $(\xbar{3},1,+\tfrac{1}{3})$, which in realistic settings should mix with extra components in order to allow for a phenomenologically viable Yukawa sector. In the minimal version, such extra degrees of freedom come from an additional $\mathbf{10}$ in the scalar sector. Consequently, the doublet and triplet mass matrices we have been working with in the $\mathbf{45}\oplus\mathbf{126}$ context are incomplete.    
Nevertheless, as described in Sec.~\ref{sec:Tachyonicity}, even in such a situation the non-tachyonicity conditions can be applied using Sylvester's criterion, and the datasets that we have been working with so far (e.g.,~$B_{+}$ and $R_{+}$; cf.~Table~\ref{tab:datasets}) were all derived with these constraints in play.

It is very interesting though to see what happens if these constraints are not taken into account.\footnote{The tachyonicity of the $(1,2,+\tfrac{1}{2})$ and $(\xbar{3},1,+\tfrac{1}{3})$ multiplets was not discussed in detail in previous attempts~\cite{Bertolini:2013vta,Kolesova:2014mfa}.} For this purpose, special datasets denoted by $B'_{+}$ and $R'_{+}$ have been produced. Technically, these satisfy the same requirements as $B_{+}$ and $R_{+}$, but without imposing non-tachyonicity on the doublet and triplet mass matrices. The effect of this change is best seen in the $a_{2}$--$|\gamma_{2}|$ correlation plot in Fig.~\ref{fig:Sylvester}, where viable regions with and without Sylvester's criterion are compared.

One can see that the impact of Sylvester's criterion is much bigger in the $\omega_{R}\to 0$ regime (the right panel in Fig.~\ref{fig:Sylvester}) where it leads to a significant reduction of the viable parameter space. In particular, positive $a_{2}$ is no longer available in this case (in fact, $a_{2}\lesssim -0.01$). At the same time, the lower bound on $|\gamma_2|$ (expected on the analytical grounds in Sec.~\ref{sec:analytical}) is pushed even higher, thus excluding the interesting low-$|\gamma_2|$ regions corresponding to the most favourable values of $\bar{t}$ of the RG-perturbativity measure. 

Note that this is not the case for $\omega_{BL}\to 0$ (the left panel in Fig.~\ref{fig:Sylvester}) where the doublet and triplet non-tachyonicity criterion does not affect the lower limit on $|\gamma_2|$ at all. 
This can be understood analytically by noticing that in such a limit the critical doublet and triplet fields become members of larger representations\footnote{For instance, the doublet becomes a member of the $(15,2,+\tfrac{1}{2})$ multiplet along with two other propagating fields transforming as $(3,2,+\tfrac{7}{6})$ and $(8,2,+\tfrac{1}{2})$.} of the intermediate \BLzero symmetry (cf.~Appendix~\ref{App:Masses} and Table~\ref{tab:decomposition-126}). Their masses must thus be identical (up to subdominant corrections from $\sigma$) to the companion fields whose non-tachyonicity is always checked. 

Hence, one can conclude that the non-tachyonicity constraints imposed on the triplet and doublet scalars play a very important role in determining the shape of the viable parameter space, and they are at the core of the aforementioned preference of the $\omega_{BL}\to 0$ scenario with respect to the $\omega_{R}\to 0$ one.

%======================================================
\subsection{Results for the mass spectrum \label{sec:spectrum}}
%======================================================

\begin{figure*}[htb]
    \centering
    \mbox{
   \includegraphics[width=9cm]{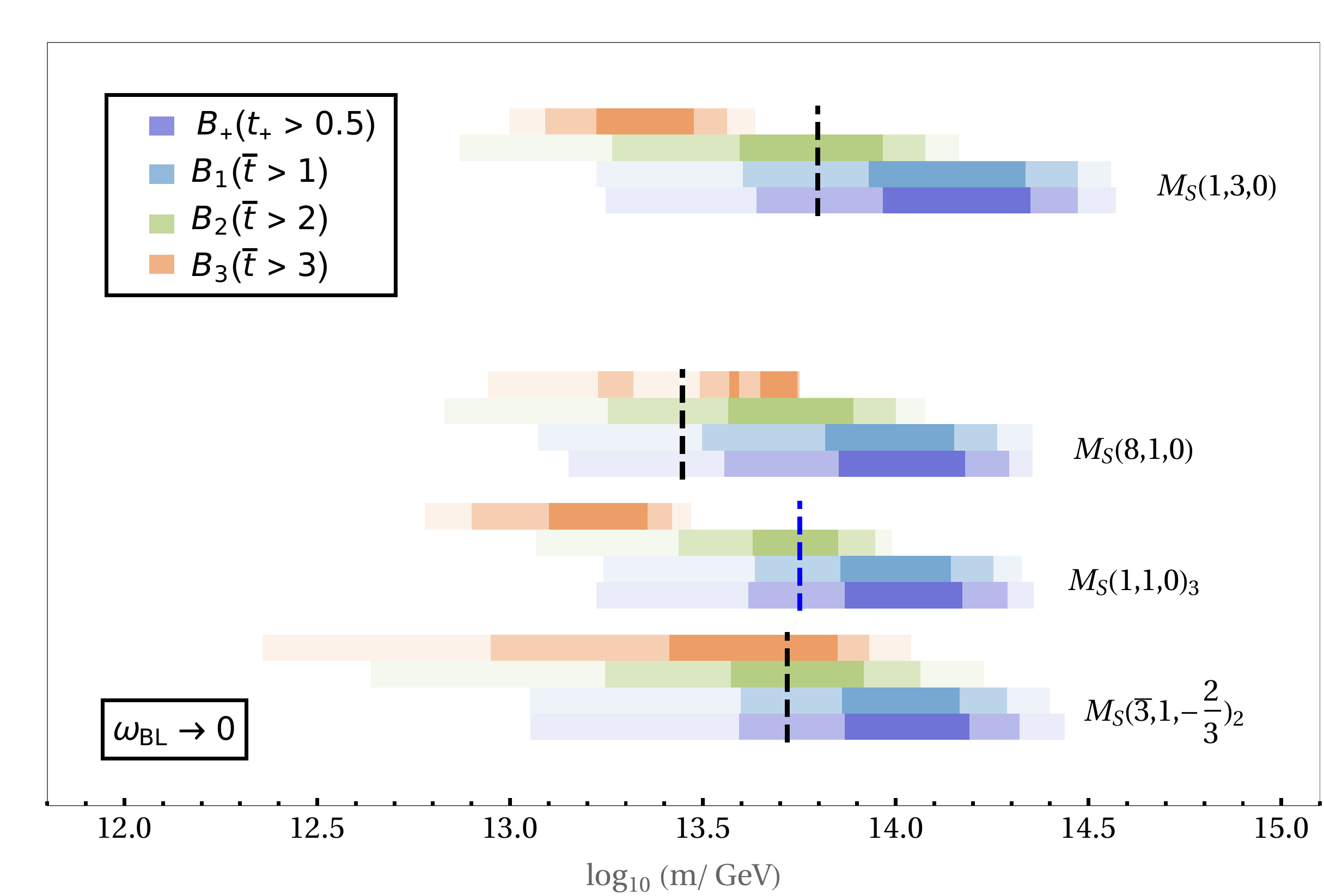}
    \includegraphics[width=9cm]{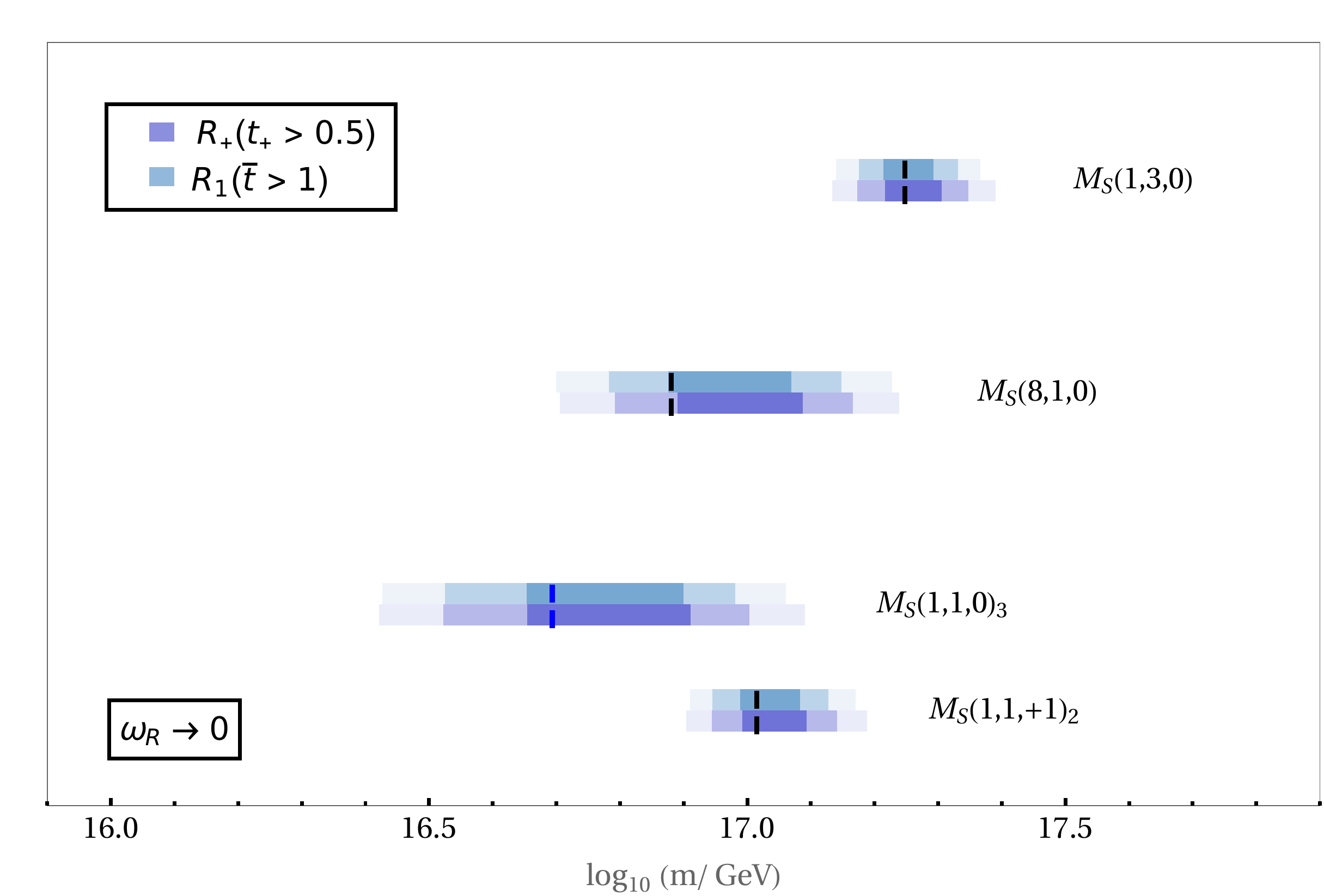}
    }
    \caption{
    The allowed $1$-, $2$-, and $3\sigma$ HDI ranges (corresponding to decreasing opacity) for the PGB masses in the $\omega_{BL}\to 0$ (left) and $\omega_{R}\to 0$ cases (right). Colors (see, e.g.,~Fig.~\ref{fig:ScalarParameters} for the legend) represent different datasets in Table~\ref{tab:datasets} obtained for different levels of strictness of the RG-perturbativity measure $\bar{t}$ defined in Appendix~\ref{app:perturbativity}. The dashed lines denote the sample points used in Fig.~\ref{fig:Unification}.
   \label{fig:PGBFields}}
\end{figure*}

\begin{figure*}[htb]
    \centering
    \mbox{
    \includegraphics[width=8cm]{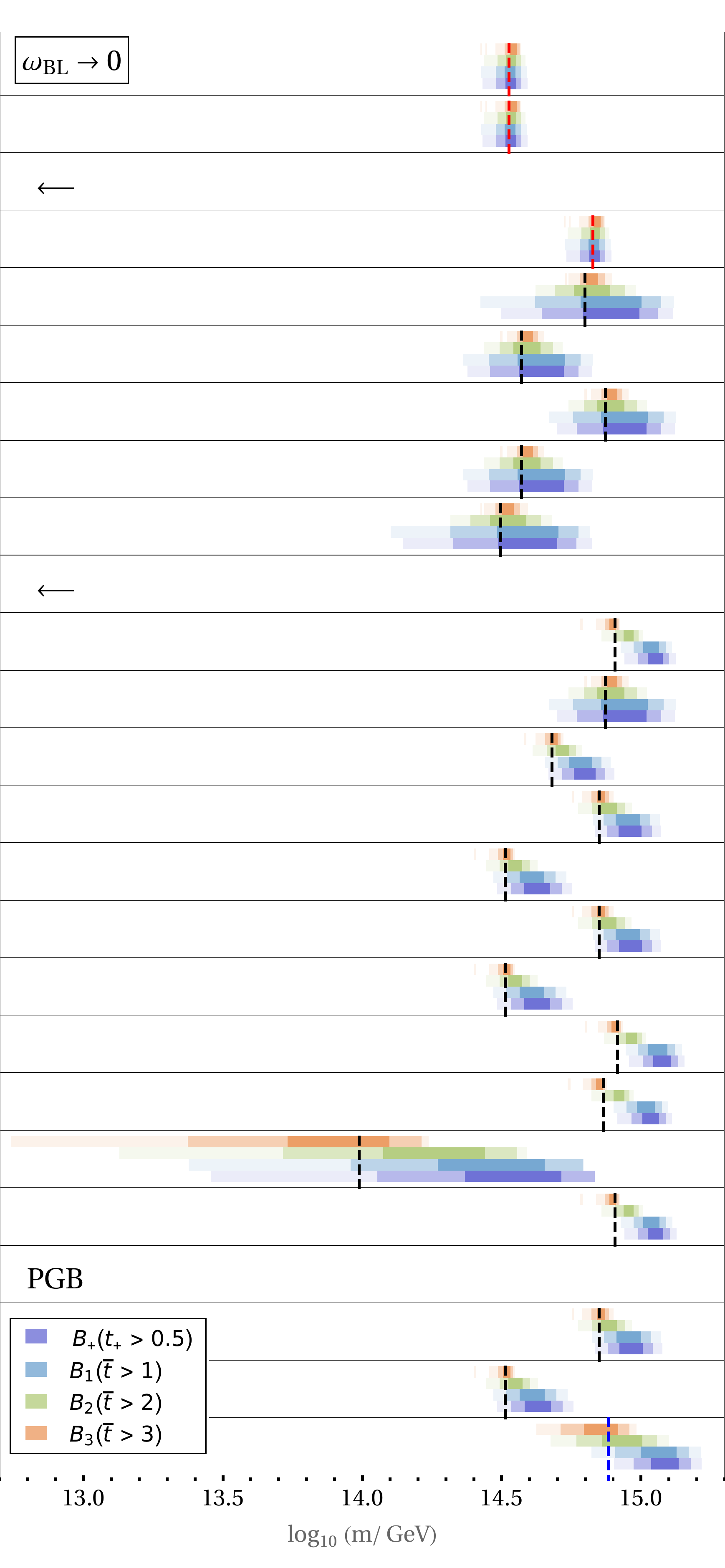}
    \includegraphics[width=8cm]{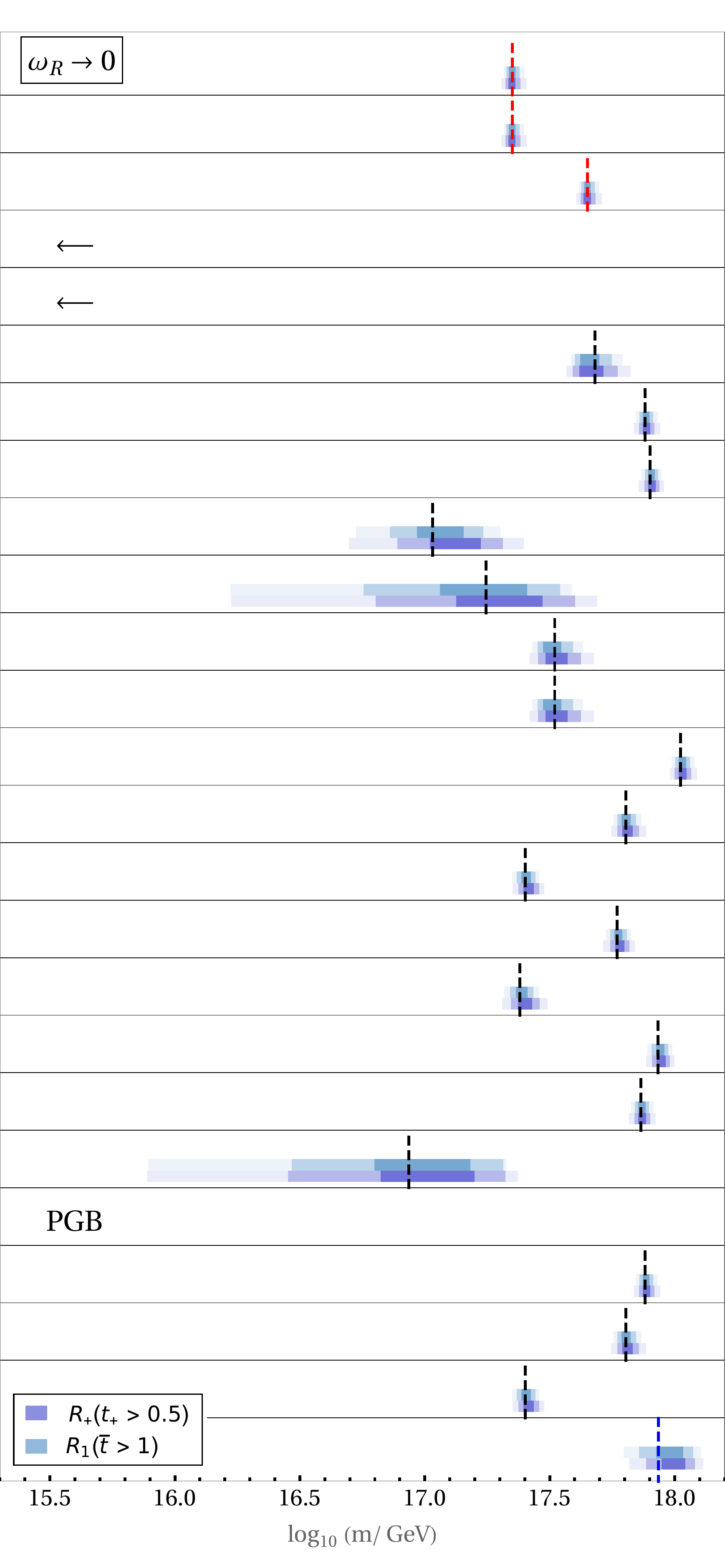}
     \includegraphics[width=2cm]{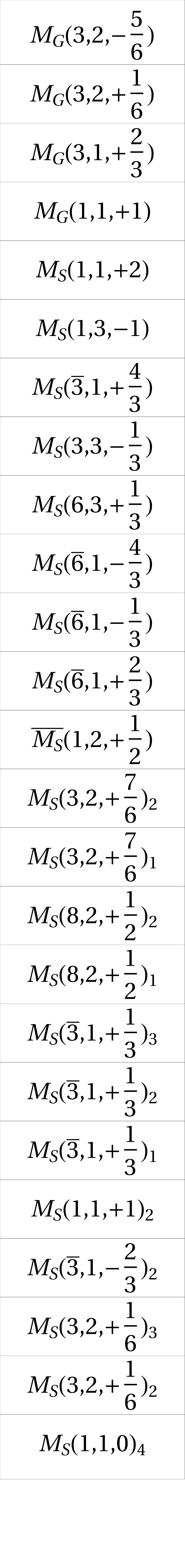}
    }
    \caption{
    The predicted $1$-, $2$-, and $3\sigma$ ranges (corresponding to decreasing opacity) for the \emph{heavy} gauge ($M_G$) and scalar ($M_S$) masses governed by $\omega_{R}$ (in the $\omega_{BL}\to 0$ case, left panel) or by $\omega_{BL}$ (in the $\omega_{R}\to 0$ case, right panel). As before (cf.~Fig.~\ref{fig:ScalarParameters}), the color code represents different datasets of Table~\ref{tab:datasets} corresponding to different levels of strictness of the RG-perturbativity measure $\bar{t}$ defined in Appendix~\ref{app:perturbativity}. Left arrows point to masses which fall outside the displayed ranges (see Fig.~\ref{fig:LightFields}), and PGB labels the pseudo-Goldstone field whose mass is plotted in Fig.~\ref{fig:PGBFields}. 
    The dashed lines denote the sample points used in Fig.~\ref{fig:Unification}.
   \label{fig:HeavyFields}}
\end{figure*}

\begin{figure*}[htb]
    \centering
    \mbox{
  \includegraphics[width=9cm]{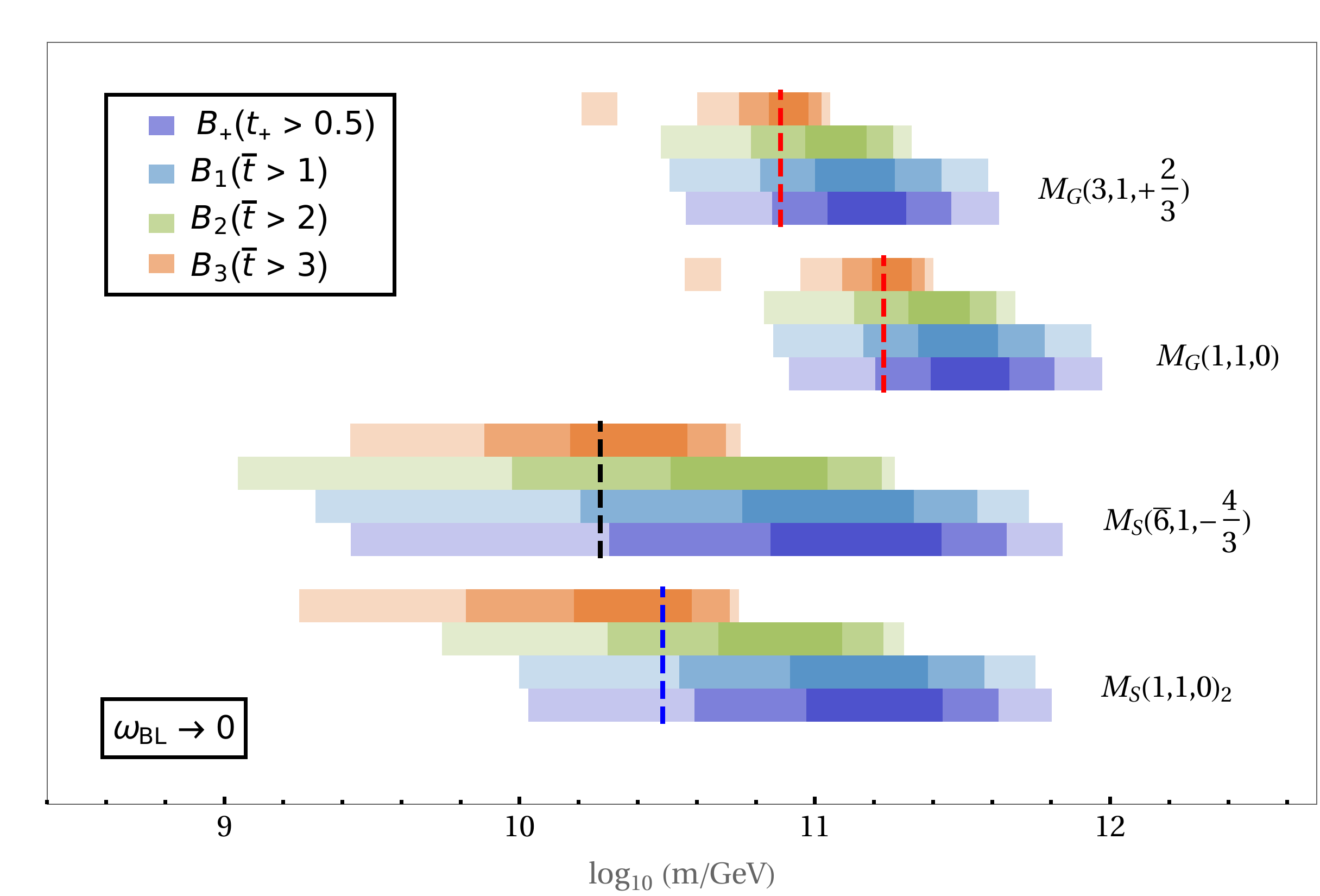}
  \includegraphics[width=9cm]{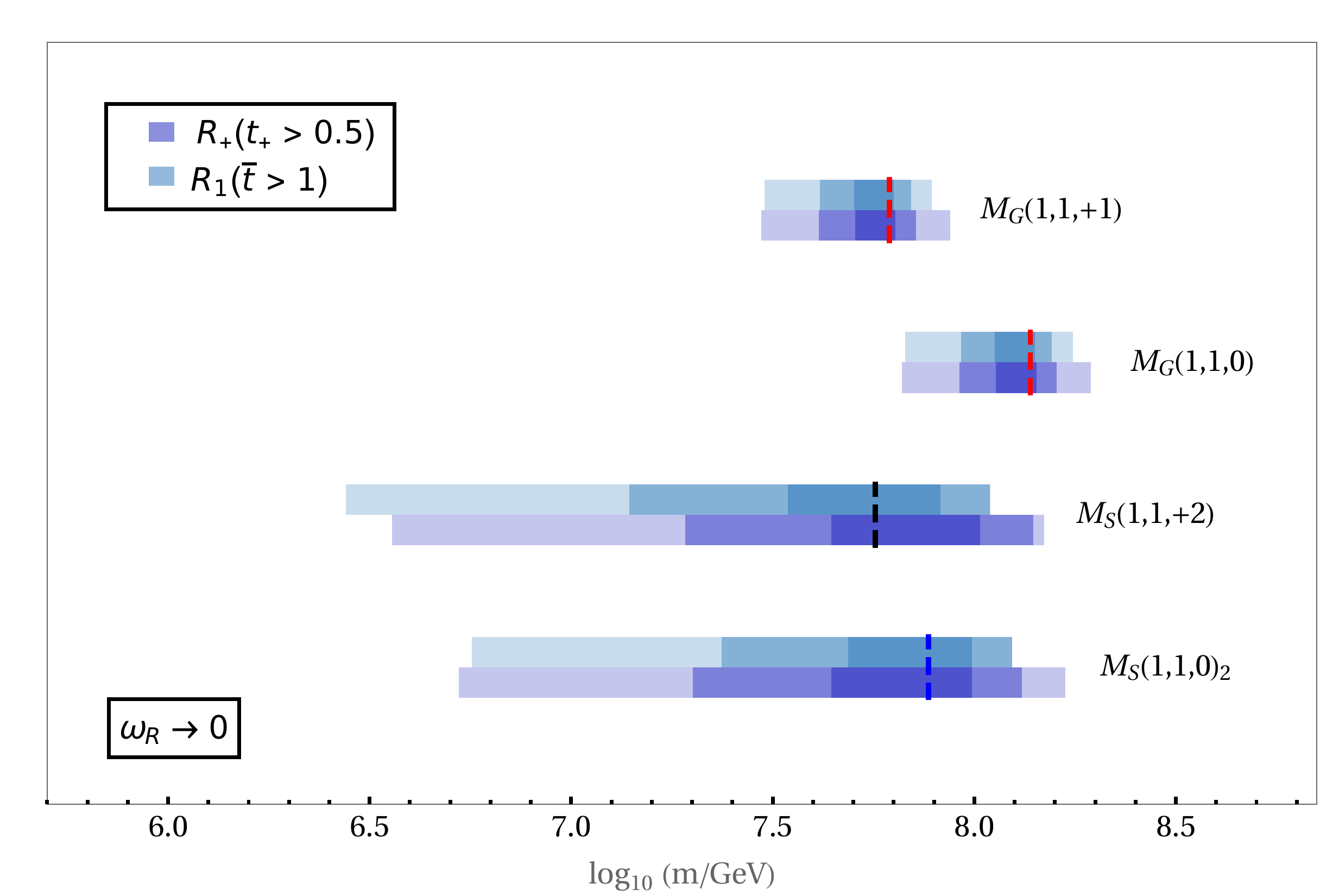}
    }
    \caption{\label{fig:LightFields}
    The mass ranges for the fields associated with the intermediate-symmetry breaking driven by the $\sigma$ scale. The conventions are the same as in Figs.~\ref{fig:PGBFields} and~\ref{fig:HeavyFields}.
   }
\end{figure*}

Next, let us turn our attention to the bosonic (i.e,.~scalar and vector) spectrum of the model. As we have already seen in Sec.~\ref{sec:parameter-space}, the criteria of non-tachyonicity, gauge-coupling unification, and perturbativity (cf.~Sec.~\ref{Sec:analysis}) shrink the viable parameter space to rather small patches, and the resulting mass ranges typically turn out to be quite narrow as well.
The results are given in a series of Figs.~\ref{fig:PGBFields}--\ref{fig:LightFields}, which correspond to three distinct classes of fields with respect to their characteristic mass scales:
\begin{enumerate}
    \item The masses of the PGB scalars (see Sec.~\ref{sec:PSG-singlets}) and those of the associated fields\footnote{Associated fields are those that in the two limits of interest belong to the same multiplet as one of the ``genuine'' PGBs discussed in  Sec.~\ref{sec:PSG-singlets}.} in the two limits of our interest ($\omega_{BL}\to 0$ and $\omega_{R}\to 0$) are covered in Fig.~\ref{fig:PGBFields}. This class of fields is especially prone to tachyonic instabilities and thus the main 
    motivation behind the one-loop analysis carried out in this study. 
    \item The spectrum of the \emph{heavy} GUT-scale fields (both scalars and vectors), i.e.,~those associated with the first stage\footnote{Let us use this simplified terminology here despite the fact that we envision the breaking to occur in a single step, albeit with a hierarchy of VEVs, rather than a true multi-stage breaking due to a dynamical mechanism.} of the unified symmetry breaking, are shown in Fig.~\ref{fig:HeavyFields}. 
    \item The masses of the intermediate $\sigma$-scale fields associated with the $\mathrm{U}(1)_{B-L}$ (i.e.,~second stage) symmetry breaking are displayed in Fig.~\ref{fig:LightFields}.
\end{enumerate}

In all these figures, $M_S$ or $M_G$ indicate the scalar or vector (gauge) boson nature of the multiplet whose SM transformation properties are given in the adjacent bracket (in the case of degeneracy, the multiplicity subscript follows an ascending mass order).
For completeness, the masses of the $(1,2,+\tfrac{1}{2})$ and $(\xbar{3},1,+\tfrac{1}{3})$ scalars are also included here, despite the fact that these may be subject to further changes in Yukawa-realistic scenarios with an additional $\mathbf{10}$ in the scalar sector (cf.~Secs.~\ref{sec:Tachyonicity} and~\ref{sec:unification-constraints}). 
Moreover, one doublet mass (the SM Higgs doublet) must be fine-tuned to the EW scale. We simulate this effect in our spectrum by replacing \emph{ad hoc} the two doublets of the model with the SM Higgs doublet and a heavy companion, whose mass $\overline{M_{S}}$ is computed as the geometric mean of the two eigenvalues of the $(1,2,+\tfrac{1}{2})$ mass matrix.

There are several points and observations of the results worth making here.
\begin{enumerate}
    \item In both cases of interest, i.e.,~in the $\omega_{BL}\to 0$ and $\omega_{R}\to 0$ limits, the shapes of the bosonic spectra confirm the expectations based on the structure of the associated symmetry-breaking patterns: 
        \begin{itemize}
        \item For a given limit scenario, the predicted ranges of different SM states sometimes closely resemble each other. These near degeneracies correspond to sets of SM representations belonging to the same intermediate-symmetry representation, where only their $|\sigma|$-proportional mass contributions originating from the second stage of symmetry breaking split degeneracy. These patterns are consistent with the tree-level expressions in Appendices~\ref{app:gauge masses} and~\ref{app:scalar masses}. 
        As an example, compare the mass ranges of the \emph{heavy} $(1,1,+2)$, $(\xbar{3},1,+\tfrac{4}{3})$, and $(\xbar{6},1,+\tfrac{2}{3})$ scalars in the $\omega_{BL}\to 0$ case. They are similar due to belonging to the same $(\xbar{10},1,+1)$ representation of the intermediate \BLzero symmetry; cf.~Table~\ref{tab:decomposition-126} in Appendix~\ref{app:decompositions}. 
        \item A direct consequence of the existence of an effective intermediate symmetry is that for either of the two scenarios an additional state joins the ranks of PGBs; cf.~Fig.~\ref{fig:PGBFields}. In particular, a complex $(\xbar{3},1,-\tfrac{2}{3})$ groups together with the singlet and octet PGBs in the $(15,1,0)$ representation of the \BLzero intermediate symmetry attained in the $\omega_{BL}\to 0$ scenario, while a complex $(1,1,-1)$ scalar joins the singlet in the  $(1,1,3,0)$ representation of the intermediate \LR in the $\omega_{R}\to 0$ case, as indicated by the decompositions of Table~\ref{tab:decomposition-45} in Appendix~\ref{app:decompositions}. These features are illustrated in Fig.~\ref{fig:PGBFields} by grouping the additional states with the associated PGBs, where the vertical spacing between them signifies the decomposition under the intermediate symmetry.
        \end{itemize}
    \item Interestingly, the GUT-scale bosonic spectrum of the $\omega_{BL}\to 0$ scenario is significantly lighter than that of the $\omega_{R}\to 0$ case (see Fig.~\ref{fig:HeavyFields}), while the opposite holds true for the $\sigma$-associated masses in Fig.~\ref{fig:LightFields}. This is in accordance with the VEV hierarchy given in Fig.~\ref{fig:Scales}. The gap between the GUT and the seesaw scale is thus much more pronounced in the latter case, amounting to about $10$ orders of magnitude, than in the $\omega_{BL}\to 0$ setting, where it is just about $4$ orders of magnitude. Note that this behaviour is in accordance with the previous estimates based on the minimal survival hypothesis; cf.~\cite{Bertolini:2009qj}. From a model-building perspective, the $\omega_{BL}\to 0$ scenario is therefore \emph{again} far more attractive, as one does not need to resort to large fine-tunings to attain potentially realistic flavour patterns (including realistic neutrino masses). Moreover, the proximity of $\omega_{BL}$ to the Planck scale in the $\omega_{R}\to 0$ case raises issues with  theoretical uncertainties due to enhanced contributions from $D > 4$ operators.    
    \item Concerning the relative positions and widths of the ranges corresponding to different datasets, one can see several effects in Figs.~\ref{fig:PGBFields}--\ref{fig:LightFields}:
        \begin{itemize}
            \item Enhancing RG perturbativity typically lowers the allowed ranges of the scalar masses, especially those of the PGBs in Fig.~\ref{fig:PGBFields}. This happens mostly due to the preference for smaller scalar couplings as indicated, e.g.,~in Figs.~\ref{fig:ScatterPlotsVSBL} and~\ref{fig:ScatterPlotsVSR}, where perturbativity generally improves toward the origin.
            \item The mass ranges for the heaviest fields are relatively narrow; cf.~Fig.~\ref{fig:HeavyFields}. For the gauge fields, 
            this is due to gauge unification constraining the values of the GUT-scale VEV and $\mathrm{SO}(10)$ gauge coupling; see Secs.~\ref{sec:VEVs-and-muR} and~\ref{sec:unification}. As for the \emph{heavy} scalars, the effect can be attributed to the structure of their mass formulae, which are often dominated by a coupling that is significantly constrained by the perturbativity criteria of Sec.~\ref{sec:parameter-space}. 
            \item For the fields whose mass origin is less definite (such as the PGBs, for which the tree-level mass contributions often compete with the loop effects), the main effect of increasing the RG-perturbativity strictness often corresponds to a shift rather than a compression of their mass ranges (the orange or green bars are just as wide as the blue or light blue ones).
        \end{itemize}
    \item Finally, we caution the eager reader against the temptation of making ballpark predictions for proton lifetime based on the presented gauge boson masses, since this requires a far more elaborate two-loop running analysis of gauge couplings in a Yukawa-realistic scenario (as opposed to the one-loop running analysis in a simplified Higgs model given here) along with a dedicated analysis of all other relevant theoretical uncertainties. These, altogether, can potentially change the naive gauge-boson-mass-based proton lifetime estimates by orders of magnitude.
    The same applies to the (usually subdominant) scalar-driven contributions --- not only are we missing the complete information about one of the key mediators [the $(\xbar{3},1,+\tfrac{1}{3})_S$ scalar leptoquark $S_1$], but also the mass ranges of other potentially relevant $\SU(3)_c$-triplets like $\tilde{S}_1 \equiv (\xbar{3},1,+\tfrac{4}{3})_S$ and $S_3 \equiv (3,3,-\tfrac{1}{3})_S$ are relatively wide; cf.~Fig.~\ref{fig:HeavyFields}. This, however, is beyond the scope of the current study and will be elaborated on elsewhere. Nonetheless, it is reassuring that in the obtained spectra all potentially harmful states [including the $(3,2,-\tfrac{5}{6})_G$ and $(3,2,+\tfrac{1}{6})_G$ vector leptoquarks] have masses well above $10^{14}\GeV$, and thus, they do not trivially violate any direct phenomenological bounds.
\end{enumerate}
%

%======================================================
\subsection{Gauge-coupling unification \label{sec:unification}}
%======================================================

\begin{figure*}[t]
    \centering
    \mbox{
    \includegraphics[width=9cm]{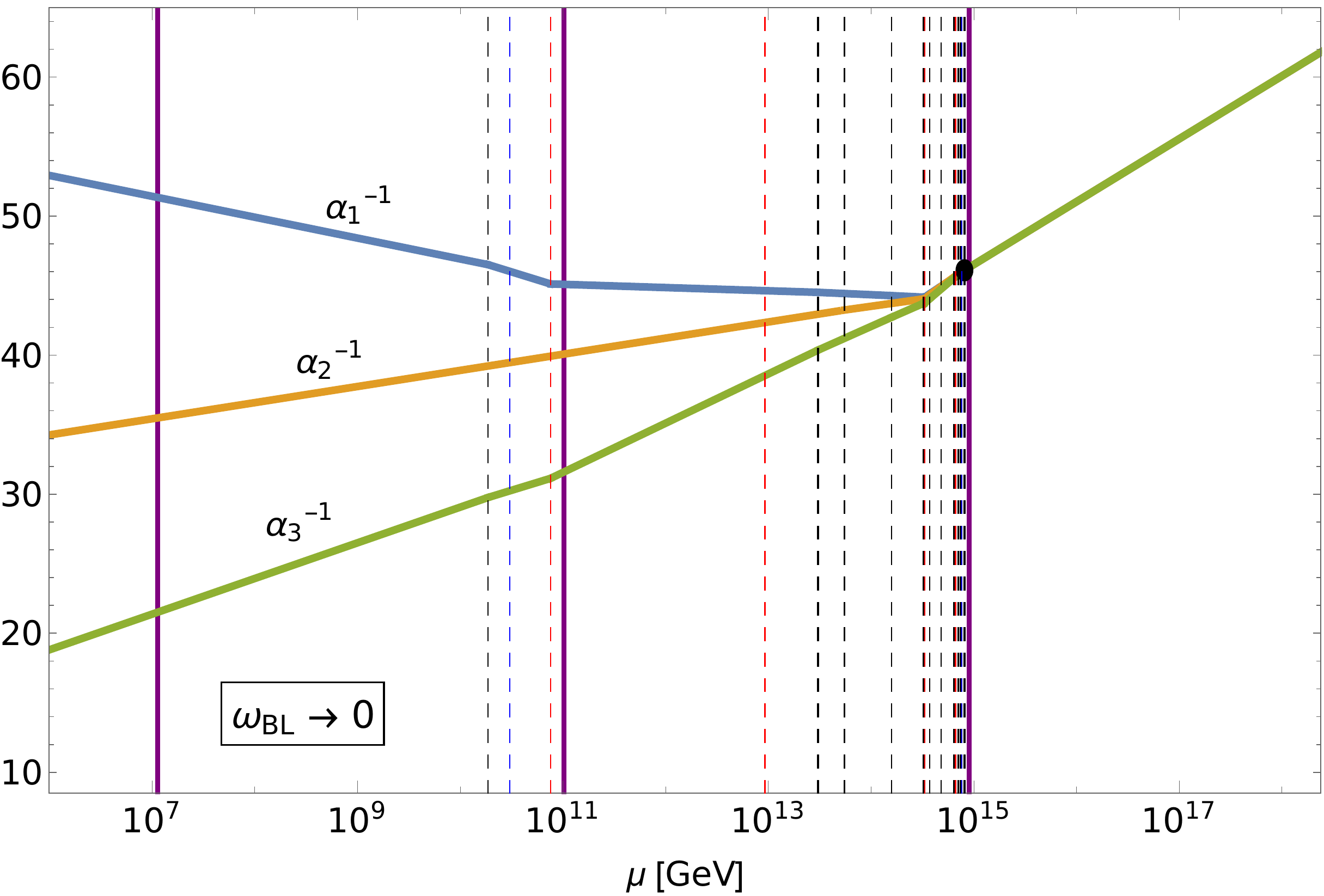}
    \includegraphics[width=9cm]{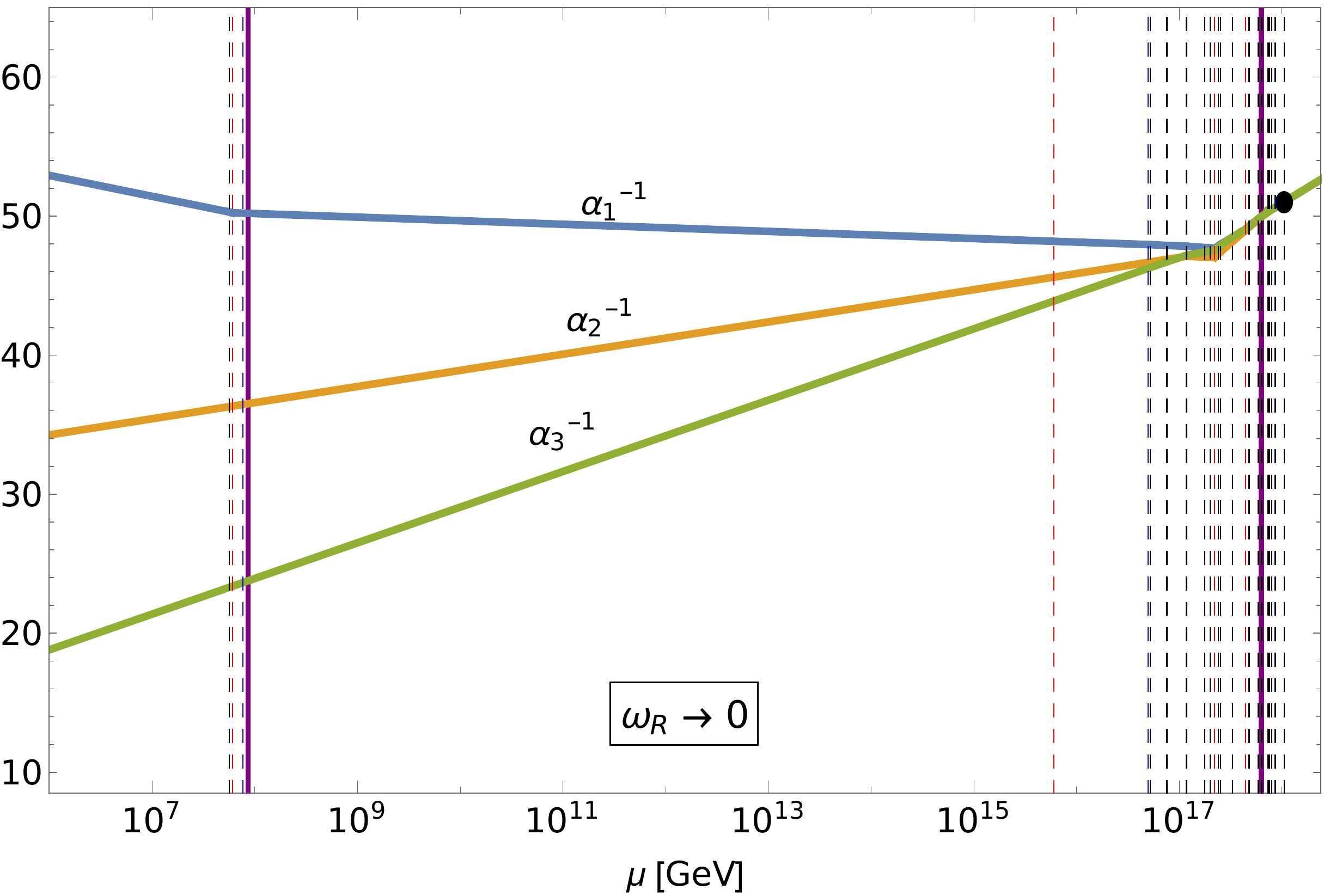}
    }
    \caption{Sample gauge-coupling-unification patterns for two viable parameter-space points specified in Table~\ref{tab:points-for-unification} for the $\omega_{BL}\to 0$ (left) and $\omega_{R}\to 0$ (right) cases of interest. The dashed vertical lines at different scales correspond to changes of the one-loop $\beta$ coefficients due to the presence of various fields' contributions, while the solid purple lines denote positions of the involved VEVs [namely, the GUT-scale VEV corresponding to $\max[|\omega_{BL}|,|\omega_{R}|]$, the intermediate-scale $|\sigma|$, and in the left panel also the induced VEV ($\omega_{BL}$ in this case)]. The \textbf{black}, \textcolor{blue}{blue}, and \textcolor{red}{red} dashed vertical lines indicate masses of SM non-singlet scalars, SM-singlet scalars, and gauge bosons, respectively (corresponding to vertical dashed lines in Figs.~\ref{fig:PGBFields}--\ref{fig:LightFields}). Finally, the \textbf{black dots} denote the point of gauge-coupling unification. Their horizontal positions correspond to the masses of the heaviest SM non-singlet thresholds (scalar or gauge) in each scenario.
   \label{fig:Unification}}
\end{figure*}

Finally, let us present a couple of examples of how the mass patterns described in previous sections satisfy the gauge unification criterion. For this purpose, we select two representative points
from the $B_{\text{RG}}$ and $R_{\text{RG}}$ datasets of Table~\ref{tab:datasets} that correspond to the $\omega_{BL}\to 0$ and $\omega_{R}\to 0$ limits.
The relevant parameter-space points are specified in Table~\ref{tab:points-for-unification}, and the one-loop gauge running (and unification) patterns are depicted in Fig.~\ref{fig:Unification}.
\begin{table*}[tb]
    \centering
    \caption{
        The input parameters for the two example points, where $\slog(x):=\log_{10}(x/\GeV)$. The signs of $\omega_{BL}$ and $\omega_{R}$ are positive. Strictly speaking, the scale $\mu_{R}$ is not a free input and it is computed; cf.~Appendix~\ref{App:viability-considerations}. \label{tab:points-for-unification}}
    \begin{tabular}{r @{$\quad$} r r r r r r r r r} % 1 + 9
        \hline
         \multicolumn{1}{c}{case}&
         \TBOX{$10^{2}\,a_2$}  &  \TBOX{$10^{2}\,a_0$} &
         \TBOX{$10^{1}\,\lambda_0$} & \TBOX{$10^{1}\,\lambda_2$} &
         \TBOX{$10^{2}\,\lambda_4$} & \TBOX{$10^{2}\,\lambda_4'$} & 
         \TBOX{$10^{1}\,\alpha$}    &  \TBOX{$10^{1}\,\beta_4$}   &  \TBOX{$10^{2}\,\beta_4'$} \\
         \hline
         $\omega_{BL}\to 0$& 
         $0.189$  & $4.81$ & $0.647$ & 
         $-0.704$  & $3.69\ZZ$ & $0.965$ &
         $-0.616$  & $3.41$  & $-4.27$ \\
         $\omega_{R}\to 0$ & 
         $-2.74\ZZ$  & $5.98$ & $1.38\ZZ$ &
         $-1.58$  & $2.51$ & $4.56\ZZ$ & 
         $-0.634\ZZ$ & $8.56$ & $-1.80$\\
         \hline
         \multicolumn{1}{c}{case}& 
         $10^{1}\,|\gamma_2|$  & $10^{2}\,|\eta_2|$ & $\arg\gamma_2$ & 
         $\arg\eta_2$          & $g\quad$ & $\slog|\omega_{BL}|$ & 
         $\slog|\omega_{R}|$   & $\slog |\sigma|$   & $\slog\mu_{R}$ \\
         \hline
         $\omega_{BL}\to 0$& 
         $0.917$  & $0.696$  & $3.38$ &
         $6.22$    & $0.526$    & $7.056$ & 
         $14.96\ZZ$   & $11.01\ZZ$    & $14.75$ \\
         $\omega_{R}\to 0$& 
         $2.28\ZZ$   & $2.54$  & $4.78$ &
         $3.10$    & $0.502$    & $17.80\ZZ$ & 
         $-2.200$  & $7.938$    & $17.76$ \\
         \hline
    \end{tabular}
\end{table*}

Several remarks are perhaps worth making here:
\begin{itemize}
    \item There is a clear qualitative difference between the $\omega_{BL}\to 0$ and $\omega_{R}\to 0$ scenarios in the positions of the two characteristic scales (namely, the $M_{\text{GUT}}$ and seesaw scale) and in the clustering of the relevant states around these. This is in accord with the discussion in Secs.~\ref{sec:VEVs-and-muR} and~\ref{sec:spectrum}. 
    It should also be pointed out that besides perturbativity, unification represents another important argument in favor of considering only the symmetry-breaking chains
    along the two special ``maximally hierarchical'' directions.
    Assuming only a single (non-SM) light threshold admitted in the bulk that can aid the (one-loop) unification, there
    are then just two viable possibilities\footnote{Note that the contribution of vector states was crucial for unification even in scenarios with either the $(6,3,+\tfrac{1}{3})$ scalar in the desert~\cite{Kolesova:2014mfa} or the exceptionally light scalar $(8,2,+\tfrac{1}{2})$~\cite{Bertolini:2013vta}.}: either having a $(3,1,+\tfrac{2}{3})$ gauge boson at $\approx 10^{12.5} \GeV$ with couplings unifying at $M_{\text{GUT}}\approx 10^{15}\GeV$, or a $(1,1,+1)$ gauge boson of mass $\approx 10^{10}\GeV$ and unification achieved at $M_{\text{GUT}}\approx 10^{17} \GeV$. This agrees reasonably well with the results in Fig.~\ref{fig:LightFields}.
    If at the same time we require that the proton-decay-mediating $(3,2,+\tfrac{1}{6})$ vector leptoquark remains \emph{heavy}, that implies a very strong preference for either the $|\omega_{BL}|, |\sigma|\ll |\omega_R|$- or $|\omega_R|, |\sigma|\ll |\omega_{BL}|$-breaking pattern; cf.~Table~\ref{tab:tree-level-gauge-boson-masses} for gauge boson masses. The produced scales $\omega_{BL}$, $\omega_R$, and $\sigma$ are in very good agreement with the results of~\cite{Bertolini:2009qj}.
    \item Given the relatively shallow angle under which the three gauge couplings eventually unify,\footnote{Interestingly, the two non-Abelian couplings actually intersect twice in the $\omega_{R}\to 0$ scenario.}
    one can expect that the two-loop effects (including contributions from the Yukawa couplings that we ignore here\footnote{A rough estimate of the size of the two-loop Yukawa contributions to the relevant beta functions can be found, e.g.,~in~\cite{Bertolini:2009qj}.}) may cause significant shifts in both GUT and seesaw scales. The figures given in this study should thus be understood as a mere first approximation to the fully physical picture.
\end{itemize}
%

%%%%%%%%%%%%%%%%%%%%%%%%%%%%%%%%%%%%%%%%%%%%%%%%%%%%%%%
\newpage
\section{Conclusions and outlook\label{Sec:Conclusions}}
%%%%%%%%%%%%%%%%%%%%%%%%%%%%%%%%%%%%%%%%%%%%%%%%%%%%%%%

The minimal renormalizable $\mathrm{SO}(10)$ Higgs model with $\mathbf{45}\oplus\mathbf{126}$ has attracted much attention in the past decade~\cite{Bertolini:2013vta,Kolesova:2014mfa,Graf:2016znk}. It is well known that at tree level, possible SM-like vacua suffer from tachyonic instabilities either in the color-octet or weak-triplet directions. We refer to these states as PGBs, since their tree-level masses are proportional to only a single scalar-potential coupling ($a_2$). The tree-level tachyonic instabilities thus force us to consider the model at the quantum level, significantly complicating the analysis. 

While the previous state of the art was the derivation of analytic PGB singlet, octet, and triplet one-loop mass formulas in the $a_2 \to 0,~ \gamma_2\to 0,~ \sigma \to 0$ regime~\cite{Graf:2016znk,JarkovskaThesis}, in this work we have developed a numerical procedure for the computation of one-loop masses of \emph{all} scalar fields associated with the GUT scale. Another important result is the analytic formulae for the one-loop beta functions of all dimensionless parameters in the scalar potential.

The new computational tools have allowed us to perform a comprehensive analysis of the Higgs model taking into account the following considerations:
\begin{enumerate}
\item \emph{Non-tachyonicity:} \par This criterion is a rigorous requirement for the consistency of the theory. Most prone to develop a tachyonic instability are the PGB states: the octet, the triplet, and the SM singlet.
\item \emph{Perturbativity:} \par In order for the perturbative calculation in a given parameter point to be valid, loop corrections to masses, as well as the coupling values under RG running, need to be under perturbative control. The developed numerical tools have allowed us to consider both issues by constructing appropriate perturbativity measures. 
\item \emph{Gauge-coupling unification:} \par This last criterion is phenomenological and puts requirements on the mass spectrum of the theory. We consider one-loop unification only.
\end{enumerate}

The results of our analysis are as follows:
\begin{itemize}
\item We have argued in Sec.~\ref{sec:quantum-level} that perturbativity requires the structure~\eqref{eq:chi-definition} 
to be kept under control, concluding that the model is perturbative only in a regime where the VEV of the $\mathbf{45}$ is dominantly aligned along the $\mathrm{SU}(4)_C\times\mathrm{SU}(2)_L\times\mathrm{U}(1)_R$ or $\mathrm{SU}(3)_c\times\mathrm{SU}(2)_L\times\mathrm{SU}(2)_R\times\mathrm{U}(1)_{B-L}$ direction. These two scenarios are referred to as $\omega_{BL}\to 0$ and $\omega_{R}\to 0$, respectively, indicating which of the two VEVs in the adjoint gets merely an induced value. The same two scenarios are preferred from the point of view of unification and proton lifetime; cf.~Sec.~\ref{sec:unification}.
\item Although we have found viable parameter points in both scenarios, there seems to be a preference for the $\omega_{BL}\to 0$ case. This holds true both from the perturbativity point of view, since the more stable points (especially with respect to RG perturbativity) were found in that scenario, as well as phenomenologically due to the better-suited seesaw and GUT scales at $10^{11}$ and $10^{15}\GeV$ compared to $10^{8}$ and $10^{18}\GeV$ in the $\omega_{R}\to 0$ case.
\item The viable part of parameter space does not admit all parameter values to vanish. In particular, a non-tachyonic spectrum requires that $\gamma_{2}$ is not close to zero in either scenario (cf.~Sec.~\ref{sec:analytical}), contrary to assumptions in previous studies~\cite{Bertolini:2013vta,Kolesova:2014mfa,Graf:2016znk}. This shows that an implementation of the one-loop PGB mass calculation crucially requires this parameter to be present.
\end{itemize}

This work represents significant progress in the efforts to elucidate the viability and consequences of the $\mathrm{SO}(10)$ Higgs model containing $\mathbf{45}\oplus\mathbf{126}$. Having identified the viable parameter regions, the obvious next step would be to implement the full model with a realistic Yukawa sector by adding a scalar representation $\mathbf{10}$ and upgrade its unification analysis to two-loop order. The additional states would also allow us to treat properly the EW symmetry breaking. Because of the small number of added fields and numerous new parameters in the scalar potential, the Higgs-model results presented here are expected to be a good approximation for the fully realistic case. However, the addition of fermions allows for the implementation of further phenomenological constraints, e.g.,~a proton decay rate prediction and computation of neutrino masses (with both type-I and type-II seesaw contributions present in general).

%%%%%%%%%%%%%%%%%%%%%%%%%%%%%%%%%%%%%%%%%%%%%%%%%%%%%%%
\section*{Acknowledgments}
%%%%%%%%%%%%%%%%%%%%%%%%%%%%%%%%%%%%%%%%%%%%%%%%%%%%%%%

K.J. and M.M. acknowledge the financial support from the Grant Agency of the
Czech Republic (GA\v{C}R) through Contract No.~20-17490S and from
the Charles University Research Center Grant No.~UNCE/SCI/013. K.J.  was also supported by the Charles University Grant Agency (GAUK) as part of Project No.~1558119.
The work of T.M. has been supported by the Slovenian Research Agency under the Ad-futura Research Grant No.~11013-28/2018.
T.M. would also like to thank the IPNP in Prague for their warm hospitality, cold beer and partial financial support during his stay there
and acknowledge the support and hospitality of the Theoretical Physics Department at the Jo\v{z}ef Stefan Institute, Ljubljana, where
the early part of this study was conducted.
The work of V.S. was supported by the Swiss National Science Foundation (SNF).

\appendix

%%%%%%%%%%%%%%%%%%%%%%%%%%%%%%%%%%%%%%%%%%%%%%%%%%%%%%%
\section{Technical details of the one-loop mass computation\label{App:one-loop masses}}
%%%%%%%%%%%%%%%%%%%%%%%%%%%%%%%%%%%%%%%%%%%%%%%%%%%%%%%

Many of the results throughout this paper require the computation of quantum corrections to either the vacuum of the theory or the masses of the particles. This appendix section provides the reader a detailed description and commentary of the procedure we use for their calculations.

%======================================================
\subsection{The general procedure\label{app:general}}
%======================================================

\def\CW{\mathcal{V}}
\def\hA{\mathbf{\hat{A}}}

The one-loop contribution to the effective potential \emph{\`a la} Coleman and Weinberg in the \emph{zero-momentum} scheme has the following compact form (cf.~\cite{Coleman:1973jx}):
\begin{align}
V_1 (\Phi) &= \frac{1}{64 \pi^2} \mathrm{Tr}\left[\mathbf{M}^4_S(\Phi) \left(\log\left[\frac{\mathbf{M}^2_S(\Phi)}{\mu^2_R}\right] - \frac{3}{2}\right)\right] %+ \nonumber\\
%&\quad 
+ \frac{3}{64 \pi^2} \mathrm{Tr}\left[\mathbf{M}^4_G(\Phi) \left(\log \left[\frac{\mathbf{M}^2_G(\Phi)}{\mu^2_R}\right] - \frac{5}{6}\right)\right]\nonumber\\[3pt]
&\equiv \CW\left(\mathbf{M}^2_{S}(\Phi),-\tfrac{3}{2},1\right)+\CW\left(\mathbf{M}^{2}_{G}(\Phi),-\tfrac{5}{2},3\right), \label{eq:EffectivePotential}
\end{align}
where the expression $\CW$ was defined for later convenience via
\begin{align}
\CW(\hA,c_1,c_2):=\frac{1}{64\pi^2}\left(c_1\,\mathrm{Tr}\big(\hA^2 \big)+ c_2\, \mathrm{Tr}\big(\hA^2\,\log[\hA/\mu_R^2]\big)\right) \label{eq:CW-definition}
\end{align}
for a matrix $\hA$ and numeric coefficients $c_1$ and $c_2$ with the hat on $\mathbf{A}$ used to denote its field dependence. The bold font is adopted for all matrix quantities. We use $\Phi$ as a generic label for the vector of all scalar fields in the theory, $\mu_R$ is the renormalization scale, and $\mathbf{M}_S^{2}(\Phi)$ and $\mathbf{M}_{G}^{2}(\Phi)$ are the tree-level field-dependent mass matrices of scalars and gauge bosons, respectively. 
The explicit expressions for their entries are given by
\begin{align}
\left[\mathbf{M}^2_S(\Phi)\right]_{ij} & = \frac{\partial^2 V_0}{\partial \Phi_i \partial \Phi^*_j} ,\label{eq:field-dependent-MS2}\\
\left[\mathbf{M}^2_G(\Phi)\right]_{ab} & = g^2 \left[(\hat{T}^a \Phi)^{\dagger} (\hat{T}^b \Phi)\right]_{(a\leftrightarrow b)}, \label{eq:field-dependent-MG2}
\end{align}
where the indices $i$ and $j$ run over all the fields in the scalar sector, $\hat{T}^{a}\Phi$ represents the action of the $a$-th $\mathrm{SO}(10)$ generator $T^{a}$ on the (reducible) scalar representation $\Phi$, and $(a\leftrightarrow b)$ symbolizes the symmetric part of the expression with respect to the indices $a$ and $b$, i.e.,~$[X_{ab}]_{(a\leftrightarrow b)}:=\tfrac{1}{2}(X_{ab}+X_{ba})$. Before any further discussion of our procedure, some important technical considerations for the explicit computation of these expressions are given below.
\begin{itemize}
    \item We reiterate that the matrices $\mathbf{M}^{2}_{S,G}(\Phi)$ are field dependent, which means that they have not been evaluated in vacuum; i.e.,~no expectation values of the fields have been inserted.
    \item The usual way to write the scalar-mass-square matrix $\mathbf{M}_{S}^{2}(\Phi)$ is in the basis of all real scalar degrees of freedom, so we need to consider the real and imaginary components of complex fields separately. In the above expression, we have instead written the scalar-mass matrix in a more convenient holomorphic and anti-holomorphic basis. This implies that $\Phi_i$ in the derivative first runs over all holomorphic fields and then over all anti-holomorphic fields. Conversely, $\Phi^\ast_j$ are conjugates of all the fields in $\Phi_j$; thus, they first run over the anti-holomorphic fields and then over holomorphic ones. The used expression is valid also for the special case of real scalar fields: They need to be counted only once, and since $\Phi_j=\Phi_j^{\ast}$, factors of $1/2$ for real mass matrices are correctly taken into account. In our particular model, the fields consist of
    \begin{align}
        &
        \Phi=(\phi,\Sigma,\Sigma^\ast) 
        \quad\text{and}\quad 
        \Phi^{\ast}=(\phi,\Sigma^\ast,\Sigma).
    \end{align}
    The number of (real) degrees of freedom over which the indices $i$ and $j$ run is $45+2\times 126=297$. 
    \item If $\Phi$ is a reducible representation, as is the case in this paper, the expression in Eq.~\eqref{eq:field-dependent-MG2} involves a sum over irreducible representations:
    \begin{align}
        \left[\mathbf{M}^2_G(\Phi)\right]_{ab} & = g^2 \left[\tfrac{1}{2}(\hat{T}^a \phi)^T (\hat{T}^b \phi)+(\hat{T}^{a}\Sigma)^\dagger(\hat{T}^{b}\Sigma)\right]_{(a\leftrightarrow b)}.
    \end{align}
    In contrast to the scalar-mass case, the complex degrees of freedom are taken into account only once, and a factor $1/2$ is inserted for the real representation $\phi$. The symmetrization in Eq.~\eqref{eq:field-dependent-MG2} also assumes that the basis of the generators is real, i.e.,~that all matrices $T^a$ are Hermitian matrices.\footnote{Crucially, the raising/lowering operators do not provide a suitable real basis for the Lie algebra.}
    Lastly, we use the standard GUT normalization of generators, in which the Dynkin index of the representation $\mathbf{10}$ of $\mathrm{SO}(10)$ equals $1$. We do not write further technical details on the tensor methods or conventions in this appendix, but invite the interested reader to check the appendices of~\cite{Graf:2016znk}, to which we adhere in this paper. Also, the appendices of~\cite{Antusch:2019avd} elaborate on different bases one can use for the representation $\mathbf{10}$ of $\mathrm{SO}(10)$.
    \item The field-dependent mass-square matrices are manifestly Hermitian, as is obvious from the expression in Eqs.~\eqref{eq:field-dependent-MS2} and~\eqref{eq:field-dependent-MG2}. Furthermore, the Hermitian expression for $\mathbf{M}_{G}^{2}(\Phi)$ is symmetric with respect to $a$ and $b$, thus resulting in a real matrix. The situation in our $\mathrm{SO}(10)$ Higgs model for any value of $\Phi$ is thus the following: $\mathbf{M}_{S}^{2}(\Phi)$ is a $297\times 297$ Hermitian matrix and $\mathbf{M}_{G}^{2}(\Phi)$ is a $45\times 45$ real symmetric matrix. 
\end{itemize}

Our initial procedure for loop corrections follows~\cite{Graf:2016znk}: Expanding in powers of $\hbar$, the one-loop expansion of the potential and the fields' VEVs read $V=V_0+V_1$ and $v=v_0+v_1$, respectively, where we used units with $\hbar=1$. The stationarity condition at tree level is
\begin{align}
    \partial_{x}V_{0}\,\Big|_{v_0}&=0,\label{eq:DV-tree}
\end{align}
and the one-loop stationarity condition $\partial_{x} V\,|_v=0$ is simplified into
\begin{align}
    \partial_{x} V_0\,\Big|_{v}+\partial_{x}V_{1}\,\Big|_{v_0}&=0,\label{eq:DV-1loop}
\end{align}
where we ignored the $\mathcal{O}(\hbar^2)$ terms, which formally contribute to two-loop order in the $\hbar$-expansion. We used an abbreviated notation where $\partial_{x}$ denotes $\partial/\partial\Phi_{x}$, and $\big|_{v}$ represents the insertion of the VEV $v$ for the fields, i.e.,~taking $\Phi=v$ in the result. Furthermore, the one-loop corrected scalar-mass matrix is determined by the second derivative. Expanding up to $\mathcal{O}(\hbar^1)$ yields
\begin{align}
    \left[\mathbf{M}^2_{S,\text{one-loop}}\right]_{xy}&=\partial_x\partial^{\ast}_{y}V_0\,\Big|_{v}+\partial_x\partial^{\ast}_{y}V_1\,\Big|_{v_0}.\label{eq:DDV-1loop}
\end{align}

An alternative route of applying vacuum conditions is to fix the VEVs as input parameters and solve the conditions for Lagrangian parameters. In our Higgs model, we choose that stationarity conditions solve for the parameters $\mu^2$, $\nu^2$, and $\tau$, while we take the VEVs $\omega_{BL}$, $\omega_{R}$, and $\sigma$ as inputs. This same approach was implicitly used in Sec.~\ref{sec:PSG-singlets}, as well as in previous work on the model~\cite{Graf:2016znk}. In this context, the $\hbar$-expansion is applied to the $\{\mu^{2},\nu^{2},\tau\}$ parameters instead of the VEVs.

The step-by-step procedure to compute the one-loop mass matrices for all scalar particles is thus the following:
\begin{enumerate}
    \item Solve the vacuum conditions in Eq.~\eqref{eq:DV-tree} for tree-level $\{\mu^2,\nu^2,\tau\}$; cf.~Eqs.~\eqref{eq:mu}--\eqref{eq:tau}.
    \item Insert the tree-level solution of $\{\mu^2,\nu^2,\tau\}$ into the 2nd term of Eq.~\eqref{eq:DV-1loop}, and solve for one-loop values of $\{\mu^2,\nu^2,\tau\}$ that are present as a linear combination in the 1st term.
    \item Insert the tree-level and one-loop vacuum into Eq.~\eqref{eq:DDV-1loop} to obtain the one-loop mass matrix (entry by entry).
\end{enumerate}
The challenging parts of this procedure are the last two steps. Assuming one has the full potential $V_0$ implemented in terms of all fields $\Phi$, it is straightforward to obtain the tree-level field-dependent mass matrices from Eqs.~\eqref{eq:field-dependent-MS2} and~\eqref{eq:field-dependent-MG2} and insert them into the Coleman-Weinberg expression for $V_{1}$ in Eq.~\eqref{eq:EffectivePotential}. The difficult task is evaluating the derivatives $\partial_{x}V_{1}$ and $\partial_{x}\partial^\ast_{y}V_{1}$ due to the matrix logarithm in the Coleman-Weinberg expression. Therefore, we ultimately seek a method to efficiently evaluate the derivatives $\partial_{x}\mathcal{V}$ and $\partial_{x}\partial_{y}^{\ast} \mathcal{V}$. 
One possible route to dealing with the matrix logarithm in $\mathcal{V}$ is to expand it into a power series of matrices around the identity and then apply derivatives to the series. This leads to an infinite series of nested commutators for the mass matrix. This approach was used in~\cite{Graf:2016znk} and can provide partial analytic insights, e.g.,~when the series of commutators terminates. However, it is not suitable for an efficient numeric calculation in the general regime.  

Notice that the evaluation of the expression $\CW$ is greatly simplified assuming that the $\Phi$-dependent matrix $\hA$ can be smoothly diagonalized via
\def\hLambda{\mathbf{\hat{\Lambda}}}
\def\hlambda{\hat{\lambda}}
\def\hR{\mathbf{\hat{R}}}
\begin{align}
    \hA=\hR\,\hLambda\,\hR^{-1},\label{eq:diagonalize-A}
\end{align}
where $(\hLambda)_{ij}=\hlambda_i\,\delta_{ij}$ (no sum over $i$) is the field-dependent diagonal matrix, and $\hR$ is the field-dependent transition matrix. We again remind the reader that throughout this appendix a hat on top of any quantity indicates its field dependence. The expression $\CW$ and its derivatives then yield

\begin{widetext}
\begin{align}
	\CW(\hA,c_1,c_2)&=\frac{1}{64\pi^2}\sum_{i}\left(c_1 \hlambda_i^2 +c_2 \hlambda_i^2\,\log[\hlambda_i/\mu_R^2]\right),\label{eq:CW-with-eigenvalues-begin}\\
	\partial_{x}\CW(\hA,c_1,c_2)&=\frac{1}{64\pi^2}\sum_{i}\left(2c_1+c_2+2c_2\,\log[\hlambda_i/\mu^2_{R}]\right) \hlambda_{i}\,\hlambda_{i,x},\\
	\begin{split}
		\partial_{x}\partial_{y}\CW(\hA,c_1,c_2)&=\frac{1}{64\pi^2} \sum_{i}
		\bigg[\left(2c_1+c_2+2c_2\,\log[\hlambda_i/\mu^2_{R}]\right) \hlambda_{i}\,\hlambda_{i,xy} +  
		%\\
		%&\qquad + 
		\left(2c_1+3c_2+2c_2\,\log[\hlambda_i/\mu^2_{R}]\right) \hlambda_{i,x}\,\hlambda_{i,y}\bigg]. \label{eq:CW-with-eigenvalues-end}
	\end{split}
\end{align}	
\end{widetext}
Note that we write $\partial_{y}$ instead of $\partial_{y}^{\ast}$ for simplicity --- the reader should perform this trivial replacement in all formulas involving the index $y$ as well, so as to obtain $\partial_{x}\partial_{y}^{\ast}\CW$ instead of $\partial_{x}\partial_{y}\CW$. 

Luckily, the procedure in Eqs.~\eqref{eq:DV-1loop} and~\eqref{eq:DDV-1loop} requires merely  $\CW$, $\partial_x\CW$ and $\partial_x\partial_y^\ast \CW$ evaluated in the tree-level vacuum $\Phi=v_0$, so we only need to know the eigenvalues and its derivatives in vacuum; i.e.,~we require just $\lambda_{i}=\hlambda_i(v_0)$, $\lambda_{i,x}=\hlambda_{i,x}(v_0)$, and $\lambda_{i,xy}=\hlambda_{i,xy}(v_0)$, assuming that the diagonalization of Eq.~\eqref{eq:diagonalize-A} is smooth in a neighborhood of $\Phi=v_0$. Notice that we use a notation where all non-hatted symbols denote quantities evaluated in vacuum, i.e.,~numeric quantities. We specify a numeric algorithm for calculating $\lambda_i$, $\lambda_{i,x}$, and $\lambda_{i,xy}$ in Appendix~\ref{app:numerical}.

Once the numeric values for $\lambda_i$, $\lambda_{i,x}$, and $\lambda_{i,xy}$ are available, their use in vacuum-evaluated  Eqs.~\eqref{eq:CW-with-eigenvalues-begin}--\eqref{eq:CW-with-eigenvalues-end}
still involves certain subtleties. Since the field-dependent matrices $\mathbf{M}^{2}_{S,G}(\Phi)$ are Hermitian, the eigenvalues $\lambda_{i}$ are always real. This is expected, as they correspond to mass squares of particles. However, the eigenvalues come as arguments into logs, so questions arise on how to properly deal with cases when they vanish or are negative. There is an issue even when they are positive and very small compared to $\mu_{R}^{2}$, since log contributions are then ``unphysically'' enhanced. We address how to deal with these log cases in Appendix~\ref{app:logs}.

Having established the calculational procedure, we now reflect on how to optimize the calculation. Notice that the expression $\CW$ defined in Eq.~\eqref{eq:CW-definition} and its derivatives are invariant under a basis change of $\hA$, and if $\hA$ is block diagonal, they can be evaluated on each diagonal block of $\hA$ separately. The matrix $\hA$ stands for either $\mathbf{M}^{2}_{S}(\Phi)$ or $\mathbf{M}^{2}_{G}(\Phi)$. Vacuum can already be inserted for all fields except for the $\Phi_x$ and $\Phi_y^\ast$ states due to derivatives, so we need a list of matrices $\mathbf{M}^{2}_{S,G}(\Phi_{x},\Phi_{y}^\ast)$ for all relevant pairs of indices $x,y$ compatible with gauge symmetry. We permute a basis of each $\mathbf{M}^{2}_{S,G}(\Phi_{x},\Phi_{y}^\ast)$ in such a way that it is block diagonal. Then, we traverse the entire list of blocks for all $x,y$ and collect only those which are formally different (treating field labels $\Phi_x$ and $\Phi_y^\ast$ as dummy variables). Hence, the evaluation of $\CW$ on each formal block needs to be computed only once, provided we keep track of their multiplicities for each fixed $x$ and~$y$. 

To recapitulate, our calculation of one-loop masses for a particular parameter point thus involves the computation of $\partial_{x}\CW(v_0)$ and $\partial_x\partial_{y}^{\ast}\CW (v_0)$ for a few thousand formally different blocks of various sizes. If the diagonalization of Eq.~\eqref{eq:diagonalize-A} for that block is smooth at $\Phi=v_0$, we compute $\lambda_{i}$, $\lambda_{i,x}$, and $\lambda_{i,xy}$ with the algorithm described in Appendix~\ref{app:numerical}. They are then inserted into Eqs.~\eqref{eq:CW-with-eigenvalues-begin}--\eqref{eq:CW-with-eigenvalues-end} evaluated at $v_{0}$, with proper logarithm treatment described in Appendix~\ref{app:logs}. 

Only a few small blocks were found not to be smoothly diagonalizable at $v_0$, all of them located in the gauge mass matrix $\mathbf{M}^{2}_{G}(\Phi_{x},\Phi_{y}^{\ast})$. Fortunately, it was possible to compute the expression $\CW$ for these blocks analytically.\footnote{There exists an analytically computable smooth Jordan decomposition for them.}

%======================================================
\subsection{Details of the numerical procedure\label{app:numerical}}
%======================================================

%......................................................
\subsubsection{Preliminaries and notation \label{app:numerical-notation}}
%......................................................

As discussed in Appendix~\ref{app:general}, we are interested in finding a 
procedure where the starting point is a $\Phi$-dependent matrix $\mathbf{\hat{A}}$, and the desired result are the eigenvalues and their first- and second-order derivatives evaluated in the tree-level vacuum $v_0$: $\lambda_{i}$, $\lambda_{i,x}$, and $\lambda_{i,xy}$. Equivalently, expanding the diagonal field-dependent matrix $\mathbf{\hLambda}$ in Eq.~\eqref{eq:diagonalize-A} into a Taylor series around vacuum via
{\small
\begin{align}
    \hLambda=\mathbf{\Lambda}+\mathbf{\Lambda}_{,x}\,(\Phi-v_0)_x+\tfrac{1}{2!}\mathbf{\Lambda}_{,xy}\,(\Phi-v_0)_x(\Phi-v_0)_y+\ldots,\label{eq:expand-lambda}
\end{align}
}
we seek the diagonal-matrix coefficients $\mathbf{\Lambda}$, $\mathbf{\Lambda}_{,x}$, and $\mathbf{\Lambda}_{,xy}$. Note that $\Phi_x$ goes over all real degrees of freedom when the indices $x,y$, etc.,~are summed over.

Obtaining the eigenvalues $\lambda_i$ is not difficult, since the tree-level vacuum can simply be inserted into $\hA$ and then the numeric matrix $\hA(v_{0})$ can easily be diagonalized. Finding the eigenvalue derivatives, however, is complicated in the most general case by possible degeneracies. In particular, suppose we have an eigenspace of $\mathbf{\Lambda}$ that corresponds to a particular eigenvalue. While any basis of this eigenspace can be used for diagonalizing $\mathbf{\Lambda}$, derivatives such as $\mathbf{\Lambda}_{,x}$ may impose a particular preferred basis in this eigenspace --- in other words, the transition matrix $\hR$ is field dependent and higher-order coefficients in its $\Phi$-expansion matter. The proper ($\Phi$-dependent) eigenbasis therefore gets progressively resolved only when higher derivatives of degenerate eigenvalues are taken into account. For a more detailed discussion, we refer the interested reader to the literature~\cite{Murthy:1988,Andrew:1993,Aa:2007}. We mimic the procedure from~\cite{Aa:2007}, in particular to derive the algorithm for the numeric evaluation of the expansion coefficients in Eq.~\eqref{eq:expand-lambda} up to 2nd order.

Before describing the algorithm and providing a quick derivation, we set up our compact but convenient notation and definitions. As part of the algorithm, we shall obtain a sequence of progressive basis transformations $\mathbf{P}_n$. We denote the cumulative transformation up to the $n$-th level by $\mathbf{R}_n$:
\begin{align}
    \mathbf{R}_{n}&:=\mathbf{P}_{1}\ldots \mathbf{P}_{n}.
\end{align}
For any matrix $\mathbf{X}$ from the set
\begin{align}
    \mathbf{X}&\in\{\mathbf{A},\mathbf{A}_{,x}, \mathbf{A}_{,y}, \mathbf{A}_{,xy}\},
\end{align}
we define its form in a different basis by
\begin{align}
    \mathbf{X}_{0}&:=\mathbf{\hat{X}}(v_{0}),\\
    \mathbf{X}_{n}&:=\mathbf{R}_{n}^{-1}\mathbf{X}_{0}\mathbf{R}_n=\mathbf{P}_{n}^{-1} \mathbf{X}_{n-1} \mathbf{P}_{n}.\label{eq:definition-Xn}
\end{align}
We combine the basis and derivative labels in the subscript via, e.g.,~$\mathbf{A}_{n,x}$, and refer to such a matrix as being in the \textbf{$n$-basis}. The matrix $\mathbf{P}_n$ transitions from the $n$-basis to the $(n-1)$-basis.

It turns out each matrix $\mathbf{P}_{n}$ is obtained in the algorithm by diagonalizing some particular numeric matrix; cf.~Eqs.~\eqref{eq:algorithm-P1} and~\eqref{eq:algorithm-P2}--\eqref{eq:algorithm-lambda-xy}. We write this diagonalization generically as
\begin{align}
        \mathcal{A}_{n-1}&=\mathbf{P}_n \mathcal{D}_n \mathbf{P}_{n}^{-1},\label{eq:recursive-structure}
\end{align}
where book-keeping subscripts under $\mathcal{A}$ and $\mathcal{D}$ correspond also to the underlying basis of the matrix, but the transformations among them analogous to Eq.~\eqref{eq:definition-Xn} do not apply. The diagonal matrices $\mathcal{D}_{n}$ are the matrix coefficients of interest in the expansion of $\hLambda$. Explicitly, for fixed $x$ and $y$ we have 
\begin{align}
    \mathcal{D}_{1}&=\mathbf{\Lambda},&
    \mathcal{D}_{2}&=\mathbf{\Lambda}_{,x},&
    \mathcal{D}_{3}&=\mathbf{\Lambda}_{,y},&
    \mathcal{D}_{4}&=\mathbf{\Lambda}_{,xy}.\label{eq:D-label}
\end{align}

Crucially, at each step $n$ there is an associated structure that $\mathcal{A}_{n-1}$ possesses.
\begin{itemize}
    \item There exists a partition of the $(n-1)$-basis already from the previous step. It is referred to as the $(n-1)$-\textbf{partition}, and the associated block structure is composed of $(n-1)$-\textbf{blocks}. Edge case: The initial $0$-basis has the trivial partition corresponding to a single block.
    \item The matrix $\mathcal{A}_{n-1}$ turns out to be $(n-1)$-block diagonal. Consequently, the transition matrix $\mathbf{P}_{n}$ that diagonalizes it can also be taken as $(n-1)$-block diagonal. 
    \item The transition matrix $\mathbf{P}_{n}$ defines the $n$-basis. Since $\mathbf{P}_{n}$ is $(n-1)$-block diagonal, the $(n-1)$-partition also applies to the $n$-basis. We order the $n$-basis so that within each $(n-1)$-block the eigenvectors belonging to the same eigenspace of $\mathcal{D}_{n}$ are grouped together. This subdivision of the $n$-basis defines the $n$-partition, which is clearly a refinement of the $(n-1)$-partition of the same basis. The transition matrix $\mathbf{P}_{n}$ is arbitrary only up to basis changes within $n$-blocks, within which $\mathbf{P}_{m}$ for $m>n$ can make changes to the basis in later steps. 
    \item Incidentally, knowledge of $\mathbf{P}_{n}$ suffices to obtain the full transition matrix $\mathbf{R}_{n}=\mathbf{R}_{n-1}\mathbf{P}_{n}$ transforming the $n$-basis to the initial $0$-basis, where for $n=1$ we have $\mathbf{R}_{1}=\mathbf{P}_{1}$. Intuitively, $\mathbf{R}_{n}$ simultaneously diagonalizes $\mathcal{A}_{k}$ for all $k=0,\ldots,(n-1)$ if the $\mathcal{A}$ matrices are rewritten in the $0$-basis. Note that $\mathbf{R}_{n}$ is in general not block diagonal.  
\end{itemize}
The above underlying structure therefore yields a progressive series of basis partitions, where the $n$-partition acts on the $n$-basis. The $n$-partition is obtained in the diagonalization procedure of $\mathcal{A}_{n-1}$ and is subordinate to the $m$-partition for any $m<n$. Consequently, $\mathbf{P}_n$ is $m$-block diagonal for $m<n$. The inductively defined structure is self-perpetuating from step to step as long as $\mathcal{A}_{n}$ is indeed $n$-block diagonal, which is true by construction.  

For any matrix $\mathbf{X}_n$ in the $n$-basis, we denote the block-diagonal and block-off-diagonal parts of the $n$-partition in the matrix by $[n]$ and $(n)$ in the superscript, respectively,
\begin{align}
    (\mathbf{X}_{n}^{[n]})_{ij}&:=
    \begin{cases}
            (\mathbf{X}_n)_{ij}&\mbox{if } i\in I, j\in J, I=J\\
            0&\mbox{if } i\in I, j\in J, I\neq J
    \end{cases},\\
    \mathbf{X}_{n}^{(n)}&:=\mathbf{X}_n-\mathbf{X}_{n}^{[n]},
\end{align}
where capital indices $I$ and $J$ are $n$-block labels. For $m>n$ we further define the notation
\begin{align}
    \mathbf{X}_{m}^{[n]}&:=(\mathbf{R}_{n}^{-1}\mathbf{R}_m)^{-1}\;\mathbf{X}_{n}^{[n]}\;(\mathbf{R}_{n}^{-1}\mathbf{R}_m),\label{eq:block-diagonal-part}\\   
    \mathbf{X}_{m}^{(n)}&:=(\mathbf{R}_{n}^{-1}\mathbf{R}_m)^{-1}\;\mathbf{X}_{n}^{(n)}\;(\mathbf{R}_{n}^{-1}\mathbf{R}_m),\label{eq:off-block-diagonal-part}
\end{align}
which represents taking the $n$-block (off-)diagonal part of $\mathbf{X}_{n}$ and further transforming it from the $n$-basis into the $m$-basis. Equivalently, we can obtain $\mathbf{X}^{[n]}_{m}$ or $\mathbf{X}^{(n)}_{m}$ by first fully rotating $\mathbf{X}$ to the $m$-basis, and taking the (off)-diagonal part of the $n$-partition only in the end. This works due to the hierarchy in block structure; i.e.,~the $n$-partition can be applied to any $m$-basis for $m>n$.

%......................................................
\subsubsection{The algorithm \label{app:algorithm}}
%......................................................

Starting with a field-dependent matrix $\hA\equiv \mathbf{A}(\Phi)$, we want to obtain $\lambda_i$, $\lambda_{i,x}$, $\lambda_{i,y}$, and $\lambda_{i,xy}$ for fixed $x$ and $y$. Following the notation introduced in Appendix~\ref{app:numerical-notation}, perform the following steps:
\begin{enumerate}
    \item Given the matrix $\hA$, compute (in the tree-level vacuum) the numeric matrices $\mathbf{A}_{0}$, $\mathbf{A}_{0,x}$, $\mathbf{A}_{0,y}$, $\mathbf{A}_{0,xy}$.
    \item Diagonalize $\mathbf{A}_0$ to obtain $(\mathbf{\Lambda},\mathbf{P}_1)$:
        \begin{align}
            \mathbf{A}_{0}&=\mathbf{P}_{1}\mathbf{\Lambda}\mathbf{P}_{1}^{-1}.\label{eq:algorithm-P1}
        \end{align}
    This gives us $\lambda_{i}$ and the $1$-partition, as well as \hbox{$\mathbf{R}_1 = \mathbf{P}_1$.}
    \item Compute an auxiliary matrix $\mathbf{\Omega}$ from $\mathbf{\Lambda}$:
        \begin{align}
            (\mathbf{\Omega})_{ij}&=
            \begin{cases}
            (\lambda_{j}-\lambda_{i})^{-1}&\mbox{if } \lambda_{i}\neq \lambda_{j}\\
            0&\mbox{if } \lambda_{i}=\lambda_{j}
            \end{cases}.\label{eq:definition-Omega}
        \end{align}
    Clearly, $\mathbf{\Omega}=\mathbf{\Omega}^{(1)}$.
    \item Compute $\mathbf{A}_{1,x}=\mathbf{R}_{1}^{-1}\mathbf{A}_{0,x}\mathbf{R}_{1}$ and diagonalize its diagonal $1$-blocks to obtain $(\mathbf{\Lambda}_{,x},\mathbf{P}_{2})$:
        \begin{align}
            \mathbf{A}^{[1]}_{1,x}&=\mathbf{P}_{2} \mathbf{\Lambda}_{,x}\mathbf{P}_{2}^{-1}.\label{eq:algorithm-P2}
        \end{align}
    This gives us $\lambda_{i,x}$ and the $2$-partition.
    \item Compute $\mathbf{A}_{2,y}=\mathbf{R}_{2}^{-1}\mathbf{A}_{0,y}\mathbf{R}_{2}$, where $\mathbf{R}_{2}=\mathbf{P}_{1}\mathbf{P}_{2}$. Take its diagonal $2$-blocks and diagonalize to get $(\mathbf{\Lambda}_{,y},\mathbf{P}_{3})$:  
        \begin{align} 
            \mathbf{A}^{[2]}_{2,y}&=\mathbf{P}_3 \mathbf{\Lambda}_{,y}\mathbf{P}_{3}^{-1}.\label{eq:algorithm-P3}
        \end{align}
    This also gives the $3$-partition.
    \item Obtain $\mathbf{\Lambda}_{,xy}$ by diagonalizing the matrix on the left side of the expression
        {\small
        \begin{align}
            \left(\mathbf{A}_{3,xy}-(\mathbf{\Omega}\cdot \mathbf{A}_{3,x})\;\mathbf{A}_{3,y}^{(1)}-(\mathbf{\Omega}\cdot\mathbf{A}_{3,y})\;\mathbf{A}_{3,x}^{(1)}\right)^{[3]}&=\mathbf{P}_{4}\mathbf{\Lambda}_{,xy}\mathbf{P}_{4}^{-1}.\label{eq:algorithm-lambda-xy}
        \end{align}
        }
    The dot $(\cdot)$ denotes entry-wise multiplication. This gives us the sought-after $\lambda_{i,xy}$, as well as the incidental transition matrix $\mathbf{P}_{4}$ and the corresponding $4$-partition.
\end{enumerate}
%

%......................................................
\subsubsection{Derivation of the algorithm}
%......................................................

To derive the algorithm we have specified in Appendix~\ref{app:algorithm},
we start by considering the field-dependent diagonalization in Eq.~\eqref{eq:diagonalize-A}, which we rearrange into
\begin{align}
    \hA\hR&=\hR\hLambda.\label{eq:derivations-start}
\end{align}
The matrix $\hA$ is provided, while both $\hR$ and $\hLambda$ are not known explicitly, but their full form is not necessary to compute the desired $\Lambda,\Lambda_{,x},\Lambda_{,y},\Lambda_{,xy}$ for fixed $x$ and $y$. We remind the reader that all hatted quantities imply field dependence, while the non-hatted quantities are evaluated in the tree-level vacuum and are numeric for given input parameters. We assume all notation defined so far in Appendix~\ref{app:numerical-notation}, in particular, Eqs.~\eqref{eq:expand-lambda}--\eqref{eq:off-block-diagonal-part}. For later convenience, we further amend the definitions with
\def\GAM{\mathbf{\hat{\Gamma}}}
\def\hS{\mathbf{\hat{S}}}
\begin{align}
    \GAM_n&:= \mathbf{P}_n\mathbf{P}_{n+1}\ldots \mathbf{P}_{N-1}\mathbf{\hat{P}}_{N},\\
    \hS_{;X}&:=\hR^{-1}\hR_{,X},\\
    \hS_{n;X}&:=\GAM_{n+1}\hS_{;X}\GAM_{n+1}^{-1},\\
    \mathbf{S}_{n;X}&:=\hS_{n;X}(v_0)=\GAM_{n+1}(v_0)\,\hS_{;X}(v_0)\,\GAM_{n+1}^{-1}(v_0),
\end{align}
where $X$ denotes any combination of derivative variables $X\in\{x,y,xy\}$. 
It turns out we need $N=5$ progressive transformations in total, with all field dependence then pushed to $\mathbf{\hat{P}}_{5}$, while the prior $\mathbf{P}$ transformations are used in the process of determining $\Lambda,\Lambda_{,x},\Lambda_{,y},\Lambda_{xy}$. The above definitions imply
\begin{align}
    \hR=\mathbf{R}_{n}\GAM_{n+1}\label{eq:def-gamma}
\end{align}
for any $n=1,2,3,4$, along with the trivial edge case $\GAM_{1}=\hR$. 
Intuitively, $\mathbf{R}_n$ is the rotation from the initial into the $n$-basis, while the remaining part of the full $\hR$ diagonalization is represented by $\GAM_{n+1}$. 
The useful quantities $\hS_{n;X}$ represent the $X$-derivative of $\hR$ unrotated by $\hR$ from the left, with the result taken in the $n$-basis, while the unhatted $\mathbf{S}_{n;X}$ is the same quantity evaluated in the tree-level vacuum. Lastly, we define $\mathbf{P}_{5}=\mathbf{\hat{P}}_{5}(v_0)$.

To perform algebraic manipulations in the subsequent derivation, the following set of commutators will likely prove useful to the reader: 
\begin{align}
0&=[\mathbf{\Lambda},\GAM_{n\geq 2}(v_0)]=[\mathbf{\Lambda},\mathbf{P}_{n\geq 2}],\label{eq:lambda-commutators-begin}\\
0&=[\mathbf{\Lambda}_{,x},\GAM_{n\geq 3}(v_0)]=[\mathbf{\Lambda}_{,x},\mathbf{P}_{n\geq 3}],\\
0&=[\mathbf{\Lambda}_{,y},\GAM_{n\geq 4}(v_0)]=[\mathbf{\Lambda}_{,y},\mathbf{P}_{n\geq 4}],\\
0&=[\mathbf{\Lambda}_{,xy},\GAM_{n\geq 5}(v_0)]=[\mathbf{\Lambda}_{,xy},\mathbf{P}_{n\geq 5}].\label{eq:lambda-commutators-end}
\end{align}
The commutators' identities with the $\mathbf{P}_{n}$ matrices follow directly from the block structure of $\mathbf{P}_{n}$ and the hierarchies among $n$-partitions described in Appendix~\ref{app:numerical-notation}. The $\mathbf{P}_{m}$ is $n$-block diagonal for $m>n$, while the matrices $\mathcal{D}_{n}$ identified in Eq.~\eqref{eq:D-label} are diagonal with the same eigenvalues in a given $n$-block, leading to vanishing $\mathbf{P}$-commutators in Eqs.~\eqref{eq:lambda-commutators-begin}--\eqref{eq:lambda-commutators-end} in particular.

As for commutators with $\mathbf{\Gamma}_{n}$, the argument needs to be a bit more subtle, since they are used in deriving the $\mathcal{A}_{n}$ matrix at a step where $\mathbf{P}_{m}$ for $m>n$ has not yet been defined. Consider the lowest non-trivial $\mathbf{\Gamma}$, which is $\mathbf{\Gamma}_{2}$. Since $\mathcal{A}_{0}=\mathbf{A}_{0}$, we have both $\mathbf{R}=\hR(v_{0})$ and $\mathbf{R}_{1}=\mathbf{P}_{1}$ diagonalizing it according to Eq.~\eqref{eq:derivations-start} evaluated in the vacuum and the definition of $\mathbf{P}_{1}$ in Eq.~\eqref{eq:recursive-structure}, respectively. Thus,
\begin{align}
    \mathcal{A}_{0}=\mathbf{R}\mathcal{D}_{1}\mathbf{R}^{-1}=\mathbf{R}_{1}\mathcal{D}_{1}\mathbf{R}_{1}^{-1},
\end{align}
with the two rotation matrices related via $\mathbf{R}=\mathbf{R}_{1}\mathbf{\Gamma}_{2}$ due to Eq.~\eqref{eq:def-gamma}. The matrix $\mathbf{\Gamma}_{2}$ thus represents the arbitrariness in the transition matrix, which must commute with the matrix being diagonalized (cf.~\cite{Aa:2007}): $[\mathbf{\Lambda},\mathbf{\Gamma}_{2}]=0$. The matrices $\mathbf{\Gamma}_{m}$ for $m>2$ are related to $\mathbf{\Gamma}_{2}$ recursively via $\mathbf{\Gamma}_n=\mathbf{P}_{n}\mathbf{\Gamma}_{n+1}$, so their commutators with $\mathbf{\Lambda}$ can be derived from the $\mathbf{\Gamma}_{2}$- and $\mathbf{P}$-commutators.  
Analogous arguments can be used to derive $\mathbf{\Gamma}$-commutators with derivatives of $\mathbf{\Lambda}$, since $\mathbf{\Gamma}_{n}$ represents the arbitrariness in the simultaneous diagonalization of the matrices $\mathcal{A}_{m}$ for all $m$ between $0$ and $n-2$.

We now perform a sequence of steps.
\begin{enumerate}
    %%%%%%%%%%%
    %%%%%%%%%%%
    \item 
        Evaluating Eq.~\eqref{eq:derivations-start} in vacuum and taking the first transformation matrix as $\mathbf{P}_{1}=\hR(v_{0})$ yields $\mathbf{A}_{0}=\mathbf{P}_1\mathbf{\Lambda}\mathbf{P}_{1}^{-1}$. Numeric diagonalization of $\mathbf{A}_{0}$ thus gives $\mathbf{P}_{1}$ and $\mathbf{\Lambda}$. Formally, this also implies the expression $\GAM_{2}=\hR(v_0)^{-1}\hR$.
    %%%%%%%%%%%
    %%%%%%%%%%%
    \item 
        Take the derivative $\partial_{x}$ of Eq.~\eqref{eq:derivations-start}:
        \begin{align}
            \hA_{,x}\hR+\hA\hR_{,x}&=\hR_{,x}\hLambda+\hR\hLambda_{,x}.
        \end{align}
        Multiply from the left with $\hR^{-1}$, rearrange and insert suitable definitions for $\hS$ to obtain
        \begin{align}
            \hR^{-1}\hA_{,x}\hR-\hLambda_{,x}&=\hS_{;x}\hLambda-\hLambda\hS_{;x}.\label{eq:derivations-d}
        \end{align}
        We now formally multiply Eq.~\eqref{eq:derivations-d} with $\GAM_{3}$ from the left and $\GAM_{3}^{-1}$ from the right, thus undoing the field-dependent rotation $\hR$ up to the as-yet-unknown $\mathbf{R}_{2}=\mathbf{R}_{1}\mathbf{P}_{2}$. We then evaluate that expression in vacuum and multiply it with $\mathbf{P}_{2}$ from the left to obtain
        \begin{align}
            \mathbf{A}_{1,x}\mathbf{P}_{2}-\mathbf{P}_{2}\mathbf{\Lambda}_{,x}=\mathbf{P}_{2}\mathbf{S}_{2;x} \mathbf{\Lambda}-\mathbf{P}_{2}\mathbf{\Lambda}\mathbf{S}_{2;x}.\label{eq:derivations-dx}
        \end{align}
        We consider this matrix equation in terms of $1$-blocks, using labels $K_{1}$ and $L_{1}$ for $1$-block rows and $1$-block columns on each side, respectively. Since $\mathbf{P}_{2}$ is $1$-block diagonal, we can write $(\mathbf{P}_{2})_{K_{1}L_{1}}=(\mathbf{P}_{2})_{K_{1}K_{1}}\delta_{K_{1}L_{1}}$. Similarly, $(\mathbf{\Lambda})_{K_{1}L_{1}}=\lambda_{K_{1}}\,\delta_{K_{1}L_{1}}$; i.e.,~each diagonal $1$-block in $\mathbf{\Lambda}$ is proportional to the identity matrix, while the off-diagonal blocks are zero. 
        
        We first examine the $1$-block-diagonal part of Eq.~\eqref{eq:derivations-dx}.
        For $K_{1}=L_{1}$, the right-hand side vanishes, since $\mathbf{\Lambda}$ acts as an identity block multiplied with the same proportionality factor in both terms, i.e.,~$\lambda_{K_1}=\lambda_{L_{1}}$. Hence,
        \begin{align}
            (\mathbf{A}_{1,x}\mathbf{P}_{2})_{K_{1} K_{1}}&=(\mathbf{P}_2\mathbf{\Lambda}_{,x})_{K_{1} K_{1}}.
        \end{align}
        Since $\mathbf{P}_{2}$ and $\mathbf{\Lambda}_{,x}$ are already block diagonal with respect to the $1$-partition, the equation 
        can be written in full-matrix form as
        \begin{align}
            \mathbf{A}^{[1]}_{1,x}\;\mathbf{P}_{2}&=\mathbf{P}_{2}\mathbf{\Lambda}_{,x}.
        \end{align}
        Diagonalizing $\mathbf{A}_{1,x}^{[1]}$ yields $\mathbf{P}_2$ (and thus $\mathbf{R}_{2}$) and the first derivatives $\mathbf{\Lambda}_{,x}$.
        \par
        We now return to the $1$-block off-diagonal parts in Eq.~\eqref{eq:derivations-dx}, i.e.,~$K_{1}\neq L_{1}$. Since $\mathbf{P}_2$, $\mathbf{\Lambda}$, and $\mathbf{\Lambda}_{,x}$ are all $1$-block diagonal, the second term on the left-hand side vanishes, and we get
        {\small
        \begin{align}
            (\mathbf{A}_{1,x})_{K_{1}L_{1}}(\mathbf{P}_{2})_{L_{1}L_{1}}&=(\lambda_{L_{1}}-\lambda_{K_{1}}) (\mathbf{P}_{2})_{K_1 K_1}\,(\mathbf{S}_{2;x})_{K_{1}L_{1}}.
        \end{align}
        }
        Solving for $(\mathbf{S}_{2;x})_{K_1 L_1}$ gives
        {\small 
        \begin{align}
            (\mathbf{S}_{2;x})_{K_1 L_1}&=\frac{1}{(\lambda_{L_1}-\lambda_{K_1})} (\mathbf{P}_{2}^{-1})_{K_1 K_1} (\mathbf{A}_{1,x})_{K_{1}L_{1}}(\mathbf{P}_{2})_{L_1 L_1}%\nonumber \\[3pt]
            %& 
            =(\mathbf{\Omega}\cdot\mathbf{A}_{2,x})_{K_{1}L_{1}},\label{eq:derivations-Sx}
        \end{align}
        }
        where the dot $(\cdot)$ denotes entry-wise multiplication and the auxiliary matrix $\mathbf{\Omega}$ is defined in Eq.~\eqref{eq:definition-Omega}. This can be written as a full-matrix equation via
        \begin{align}
               \mathbf{S}_{2;x}^{(1)}&=\mathbf{\Omega}\cdot\mathbf{A}_{2,x}.
        \end{align}
        Note that the knowledge of the $1$-block off-diagonal parts of $\mathbf{S}_{2;x}$ is sufficient for all further steps in our derivation of the quantities of interest.
    %%%%%%%%%%%
    %%%%%%%%%%%
    \item Perform the following sequence of operations: Take the derivative $\partial_{y}$ of Eq.~\eqref{eq:derivations-start}, multiply with $\hR^{-1}$ from the left, then formally multiply with $\GAM_{4}$ from the left and $\GAM_{4}^{-1}$ from the right, and finally evaluate in vacuum. These steps give the analogue of Eq.~\eqref{eq:derivations-dx}:
        \begin{align}
            \mathbf{A}_{2,y}\mathbf{P}_{3}-\mathbf{P}_{3}\mathbf{\Lambda}_{,y}=\mathbf{P}_{3}\mathbf{S}_{3;y} \mathbf{\Lambda}-\mathbf{P}_{3}\mathbf{\Lambda}\mathbf{S}_{3;y}.\label{eq:derivations-dy}
        \end{align}
        Similar to the case of $\partial_{x}$, we consider this equation in terms of $2$-blocks. For diagonal blocks, we get 
        \begin{align}
            (\mathbf{A}_{2,y}\mathbf{P}_{3})_{K_2 K_2}=(\mathbf{P}_3\mathbf{\Lambda}_{,y})_{K_2 K_2},
        \end{align}
        yielding the full-matrix equation 
        \begin{align}
            \mathbf{A}_{2,y}^{[2]}\mathbf{P}_{3}&=\mathbf{P}_{3}\mathbf{\Lambda}_{,y}
        \end{align}
        indicating that diagonalization of $\mathbf{A}_{2,y}^{[2]}$ produces the transition matrix $\mathbf{P}_{3}$ and the eigenvalue derivatives $\mathbf{\Lambda}_{,y}$. The block off-diagonal part of Eq.~\eqref{eq:derivations-dy} analogous to the $\partial_x$ case gives
        {\small
        \begin{align}
        (\mathbf{A}_{2,y})_{K_{1}L_{1}}\,(\mathbf{P}_{3})_{L_{1}L_{1}}&=
        (\lambda_{L_{1}}-\lambda_{K_{1}})\;(\mathbf{P}_{3})_{K_{1}K_{1}}(\mathbf{S}_{3;y})_{K_{1}L_{1}}
        \end{align}
        }
        in terms of the coarser $1$-blocks with $K_{1}\neq L_{1}$,
        resulting in
        \begin{align}
        (\mathbf{S}_{3;y})_{K_{1}L_{1}}&=(\mathbf{\Omega}\cdot\mathbf{A}_{3,y})_{K_{1}L_{1}}.
        \end{align}
        In full-matrix notation, this can be written as
        \begin{align}
        \mathbf{S}_{3;y}^{(1)}&=\mathbf{\Omega}\cdot\mathbf{A}_{3,y}.\label{eq:derivations-Sy}
        \end{align}
        %
    %%%%%%%%%%%
    %%%%%%%%%%%
    \item 
        We now turn to the 2nd derivative and take $\partial_{x}\partial_{y}$ of Eq.~\eqref{eq:derivations-start}:
        \begin{align}
            &\hA_{,xy}\hR+\hA_{,x}\hR_{,y}+\hA_{,y}\hR_{,x}+\hA\hR_{,xy} %\nonumber\\
            %&\quad 
            =\hR_{,xy}\hLambda+ \hR_{,x}\hLambda_{,y}+ \hR_{,y}\hLambda_{,x}+ \hR\hLambda_{,xy}.
        \end{align}
        Multiplying it by $\hR^{-1}$ on the left and inserting the $\hS$-definitions gives
        \begin{align}
            &\hR^{-1}\hA_{,xy}\hR+
            \hR^{-1}\hA_{,x}\hR\hS_{;y}+
            \hR^{-1}\hA_{,y}\hR\hS_{;x}+
            \hLambda\hS_{;xy} %\nonumber \\
            %&\quad 
            =
            \hS_{;xy}\hLambda+
            \hS_{;x}\hLambda_{,y}+
            \hS_{;y}\hLambda_{,x}+
            \hLambda_{,xy}.
        \end{align}
        Multiplying this equation from the left with $\GAM_{5}$, from the right with $\GAM_{5}^{-1}$, evaluating it in vacuum, making use of the commutators in Eq.~\eqref{eq:lambda-commutators-end} and expressing $\mathbf{\Lambda}_{,xy}$ yields
        \begin{align}
        \begin{split}
            \mathbf{\Lambda}_{,xy}&=
            \mathbf{P}_{4}^{-1}\mathbf{A}_{3,xy}\mathbf{P}_{4}%+ \\
            %&\quad 
            +\mathbf{P}_{4}^{-1}\mathbf{A}_{3,x}\mathbf{P}_{4}\mathbf{S}_{4;y}+
            \mathbf{P}_{4}^{-1}\mathbf{A}_{3,y}\mathbf{P}_{4}\mathbf{S}_{4;x}%-\\
            %&\quad 
            - \mathbf{S}_{4;x}\,\mathbf{\Lambda}_{,y}
            - \mathbf{S}_{4;y}\,\mathbf{\Lambda}_{,x}
            +[\mathbf{\Lambda} ,\mathbf{S}_{4;xy}].\label{eq:derivations-dxy-part}
        \end{split}
        \end{align}
        Since $\mathbf{\Lambda}_{,xy}$ is diagonal (due to $\hLambda$ being diagonal by definition), knowledge of diagonal $3$-blocks in Eq.~\eqref{eq:derivations-dxy-part} suffices:
        \begin{align}
            (\mathbf{\Lambda}_{,xy})_{K_{3}K_{3}}&=
            \big(
            \mathbf{P}_{4}^{-1}\mathbf{A}_{3,xy}\mathbf{P}_{4}+            
            (\mathbf{P}_{4}^{-1}\mathbf{A}_{3,x}\mathbf{P}_{4}-\mathbf{\Lambda}_{,x})\mathbf{S}_{4;y}+ %\nonumber\\
            %&\quad +
            (\mathbf{P}_{4}^{-1}\mathbf{A}_{3,y}\mathbf{P}_{4}-\mathbf{\Lambda}_{,y})\mathbf{S}_{4;x}
            \big)_{K_{3} K_{3}} \nonumber\\[2pt]
            &=
            \big(\mathbf{P}_{4}^{-1}\mathbf{A}_{3,xy}\mathbf{P}_{4}+
            \mathbf{P}_{4}^{-1}[\mathbf{S}_{3;x},\mathbf{\Lambda}]\mathbf{S}_{3;y}\mathbf{P}_{4}+ %\nonumber\\
			%&\quad +           
            \mathbf{P}_{4}^{-1}[\mathbf{S}_{3;x},\mathbf{\Lambda}]\mathbf{S}_{3;x}\mathbf{P}_{4}
            \big)_{K_{3}K_{3}}.\label{eq:derivations-dxy-part2}
        \end{align}
        To derive the first line, we used the fact that the diagonal matrices $\mathbf{\Lambda}$, $\mathbf{\Lambda}_{,x}$, and $\mathbf{\Lambda}_{y}$ respect the $3$-partition; i.e.,~each of their diagonal blocks is proportional to an identity matrix block, and thus they commute with any matrix on the $3$-block-diagonal part. The second line was derived by taking Eqs.~\eqref{eq:derivations-dx} and~\eqref{eq:derivations-dy} into account, properly transformed to the $3$-basis.
        \par
        Eq.~\eqref{eq:derivations-dxy-part2} gives the diagonal $3$-blocks of $\mathbf{\Lambda}_{,xy}$ in terms of an expression that is sandwiched between $\mathbf{P}_{4}^{-1}$ and $\mathbf{P}_{4}$. Since $\mathbf{P}_{4}$ is $3$-block diagonal, and $(\mathbf{\Lambda})_{K_{3}L_{3}}=\lambda_{K_{3}}\,\delta_{K_{3}L_{3}}$, we can write the result more explicitly as
        \begin{align}
        \begin{split}
            (\mathbf{\Lambda}_{,xy})_{K_{3}K_{3}}&=
            (\mathbf{P}_{4}^{-1})_{K_{3}K_{3}}
            \Big((\mathbf{A}_{3,xy})_{K_{3}K_{3}}
            %+ \\
            %&\quad 
            + (\lambda_{L_{3}}-\lambda_{K_{3}}) 
            \big((\mathbf{S}_{3;x})_{K_3 L_3}(\mathbf{S}_{3;y})_{L_3 K_3}+\\
            &\quad 
            +(\mathbf{S}_{3;y})_{K_3 L_3}(\mathbf{S}_{3;x})_{L_3 K_3}\big)
            \Big)
            (\mathbf{P}_{4})_{K_{3}K_{3}},
        \end{split}    
        \end{align}
        with an implied sum over $L_{3}$ on the right-hand side. For $K_{3}$ and $L_{3}$ belonging to the same $1$-block, i.e.,~$K_{3},L_{3}\in K_{1}$, the factor $\lambda_{L_{3}}-\lambda_{K_{3}}$ becomes zero, since the eigenvalues in $1$-blocks are identical. Hence, only the $1$-block off-diagonal entries of $\mathbf{S}_{3;x}$ and $\mathbf{S}_{3;y}$ need to be considered. These are available from Eqs.~\eqref{eq:derivations-Sx} and~\eqref{eq:derivations-Sy} transformed to the $3$-basis if necessary (which is subordinate to the $1$-partition of $\mathbf{\Omega}$). This finally gives
        \begin{align}
        \begin{split}
            (\mathbf{\Lambda}_{,xy})_{K_{3}K_{3}}&=\ (\mathbf{P}_{4}^{-1})_{K_3 K_3}
            \Big((\mathbf{A}_{3,xy})_{K_{3}K_{3}}+(\lambda_{L_{3}}-\lambda_{K_{3}})
            \times %\nonumber\\
            %&\qquad \times 
            \big((\mathbf{\Omega}\cdot \mathbf{A}_{3,x})_{K_{3}L_{3}}(\mathbf{\Omega}\cdot \mathbf{A}_{3,y})_{L_{3}K_{3}} + %\nonumber
            \\
            &\quad +
            (\mathbf{\Omega}\cdot \mathbf{A}_{3,y})_{K_{3}L_{3}}(\mathbf{\Omega}\cdot \mathbf{A}_{3,x})_{L_{3}K_{3}}\big)\Big)
            (\mathbf{P}_{4})_{K_3 K_3}.
        \end{split}
        \end{align}
        The $\lambda_{L_3}-\lambda_{K_3}$ factor removes one entry-wise $\mathbf{\Omega}$ multiplication while preserving only the $1$-block off-diagonal parts. The entire equation for diagonal $3$-blocks of $\mathbf{\Lambda}_{,xy}$ can then be extended to a full-matrix equation by using the block-diagonal extraction operator $[3]$ on both sides. By taking into account that $\mathbf{\Lambda}_{,xy}$ and $\mathbf{P}_{4}$ are $3$-block diagonal already, we get
        {\small
        \begin{align}
            \mathbf{\Lambda}_{,xy}&=\mathbf{P}_{4}^{-1}\Big(\mathbf{A}_{3,xy}-(\mathbf{\Omega}\cdot \mathbf{A}_{3,x})\,\mathbf{A}_{3,y}^{(1)}-(\mathbf{\Omega}\cdot \mathbf{A}_{3,y})\,\mathbf{A}_{3,x}^{(1)}\Big)^{[3]}\mathbf{P}_{4}.
        \end{align}
        }
        We have thus derived the final result of Eq.~\eqref{eq:algorithm-lambda-xy}. Note that this expression is manifestly symmetric with respect to $x$ and $y$.
\end{enumerate}
%

%======================================================
\subsection{Detailed handling of logs\label{app:logs}}
%======================================================

The procedure of Appendix~\ref{app:general} computes one-loop corrections only to the mass parameters in the scalar potential, so the resulting quantity is the \emph{one-loop effective mass} rather than the \emph{physical mass} (also referred to as the \emph{pole mass}). The latter requires also knowledge of momentum-dependent contributions coming from self-energy diagrams, which we do not consider. It is possible, however, to improve upon the effective-mass calculation by carefully considering its logarithmic contributions. This improved version is referred to as the \emph{regularized effective mass}.   

One-loop effective scalar masses in Eq.~\eqref{eq:DDV-1loop} inevitably contain contributions of the form
\begin{align}
\label{eq:log}
\log\left[\frac{m^2_l}{\mu_R^2}\right],
\end{align}
where $m_l$ is the tree-level mass of a scalar or gauge field in the loop part of a Feynman diagram. The following situations require further attention:
\begin{enumerate}
    \item The particles in the loop are much lighter than the rest of the spectrum, e.g.,~when their masses are near the intermediate scale. In such a case, the regime $m^2_l \ll \mu_R^2$ causes logarithmic contributions to be unphysically large. This effect is cancelled in physical masses by self-energy contributions, which we do not have access to.  \label{item:1}
    \item For WGBs, $m_{l}^{2}=0$, so the logs contain an IR divergence unless their prefactor expressions vanish as well.\footnote{Some of the diverging logarithms of Eq.~\eqref{eq:log} are tamed by a $m^2_l$-proportional prefactor.} \label{item:2}
    \item At least one of the PGBs has a tree-level tachyonic instability, i.e.,~$m_{l}^{2}<0$, which causes the argument of the log to be negative. \label{item:3}
\end{enumerate}

We prepare the ground for solving these issues in Appendices~\ref{app:regularized-mass} and~\ref{app:physical-vs-regularized} and then present a comprehensive list of how to practically treat all the cases in Appendix~\ref{app:practical}.

%......................................................
\subsubsection{The regularized effective mass\label{app:regularized-mass}}
%......................................................

We define a quantity called \emph{the regularized one-loop effective scalar mass} $M^2_{S, reg}$ : We take the expression for the one-loop effective scalar mass $M^2_S$ but solve the problem of IR divergences by emulating the shift from the effective to the physical mass. 

\begin{figure}[htb]
    \centering
    \mbox{
    \includegraphics[width=0.5\columnwidth]{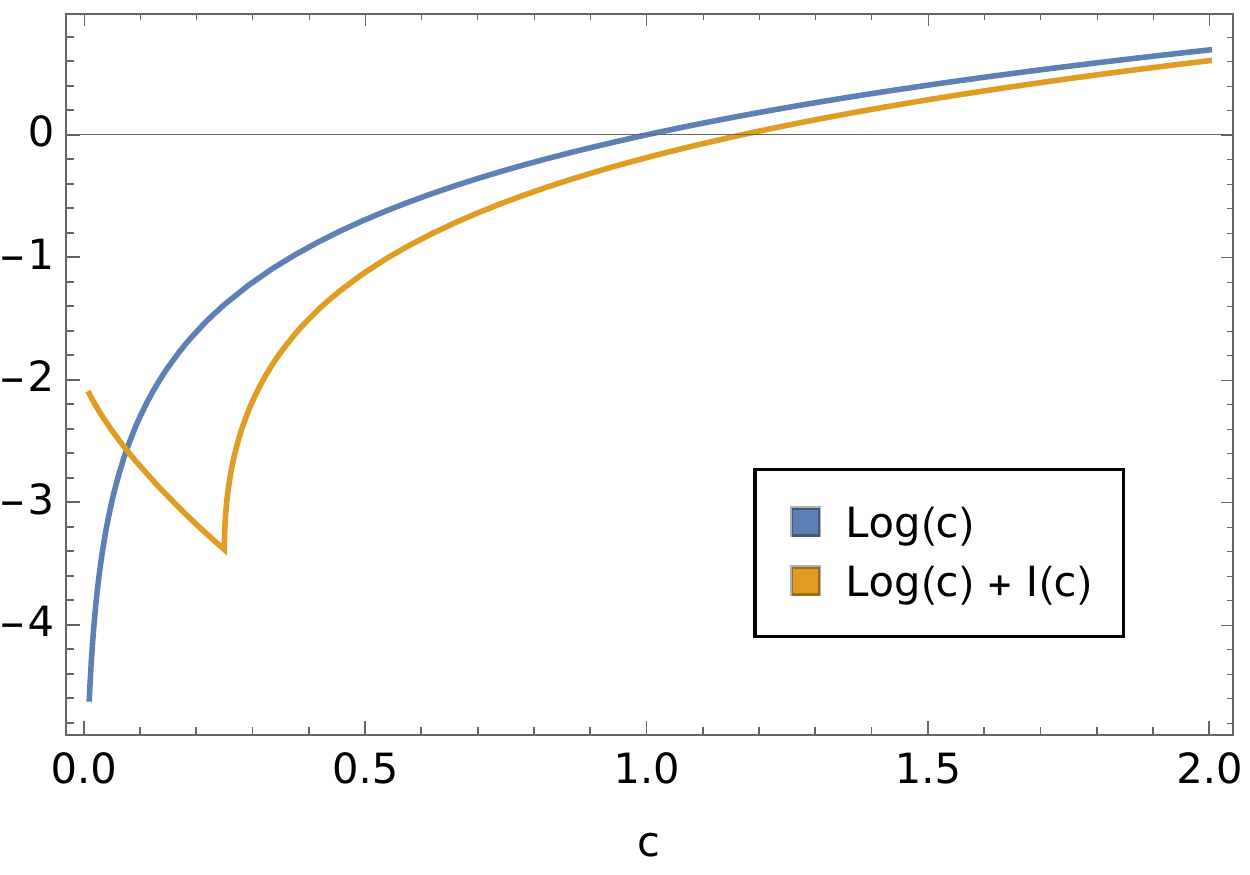}
    }
    \caption{Logarithmic contributions to the one-loop effective scalar masses with taming of the IR divergences [$\log(c) + I(c)$] and without it [$\log(c)$]. The parameter $c = \frac{m^2_l}{m^2_S}$, where $m_l$ and $m_S$ are tree-level masses of fields in the loop and on the outer legs, respectively.}
   \label{fig:TamingOfTheLogs}
\end{figure}

In particular, we take inspiration from the Abelian Higgs model~\cite{Malinsky:2012tp}, which suggests using the replacement
\begin{align}
\log \left[\frac{m^2_l}{\mu_R^2}\right] &\mapsto \int_0^1 dx \log \left[\frac{|m^2_l - m^2_S x(1-x)|}{\mu_R^2}\right] %\nonumber \\
%&\quad 
= \log \left[\frac{m^2_S}{\mu_R^2}\right] + \log\left[c \right] + I\left(c\right), \label{eq:logReplacement}
\end{align}
where $m_S$ is the tree-level mass of the particle on the outer legs (i.e.,~the particle whose one-loop mass we are computing) and $c:= \frac{m^2_l}{m^2_S}$. The term $I\left(c\right)$ includes integration and is rather time-consuming to be evaluated in all one-loop corrections. The effect of the integral $I\left(c\right)$ is shown in Fig.~\ref{fig:TamingOfTheLogs}.

We simplify the approximation in the replacement rule of Eq.~\eqref{eq:logReplacement} by following the normal log (blue curve) at high $c$ until $I\left(c_0\right) = 0$ for $c_0 \approx 0.0763$ (the intersection of the blue and orange curves), then take a constant value for $c$ smaller than $c_0$. The regularized one-loop effective scalar mass is thus defined as the one-loop effective mass with all logarithmic terms of Eq.~\eqref{eq:log} substituted by
\begin{align}
\label{eq:RegMassDef}
M^2_{S,reg}: \log \left[\frac{m^2_l}{\mu_R^2}\right] \mapsto  \begin{cases} 
\log \left[\frac{m^2_l}{\mu_R^2}\right]  &;\frac{m^2_l}{m^2_S} > c_0, \\
\log \left[c_0 \frac{m^2_S}{\mu_R^2}\right]  &;\frac{m^2_l}{m^2_S} \leq c_0.
\end{cases}
\end{align}
We refer to this procedure as \emph{taming the logs}.

%......................................................
\subsubsection{Physical vs. regularized effective mass\label{app:physical-vs-regularized}}
%......................................................

The regularized one-loop effective mass is the best possible approximation to the actual physical mass without knowledge of self-energy contributions. A crucial observation for inferring the non-tachyonicity of the physical spectrum is that in the perturbative regime, the non-tachyonicity of the regularized effective mass implies that the physical mass is also non-tachyonic. We schematically argue this point below.

Let us consider the continuous Fourier-transformed two-point one-particle irreducible (OPI) Green's function $\Gamma^{(2)}_{\overline{\text{MS}}} (p^2)$ evaluated in the $\overline{\text{MS}}$ renormalization scheme, where $p^2$ is square of the outer leg particle's four-momentum. In what follows, we suppress the matrix structure in mass expressions for simplicity. In the effective potential approach, we obtain the effective mass of a field, which is simply the zero-momentum value of the Green's function:
\begin{align}
\label{eq:effmassdef}
M^2_S = -\Gamma^{(2)}_{\overline{\text{MS}}} (p^2 = 0).
\end{align}
Subsequently,
\begin{align}
M^2_{S,reg} = M^2_S - \text{IR divergences}
\end{align}
schematically holds for the regularized effective mass. The physical mass though is defined as a solution to the equation
\begin{align}
\label{eq:physmassdef}
\Gamma^{(2)}_{\overline{\text{MS}}} (p^2 = M^2_{phys})= 0.
\end{align}
In addition, the behaviour of the two-point Green's function is constrained in such a way that
\begin{align}
\label{eq:2pGreenFuncBeh}
\left.\frac{d Z\Gamma^{(2)}_{\overline{\text{MS}}}}{d p^2} \right \vert_{p^2 = M^2_{phys}} = 1,
\end{align}
where $Z$ is the field-strength renormalization. We require that there exists only one solution to Eq.~\eqref{eq:physmassdef}, and we stay in the perturbative regime, where higher-loop corrections are subdominant with respect to the tree-level values. The latter tree-level case corresponds to
\begin{align}
M^2_{S,reg} =  M^2_{phys}, & \quad\quad Z  = 1.
\end{align}
\begin{figure}[htb]
    \centering
    \fbox{
    \includegraphics[width=0.6\columnwidth]{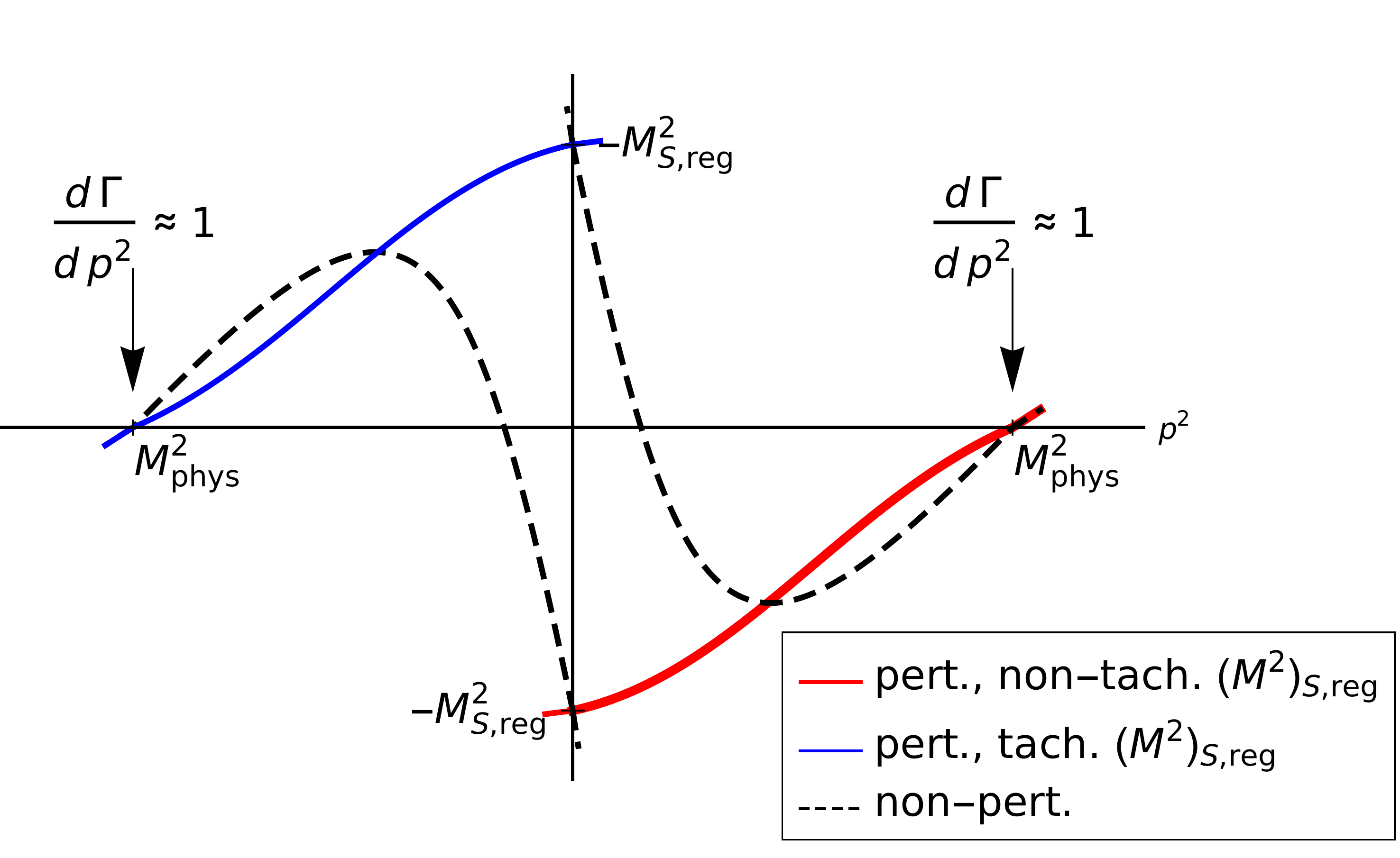}
    }
    \caption{Four different schematic depictions of two-point OPI Green's function $\Gamma^{(2)}_{\overline{\text{MS}}}$ behaviour demonstrating four combinations of tachyonic or non-tachyonic physical $M^2_{phys}$ or regularized effective $M^2_S$ masses. 
    }
   \label{fig:GreenFunctionBehaviour}
\end{figure}

Schematically, there are four different qualitative solutions that can occur; see Fig.~\ref{fig:GreenFunctionBehaviour}. The solid lines show what must qualitatively happen in perturbative scenarios, while the \textbf{black} dashed lines illustrate what might occur in non-perturbative situations significantly deviating from the tree-level case. The \textbf{\textcolor{red}{red}} color labels the case with a perturbative non-tachyonic $M^2_S$ mass, while the \textbf{\color{plotBlue2}blue} color is used for the tachyonic $M^2_S$ mass scenario. The non-perturbative cases are not of interest here since they prevent us from using perturbative expansions of OPI Green's functions. %lead to non-uniqueness of the pole mass. 
Hence, one can conclude that \textbf{in the perturbative regime}
{\small
\begin{center}
	$\mathbf{M^2_{phys}}$ is non-tachyonic $\mathbf{\Leftrightarrow}$ $\mathbf{M^2_{S,reg}}$ is non-tachyonic.\\
\end{center}
}
For valid parameter points, which pass the perturbativity and non-tachyonicity check, the situation for all mass eigenstates thus corresponds to the red-line scenario. 

%......................................................
\subsubsection{Practical aspects of the one-loop scalar-mass calculations\label{app:practical}}
%......................................................

%
\begin{table*}[htb]
\caption{
Summary of potential issues related to vastly different scales (associated with GUT-scale and intermediate-scale sectors at play) appearing in the one-loop mass calculations, and their practical treatment.}
\label{tab:MassIssueSummary}
\begin{tabular}{p{41mm}@{$\quad$}p{43mm}@{$\quad$}p{41mm}@{$\quad$}p{40mm}}
\hline
\textbf{Light field type} & \textbf{A: Potential issues related to the calculation of loop contributions to their masses} & \textbf{B: Potential issues related to their presence in other fields' loop mass corrections}
& \textbf{Practical treatment of A and/or B} \\
\hline
\hline
\textbf{WGBs} (would-be \hyphenation{Gold-sto-ne} Goldstone bosons) 
& None. WGBs are massless at all loops if calculated exactly. & WGBs produce IR-diverging logarithms, such as \hbox{$\log \left[\frac{m_{WGB}^2}{\mu_R^2}\right]$} with \hbox{$m_{WGB}=0$}, that get cancelled in physical masses.  
&  {\bf A:} The WGBs' zero masses will be recovered no matter the order of the perturbation series.\par
{\bf B:} The shift from the effective to physical masses is modelled by  substitution~\eqref{eq:RegMassDef}.\\
\hline
\textbf{Fields associated with intermediate-symmetry breaking}
& Loop corrections to the light fields should be calculated using intermediate-symmetry effective-field theory, not the full $\mathrm{SO}(10)$. & If the light mass is way lighter than $m^2_S$ (denoting the tree-level mass the loop correction is calculated to) potentially large logs may emerge. 
&{\bf A:} The tree-level mass will be used for all practical purposes. 
\par {\bf B:} Large logs will get replaced (as for WGBs) via substitution~\eqref{eq:RegMassDef}.\\
\hline
\textbf{PGBs} (pseudo-Goldstone bosons), such as those discussed in Sec.~\ref{sec:PSG-singlets} 
& 
Light fields in the loops may produce large logs as they are not properly tamed by using substitution~\eqref{eq:RegMassDef} with tree-level PGB mass inserted in.
 & The same as above for the light fields associated with intermediate-symmetry breaking. 
 & {\bf A:}
An iterative approach is employed,\textsuperscript{\emph{a}} with the PGB tree-level masses in~\eqref{eq:RegMassDef} substituted by one-loop PGB masses calculated in the previous iteration; cf.~Appendix~\ref{app:perturbativity}.
 \par
 {\bf B: } Via Eq.~\eqref{eq:RegMassDef}.
 \\
\hline
\multicolumn{4}{l}{\textsuperscript{\emph{a}}\footnotesize{Some potentially large logs vanish in the two limits of interest due to the $T_i$ and $O_i$ prefactors of Ref.~\cite{Graf:2016znk}, Appendix~B.}}
\end{tabular}
\end{table*}

All potential issues related to the presence of vastly different scales in the one-loop mass calculation (due to lighter fields in the loop), as well as methods of their resolution are identified in Table~\ref{tab:MassIssueSummary}. This resolves cases~\ref{item:1} and~\ref{item:2} in Appendix~\ref{app:logs} proper.

The arguments of the logs are always taken in \textbf{absolute values}, so that the issues with a negative log argument are avoided (case~\ref{item:3} of Appendix~\ref{app:logs} proper). Since we check the non-tachyonicity of the spectrum for every point at the one-loop level, and our perturbativity requirements demand the loop contributions to GUT-scale particles to be smaller than the tree-level values, the absolute value is relevant only for PGB fields in the loop. Furthermore, since the PGB contribution is tamed in the computation of masses for \emph{heavy} fields, we only need to worry about this problem when computing corrections to PGBs coming from PGBs in the loop. In such a case, the tree-level PGB approximation is considered bad anyway, and we perform a further iterative procedure with an associated perturbativity check; cf.~item~\ref{app:iterativePGB} in Appendix~\ref{app:perturbativity}.

%%%%%%%%%%%%%%%%%%%%%%%%%%%%%%%%%%%%%%%%%%%%%%%%%%%%%%%
\section{Details of the numerical analysis\label{App:viability-considerations}}
%%%%%%%%%%%%%%%%%%%%%%%%%%%%%%%%%%%%%%%%%%%%%%%%%%%%%%%

In this appendix, we address the technical details of the viability criteria formulated in Sec.~\ref{Sec:analysis}. In particular, we outline the implementation of various parameters describing the quality of fits presented in Sec.~\ref{sec:numerical-scan} along with the penalties associated with the potential violation of the tachyonicity and perturbativity criteria in the step-by-step numerical procedure.

%======================================================
\subsection{The viability criteria of Sec.~\ref{Sec:analysis}\label{app:viability-considerations-details}}
%======================================================

%......................................................
\subsubsection{Non-tachyonicity of the scalar spectrum\label{app:tachyonicity}}
%......................................................

Non-tachyonicity of the mass spectrum is an essential consistency criterion. To this end, we  use the regularized one-loop effective mass defined in Appendix~\ref{app:logs} for all the scalars, except for those fields which are parametrically lighter due to their association with the intermediate-symmetry scale $\sigma$. Since none of these states is a pseudo-Goldstone boson, we can afford using the tree-level formulae for their masses instead; see Table~\ref{tab:MassIssueSummary}.

The contribution to the penalty of a parameter-space point from the possible scalar spectrum tachyonicity is defined as
\begin{align}\label{eq:A1}
p_1 &= A_1 \arctan \left[ \sqrt{\sum_{j}\frac{1- \sign\left(M^2_j\right)}{2} \frac{\vert  M^2_j \vert^2}{s_j}}\;\right],
\end{align}
where $A_1\in \mathbb{R}^{+}$ is a weight parameter, $j$ runs over all the scalar fields, and $s_j$ stands for the expected scale of $M^2_j$, i.e.,~$s_j=\omega_{max}^2$ ($\omega_{max}:= \max \left[ |\omega_{BL}|,|\omega_R| \right]$) for the \emph{heavy} fields associated with the GUT scale, $s_j = \tfrac{\omega_{max}^2}{10}$ for PGBs, and $s_j= \sigma^2$ for the fields which should be naturally around the intermediate scale. The only exceptions are the SM multiplets $(3,1,-\tfrac{1}{3})$ and $(1,2, -\tfrac{1}{2})$, for which the mass and sign factors $|M^2_j|$ and $\sign\left(M^2_j\right)$ are replaced by the leading principal minors of the relevant mass matrices and their signatures, respectively\footnote{For these fields, the typical scale factor $s_j$ is defined as $\omega_{max}^{2k}$ where $k$ stands for the dimensionality of the corresponding minor.}; see Sec.~\ref{sec:Tachyonicity}. Note that the first factor in the sum above ensures that there is no penalty for non-tachyonic masses. 

%......................................................
\subsubsection{Unification of gauge couplings\label{app:unification}}
%......................................................

In this study, the SM gauge couplings are expected to be unified at the one-loop level; cf.~Sec.~\ref{sec:unification-constraints}. The evolution of the usual $\alpha$-factors
\begin{align}
\label{eq:gaugeFactors}
\alpha_i &= \frac{g_i^2}{4\pi}
\end{align}
with the EW-scale boundary conditions is then driven by the equations
\begin{align}
\label{eq:da/dt}
\frac{d}{dt} \alpha_i^{-1} &= -a_i\,.
\end{align}
The dimensionless running parameter $t$ is defined as
\begin{align}
\label{eq:t}
t &= \frac{1}{2\pi} \log \frac{\mu_{R}}{M_Z},
\end{align}
with $\mu_{R}$ denoting the running renormalization scale, and
\begin{align}
\label{eq:ai}
a_i &= -\frac{11}{3}\sum_{G} T(G_i)D(G_i) + \frac{4}{3}\sum_F \kappa_FT(F_i)D(F_i)
%+ \nonumber\\
%&\quad 
+ \frac{1}{3} \sum_S \eta_S T(S_i)D(S_i),\qquad i=1,2,3
\end{align}
correspond to the beta functions of the ``reduced'' one-loop SM gauge couplings. The three sums above run over the gauge, fermion, and scalar fields, respectively, $T(R_i)$ are the Dynkin indices of representations $R_i$ with respect to the group factor $G_i~\in~\lbrace \SU(3)_c, \SU(2)_L,\mathrm{U}(1)_Y \rbrace$, and $D(R_i)$ denote the relevant multiplicities of $R_i$. We take $\kappa_F = \tfrac{1}{2}$ or $1$ for Weyl or Dirac fermions, respectively; likewise $\eta_S = \tfrac{1}{2}$ or $1$ for real or complex scalar fields. 

The solution of Eq.~\eqref{eq:da/dt} is trivial:
\begin{align}
\label{eq:aSolution}
\alpha_i^{-1}(t) &= \alpha_i^{-1} - \left(a_i^{(SM)} t +
\sum_{t_k<t}\Delta a_i^{(k)} (t - t_k)\right),
\end{align}
where we used the abbreviation $\alpha_i^{-1}(0) = \alpha_i^{-1}$ and $t_k~=~\tfrac{1}{2\pi} \log \tfrac{M_k}{M_Z}$. In the expression above, the Standard Model contributions to Eq.~\eqref{eq:ai} have been collected in $a_i^{(SM)}$ with values $(a_3^{(SM)},a_2^{(SM)},a_1^{(SM)})=(-7,-\tfrac{19}{6},\tfrac{41}{10}),$ while the $\Delta a_i^{(k)}$ terms listed in Tables~\ref{tab:tree-level-gauge-boson-masses} and~\ref{tab:tree-level-scalar-masses} encompass the effects of gradual addition of the beyond-SM fields at scales $t$ exceeding their individual masses $t_k$.

The gauge-coupling-unification constraint thus reads $\alpha_i^{-1}(t_{GUT}) = \alpha^{-1}(t_{GUT}),~i = 1,2,3,$ where $t_{GUT}$ is any scale at or above the heaviest threshold non-trivially influencing the gauge coupling evolution. Thus, relation~\eqref{eq:aSolution} can be rewritten as
\begin{align}
\begin{split}
    \alpha_i^{-1}&= \alpha^{-1}(t_{GUT}) + a_i^{(SM)} t_{GUT} %+ \\
    %&\quad 
    + \backspace\quad \sum_{\forwardspace k \in \text{all fields}\backslash SM}\backspace \Delta a_i^{(k)} (t_{GUT}-t_k)%- \\
    %&\quad 
    -\frac{1}{4\pi} \log \frac{\alpha^{-1}(t_{\mu_R})}{\alpha^{-1} (t_{GUT})} \backspace \,\,\, \sum_{\forwardspace j \in \text{gauge bosons}\backslash SM} \backspace \Delta a_i^{(j)}, \\
\end{split}\label{eq:aSolution2}
\end{align}
where the last subleading term takes care of the fact that the gauge boson masses are calculated with the $\mathrm{SO}(10)$ gauge coupling evaluated at the renormalization scale $\mu_R$, which is optimally chosen for each point separately (to be specified in Appendix~\ref{app:procedure}).

In addition to all details of the bosonic spectrum, a good unification pattern is characterized by $3$ main parameters, namely, $t_{GUT}$, $t_\sigma~=~\tfrac{1}{2\pi} \log \tfrac{|\sigma|}{M_Z}$, and  $\alpha^{-1}$. These three quantities correspond to three basic degrees of freedom which can be manipulated by: (i) scaling all dimensionful quantities $\lbrace \omega_{BL},\omega_R,\sigma,\mu_R \rbrace$ by a common factor, (ii) shifting the intermediate-scale VEV $\sigma$, and (iii) modifying the initial condition for the $\mathrm{SO}(10)$ gauge coupling. Parametrizing the  associated shifts by
\begin{align}
\label{eq:UnificationParameters}
\lbrace\Delta t_{GUT}, \Delta t_\sigma, \Delta \alpha^{-1} \rbrace,&
\end{align}
Eqs.~\eqref{eq:aSolution2} become the following set of three independent conditions for these parameters\footnote{Here we assume that the $\mathrm{SO}(10)$ fields (except the SM field content) can be divided into two categories according to mass scale: (i) the \emph{heavy} fields and PGBs, and (ii) the intermediate-scale fields. Their masses are governed by $\omega_{max}$ and $\sigma$, respectively.}:
\begin{align}
\label{eq:UnificationCondition2}
\begin{split}
    \alpha_i^{-1}&=\phantom{+} \alpha^{-1}(t_{GUT}) + \Delta \alpha^{-1} + a_i^{(SM)} (t_{GUT} + \Delta t_{GUT}) %- \\
    %&\quad 
    -\backspace \sum_{\forwardspace k \in \text{all fields}\backslash SM}\backspace\Delta a_i^{(k)} (t_k - t_{GUT}) - \backspace\sum_{\forwardspace l \in \text{intermediate fields}} \backspace\backspace\quad \Delta a_i^{(l)} \Delta t_\sigma -\\ 
    &\quad -\frac{1}{4\pi} \log \frac{\alpha^{-1}(t_{\mu_R})}{\alpha^{-1} (t_{GUT}) + \Delta \alpha^{-1}}\backspace \;\;\; \sum_{\forwardspace j \in \text{gauge bosons}\backslash SM} \backspace\backspace\quad \Delta a_i^{(j)}.\\
\end{split}
\end{align}
These are rather easy to solve for most of the initial choices of $t_{GUT}$, $t_\sigma$, and  $\alpha^{-1}$, which together with the resulting shifts yield the desired gauge unification pattern.

%......................................................
\subsubsection{Perturbativity constraints\label{app:perturbativity}}
%......................................................

As far as the perturbativity constraints of Sec.~\ref{sec:Perturbativity} are concerned, we implement four simple tests that quantify the level of our satisfaction\footnote{Recall that perturbativity constraints are the kind of criteria which, practically by definition, require human (and hence, biased) input; cf.~Sec.~\ref{sec:Perturbativity}.} with the overall perturbativity.  The associated considerations in the order of Sec.~\ref{sec:Perturbativity} are as follows:
\begin{enumerate}
%%%%%%%%%%%%%%%%%%%%%%%%%
\item \underline{Global-mass-perturbativity test\label{app:massPerturbativity}}\\
The penalty $p_2$ associated with the Global-mass-perturbativity test is defined as
\begin{align}\label{eq:A2}
p_2 &= \frac{2}{\pi} A_2 \cdot \arctan \left[ H(\overline{\Delta}-1)\right],
\end{align}
where $A_2\in \mathbb{R}^{+}$ is a weight factor, $H(x):= x\,\theta(x)$, $\theta(x)$ is the Heaviside step function,
$\overline{\Delta}$ is defined in a repeat of Eq.~\eqref{eq:definition-delta} by
\begin{align}
    \overline{\Delta} &:= \frac{\max_{i,j \in \text{\emph{heavy} fields}}[|M^2_{ij, \text{one-loop}}-M^2_{ij,\text{tree}}|]}{\overline{M^2}_{\rm heavy}}, 
\end{align}
with $M^2_{ij,\text{one-loop}}$ and $M^2_{ij,\text{tree}}$ denoting one-loop and tree-level scalar-mass-matrix elements, respectively. The symbol $\overline{M^2}_{\rm heavy}$ denotes the average (over real degrees of freedom) of the \emph{heavy} tree-level scalar masses. In physical terms, a penalty is awarded if the maximal one-loop correction of the mass square is larger than the average tree-level mass square.

%%%%%%%%%%%%%%%%%%%%%%%%%
\item \underline{Perturbativity of the RG evolution\label{app:rgePerturbativity}}\\
For a given individual point in the parameter space, all calculations are done at a certain ``optimal'' renormalization scale $\mu_R$. For measurable quantities (such as pole masses), the specific choice of $\mu_R$ should not matter. However, due to our use of the effective potential approach and proxy quantities such as regularized effective masses, the $\mu_{R}$-dependence is technically not eliminated~\cite{Bando:1992np,Bando:1992wy}. \par

The residual $\mu_{R}$-dependence needs to be kept under control to ensure that the perturbativity constraints will not get out of hand once the renormalization scale is changed. For that purpose, we perform one-loop RG running of dimensionless parameters and check for the stability of the vacuum position under loop corrections (cf.~next point) at a renormalization scale other than $\mu_{R}$ as well (specifics are given later in Appendix~\ref{app:procedure}).

For further convenience, we introduce some auxiliary quantities, repeating here the definitions of Eqs.~\eqref{eq:definition-tplusminus} and~\eqref{eq:definition-tbar} for completeness:
\begin{align}
    t_\pm &= \log_{10} \frac{\mu_{R\pm}}{\mu_R},
\end{align}
where $\mu_R$ is the starting ``optimal'' renormalization scale specific to every point in the parameter space, and  $\mu_{R+}$ ($\mu_{R-}$) denotes the renormalization scale for which the couplings hit a Landau pole when running upward (downward). The geometric average
\begin{align}
     \overline{t}= \sqrt{t_- t_+}
\end{align}
encodes, roughly speaking, how many orders of magnitude the scalar couplings can run up and down before encountering a Landau-pole-type singularity. Finally we define an optional penalty $p_5$ in order to test levels of robustness with respect to the RG running:
\begin{align}
     p_5 = \frac{2}{\pi} A_5 \cdot \arctan \left[ H \left(\frac{1}{\bar{t}} - \frac{1}{\bar{t}_{thr}}\right) \right],
\end{align}
where $A_5 \in \mathbb{R}^{+}$ is a weight factor, $H(x)$ is defined below Eq.~\eqref{eq:A2}, and $\bar{t}_{thr}$ is an acceptance threshold for $\bar{t}$.

%%%%%%%%%%%%%%%%%%%%%%%%%
\item \underline{Stability of the vacuum position\label{app:vacuumStability}}\\
    We start by defining a vector of dimensionful scalar parameters $\vec{w} = \lbrace \mu^2,\nu^2,\tau^2 \rbrace$.
It is connected to the VEVs via one-loop stationarity conditions, which can be written in the form
\begin{align}
\label{eq:general_stat_cond}
\vec{w} = \vec{w_0} +f(\vec{w}),
\end{align}
where $\vec{w_0}$ denotes the tree-level part from Eqs.~\eqref{eq:mu}--\eqref{eq:tau}, and the function $f(\vec{w})$ represents one-loop corrections. Eq.~\eqref{eq:general_stat_cond} is solved iteratively via
\begin{align}
    \label{eq:stat_cond_iter}
    \vec{w}^{(k+1)}  = \vec{w}_0 + f(\vec{w}^{(k)}),~k\geq1 ,
\end{align}
taking $\vec{w}^{(0)} = \vec{w_0}$. The iterative procedure is stopped, and $\vec{w}$ is deemed to not have converged if
\begin{align}
    \max_{i\in \lbrace 1,2,3 \rbrace} \sqrt{\left\vert \frac{w^{(k)}_i}{w_{0i}}\right\vert } & >1+\zeta , \label{eq:stat_cond_stop}
\end{align}
where we set $\zeta = 0.3$, and the result of the iterative procedure is labelled as $\vec{w}_{iter}$. The maximal number of iterations is chosen to be $30$.

The penalization $p_3$ that assesses vacuum stability is the following:
\begin{widetext}
\begin{align}
	\nonumber 
	\text{if}&& 	\sqrt{\frac{\lVert \vec{\lambda} \rVert^2}{12}}> 10: && 
	p_3 &= \frac{1}{3} \left( A_3^{(1)} + A_3^{(2)} \right) + \frac{2}{3\pi} A_3^{(3)} \arctan \left[\sqrt{\frac{\lVert \vec{\lambda} \rVert ^2}{12}}-10 \right],\\[8pt]
	\nonumber 
	\text{else if}&& \vec{w}_{iter}\text{ does not converge}: && 
	p_3 &=\frac{1}{3} A_3^{(1)}+\frac{2}{3\pi} A_3^{(2)} \arctan \Bigg[ H\left(\frac{\lVert \vec{w}_{iter} - \vec{w}_0 \rVert }{\lVert \vec{w}_0 \rVert}-1\right) + \nonumber \\
	&&&&& \qquad +( 30 - \#_{\text{iterations}}) \Bigg],\nonumber \\[8pt]
	\nonumber 
	\text{else}&& \vec{w}_{iter} \text{ converges:}&&
	p_3 &= \frac{2}{3\pi} A_3^{(1)} \arctan \left[ H\left(\frac{\lVert \vec{w}_{iter} - \vec{w}_0 \rVert}{\lVert \vec{w}_0 \rVert}-1\right)+ \right.\\
	&&&&&\left.\quad\quad +H\left(\frac{\lVert \vec{w}_{iter} - \vec{w}^{(1)} \rVert }{\lVert \vec{w}^{(1)} \rVert}-1\right)\right],
	\label{eq:A3}
\end{align}
\end{widetext}
where $A_3^{(i)}\in \mathbb{R}^{+}$ are weight factors, $H(x)$ is defined below Eq.~\eqref{eq:A2}, and the vector of scalar couplings is
\begin{align}
    \vec{\lambda} &:= \lbrace a_2,a_0,\lambda_0,\lambda_2,\lambda_4,\lambda_4^\prime, \alpha,\beta_4,\beta_4^\prime,\gamma_2,\eta_2 \rbrace. \label{eq:vector-of-parameters}
\end{align}

The penalization in Eq.~\eqref{eq:A3} is constructed so that each parameter point falls under exactly one of three conditions. The first condition and the associated $A_3^{(3)}$ weight penalize points which fall well outside the $\mathcal{O}(1)$ circle.
The second condition and $A_{3}^{(2)}$ penalize points for which the vector of couplings $\vec{\lambda}$ is sufficiently small, but $\vec{w}_{iter}$ does not converge under the criterion in Eq.~\eqref{eq:stat_cond_stop}; i.e.,~the iterative procedure is stopped prematurely. Analogously, the third condition and weight $A_{3}^{(1)}$ penalize points if a converged $\vec{w}_{iter}$ differs too much from the tree-level $\vec{w}_0$ and the initial approximation of the one-loop result $\vec{w}^{(1)}$.
Note that the expressions are such that the penalization is smaller if a later condition applies, effectively ranking the criteria in descending order of importance. 

%%%%%%%%%%%%%%%%%%%%%%%%%
\item \underline{Iterative pseudo-Goldstone masses\label{app:iterativePGB}}\\
It turns out the computation of regularized one-loop effective masses for PGBs involves certain subtleties. As laid out in Table~\ref{tab:MassIssueSummary}, the presence of particles with masses much below the GUT scale in the loop is handled by the replacement in Eq.~\eqref{eq:RegMassDef} in order to tame unphysical large-log contributions~\cite{Martin:2014bca}. When computing mass corrections to PGBs, Eq.~\eqref{eq:RegMassDef} implies replacing log arguments 
involving the mass of lighter (e.g.,~intermediate-scale) particles with the PGB mass, leading to an equation of the form
\begin{align}
    M^{2}_{PGB}&=C_{1}+C_{2}\,\log(M^{2}_{PGB}/\mu_{R}^{2}),\label{eq:MPGB-self}
\end{align}
where $C_{1}$ and $C_{2}$ are numeric coefficients independent of the expression $M^{2}_{PGB}$. The tree-level masses of the PGBs are accidentally small and not very close to the chosen renormalization scale. This implies that the log contribution is additionally enhanced, and the $C_{1}$ term, usually dominated by the tree-level mass, is reduced. The relative importance of the $C_{2}$ term is thus increased, and the solution of Eq.~\eqref{eq:MPGB-self} can be far away from $C_{1}$.
This is problematic conceptually, since the $C_{2}$ term itself is merely an approximation arising from taming the logs. Note that the computation of \emph{heavy} masses does not suffer from the same problem, despite an analogous equation, since their tree-level masses are large and the $C_{1}$ term there dominates.

In practice, we solve Eq.~\eqref{eq:MPGB-self} iteratively via
\begin{align}
    M^{2}_{PGB,(i+1)}&=C_{1}+C_{2}\,\log(|M^{2}_{PGB,(i)}|/\mu_{R}^{2})\label{eq:MPGB-iter}
\end{align}
starting with the value $M^{2}_{PGB,(0)}$ as the tree-level mass and requiring that all iterations after $i\geq 1$ are non-tachyonic. Note the absolute value in the argument of the log placed there to deal with the possible initial tachyonic instability. We perform a maximum of $30$ iterations. We deem that convergence has been achieved and stop the process if the relative size of the shift between $M^2_{PGB,(i+1)}$ and $M^{2}_{PGB,(i)}$ of two consecutive steps is smaller than $10^{-3}$, a size corresponding to corrections at two-loop level. We denote the result of the iterative process as $M^{2}_{PGB,iter}$, while $M^{2}_{PGB,(1)}$ is what we refer to in Appendix~\ref{app:logs} as the \emph{regularized one-loop effective mass}.

Because of the issues discussed above, a point is considered valid not only if 
convergence in Eq.~\eqref{eq:MPGB-iter} is achieved, but also that the $C_{2}$-proportional log contribution is sufficiently small, ensuring the log-taming approximation works as intended and the computed regularized effective mass is close to the physical mass. Hence, we demand that the relative difference between the regularized one-loop effective PGB mass $M^2_{PGB,(1)}$ and the iterative solution $M^2_{PGB,iter}$ is less than\footnote{Smaller differences are at the level of two-loop mass corrections. Moreover, this choice assures that the replacement of $M^2_{PGB,(1)}$ with $M^2_{PGB,iter}$ in gauge-coupling unification causes changes comparable to two-loop running effects.} $10\%$. We define the penalty $p_4$ associated with these considerations, applied now to multiple PGBs, by
{\small 
\begin{align}
    \text{if } && 
    M^2_{PGB,iter}\ \text{does not converge}: &&\nonumber%\\
    &\qquad p_4 = \frac{1}{2} A_4^{(1)}+ \frac{1}{\pi} A_4^{(2)} \arctan \left[ H(\Delta M^2-10^{-1})\right. %+ \nonumber\\
    %&\qquad\qquad 
    \left.+ H(\delta M^2_{iter}-10^{-3})\right],\nonumber \\[8pt]
    \text{else } && 
    M^2_{PGB,iter}\ \text{converges}: &&%\nonumber\\
    &\qquad p_4 =  \frac{1}{\pi} A_4^{(1)} \arctan \left[H(\Delta M^2-10^{-1})\right],
    \label{eq:A4}
\end{align}
}
where $A_4^{(i)}\in \mathbb{R}^{+}$ are weight factors, $H(x)$ is defined below Eq.~\eqref{eq:A2}, and
{\small
\begin{align}
  \Delta M^2 &:=   \max_{x \in PGBs}\left \vert\frac{ \left(M^2_{PGB,(1)}\right)_x - \left(M^2_{PGB,iter}\right)_x }{\left(M^2_{PGB,(1)}\right)_x}\right \vert ,\\[8pt]
  \delta M^{2}_{iter} &:= \max_{x \in PGBs}\left \vert\frac{ \left(M^2_{PGB,(i_{max})}\right)_x - \left(M^2_{PGB,(i_{max}-1)}\right)_x }{\left(M^2_{PGB,(i_{max})}\right)_x}\right \vert ,
\end{align}
}
with $i_{max}=30$.

\end{enumerate}
%

%======================================================
\subsection{Implementation of the numerical analysis: Step-by-step  procedure\label{app:procedure}}
%======================================================

Eventually, the criteria discussed above are used to find viable points, i.e.,~those which pass all the consistency checks and hence have zero penalization. Conceptually, the numerical procedure for assessing a single parameter point consists of two steps:
\begin{enumerate}
    \item[(a)] 
    First, a point in the parameter space is selected. The VEVs and unified gauge coupling $g$ are adjusted so that the resulting one-loop scalar spectrum is consistent with gauge unification requirements.
    \item[(b)]
    Second, the penalization of the adjusted point is calculated. It is used for determining its viability.
\end{enumerate}

The procedure of choosing new candidate points in the parameter space for assessment, i.e.,~a scan of the parameter space, follows the differential evolution algorithm~\cite{diffevolalg} as the minimization algorithm of choice. We use a stochastic implementation with a mutation factor $F$ randomly sampled from the interval $[0.5,2]$. The scanning process first discovers viable points by minimizing the penalization function until it reaches zero, and then it explores the zero penalty region.
In this way, datasets of viable points from Table~\ref{tab:datasets} are produced.

Returning to part~(a), the detailed steps are the following:
\begin{enumerate}
    \item The input parameters for a point consist of 
        \begin{align}
        \vec{\lambda},\quad g,\quad \omega_{BL},\quad \omega_{R},\quad \sigma,    
        \end{align}
        where $\vec{\lambda}$ is defined in Eq.~\eqref{eq:vector-of-parameters} and $g$ is the unified gauge coupling. These parameters are deemed to be the running values at the yet to be determined ``optimal'' renormalization scale $\mu_{R}$. For later convenience, we label the $\mathrm{SO}(10)$-breaking VEV as $\omega_{max}$ ($\omega_R$ or $\omega_{BL}$ in the regime $\omega_{BL} \to 0$ or $\omega_R \to 0$, respectively), and the subdominant induced VEV as $\omega_{min}$ ($\omega_{BL}$ or $\omega_{R}$ in the $\omega_{BL} \to 0$ or $\omega_R \to 0$ limit, respectively).
    %%%%%%%%%%%%%%%%%%%%%%%%%
    \item We initially fix $g=0.5$ and $\omega_{max}=10^{16}\GeV$. The $\vec{\lambda}$, $\sigma$, and $\omega_{min}$ inputs are provided by the selection of the parameter point, for which we demand that they conform to Eq.~\eqref{eq:VEVConstraint}.
	%%%%%%%%%%%%%%%%%%%%%%%%%
	\item With all the initial input parameters now set, the tree-level gauge and scalar spectra are calculated. 
	%%%%%%%%%%%%%%%%%%%%%%%%%
	\item The ``optimal'' renormalization-scale square $\mu^{2}_R$ is determined as the arithmetic mean (over real degrees of freedom) of the \emph{heavy} tree-level scalar-mass squares.
	%%%%%%%%%%%%%%%%%%%%%%%%%
	\item With $\mu_R$ in hand, the one-loop scalar-mass spectrum for \emph{heavy} and PGB fields is computed using the method described in Appendix~\ref{App:one-loop masses}, replacing their tree-level values. The corrections to gauge fields and the intermediate-scale scalars are not calculated. We refer to this collection of our best available results for all gauge and scalar masses as the \emph{initial spectrum}.
	%%%%%%%%%%%%%%%%%%%%%%%%%
	\item The \emph{initial spectrum} is used for thresholds in the one-loop gauge evolution analysis.\footnote{For tachyonic masses, we take their absolute values. Such points are later penalized due to tachyonicity anyway. Note that the tachyonic property is generally preserved even after adjustment described in this step.} The complete SM-gauge-coupling unification with the EW-scale boundary conditions~\cite{Zyla:2020zbs}
	\begin{align}
	\label{eq:Cond1}
	\alpha_{EM}^{-1}\,(M_Z) &=
	127.952 \pm 0.009,\\
	\label{eq:Cond2}
	\sin^2\theta_W\,(M_Z) &=
	0.23121 \pm 0.00004,\\
	\label{eq:Cond3}
	\alpha_3\,(M_Z) &= 0.1179 \pm 0.0010
	\end{align}
	is achieved as described in Appendix~\ref{app:unification}.
	Technically, this is accomplished by three simultaneous actions: (i) rescaling of all dimensionful quantities $\lbrace \omega_{BL},\omega_R,\sigma,\mu_R \rbrace$ by a common factor, (ii) further adjustment\footnote{The VEVs $\sigma$ and $\omega_{min}$ are adjusted in such a way that the ratio $\chi$ of Eq.~\eqref{eq:chi-definition} stays constant.} of $\sigma$ and $\omega_{min}$, and (iii) optimizing the value of $g$. This procedure provides new values for $g$, $\mu_{R}$, and the three VEVs --- defining the \emph{adjusted parameter point}.
	%%%%%%%%%%%%%%%%%%%%%%%%%
	\item For the \emph{adjusted parameter point}, the tree-level spectrum for gauge and scalar fields is computed. Furthermore, the \emph{heavy} and PGB masses are improved with the one-loop correction. This collection of the best available mass values is referred to as the \emph{updated spectrum}. 
	%%%%%%%%%%%%%%%%%%%%%%%%%
\end{enumerate}
In part~(b), the remaining consistency criteria of Appendices~\ref{app:tachyonicity} and~\ref{app:perturbativity} (tachyonicity and perturbativity) are imposed on the \emph{adjusted parameter point} and the relevant penalizations are calculated. The detailed steps for part~(b) are as follows:
\begin{enumerate}
		\item The \textbf{stability of the vacuum position} is investigated --- the penalization $p_3^{(1)}$ is computed. For a viable point, the one-loop corrections should not shift the position of the tree-level vacuum by more than $100\%$.
		\item All the dimensionless parameters are \textbf{RG run} half-an-order of magnitude upward using one-loop beta functions (cf.~Appendix~\ref{App:BetaFunctions}). Note that this running distance in scale turns out to be consistent with the spread of $\mu_R$ values for viable points (see Sec.~\ref{Sec:Results}), implying that all viable points could be adjusted to a common scale for comparison if desired.
		\item RG-run couplings are used to \textbf{recheck the stability of the vacuum position} --- the penalization $p_3^{(2)}$ is acquired. All viable points satisfy $t_{+} > 0.5$, and their shift of the tree-level vacuum position due to the one-loop corrections is still, even after running, not more than $100\%$. 
		\item The \textbf{tachyonicity} of the \emph{updated spectrum} is inspected --- the penalization $p_1$ is obtained.
		\item \textbf{Global mass perturbativity} is examined --- the penalization $p_2$ is computed. For viability, the maximal one-loop scalar-mass correction is required to be smaller than the average of \emph{heavy} tree-level scalar masses, i.e.,~$\overline{\Delta} \leq 1$.
		\item The \textbf{iterative pseudo-Goldstone masses} are investigated --- the penalization $p_4$ is determined. For a viable point, the relative difference between the initial and final values in the iterative procedure for the PGB one-loop mass is constrained to less than $10\%$, i.e.,~$\Delta M^{2} < 0.1$.
		\item (Optional) \textbf{Stricter RG-perturbativity} criterion is imposed --- the penalization $p_5$ is added. All viable points satisfy $\bar{t}> \bar{t}_{thr}$, where $\bar{t}_{thr}$ is an acceptance threshold of a given search (see Table~\ref{tab:datasets} in Sec.~\ref{sec:numerical-scan}). 
		\item The overall penalization of a point is the following: 
		\begin{align}\label{eq:totalpenalty}
		p & = p_1 + p_2 + p_3^{(1)} + p_3^{(2)} + p_4 +p_5.   
		\end{align}
\end{enumerate}

The relative significance of a particular criterion in the minimization algorithm, such as tachyonicity or perturbativity, can be tuned by changing the value of the $A_i$ weight factors present in penalizations $p_i$. We chose the configuration $A_i = 1$, which leads to $p_i \in [0,1]$ for all $i=1,2,3,4,5$. If the point has zero penalization, it satisfies all the imposed criteria and belongs to the set of viable points obtained in the parameter-space scan. Note that the shape of the viable part of the parameter space is independent of the values $A_i$.

%%%%%%%%%%%%%%%%%%%%%%%%%%%%%%%%%%%%%%%%%%%%%%%%%%%%%%%
\section{Beta functions for scalar parameters\label{App:BetaFunctions}}
%%%%%%%%%%%%%%%%%%%%%%%%%%%%%%%%%%%%%%%%%%%%%%%%%%%%%%%

The one-loop beta functions for scalar-potential parameters can be extracted from the field-dependent effective potential~\eqref{eq:EffectivePotential} following the procedure outlined in~\cite{Coleman:1973jx}.

%======================================================
\subsection{General principles \label{app:general beta}}
%======================================================

Suppose that at tree level a scalar coupling $\lambda$ can be written as a derivative of the  scalar potential via
\begin{align}
\label{eq:Lambda}
\lambda & = \frac{\partial^4 V_0(\Phi)}{\prod_{i=1}^4 \partial \Phi_i} 
\end{align}
for a suitable choice of scalar fields $\Phi_i.$ The complete one-loop effective potential~\eqref{eq:EffectivePotential} has to satisfy the Callan–Symanzik renormalization group equation
\begin{align}
\label{eq:RGEVeff}
\left(\frac{\partial}{\partial \log \mu_R} + \beta_\lambda \frac{\partial }{\partial \lambda} + \sum_j \gamma_{\Phi_j} \Phi_j \frac{\partial}{\partial \Phi_j}\right) V(\Phi) & = 0 ,
\end{align}
where $\mu_R$ is the running renormalization scale, and the $j$ index runs over all scalar fields. Consequently,
{\small
\begin{align}
\label{eq:RGEVeff2}
\left(\frac{\partial}{\partial \log \mu_R} + \beta_\lambda \frac{\partial }{\partial \lambda} + \sum_j  \gamma_{\Phi_j}  \frac{\partial}{\partial \Phi_j}  + \sum_{i=1}^4 \gamma_{\Phi_i}\right) \frac{\partial^4 V(\Phi,\mu_R)}{\prod_{i=1}^4 \partial \Phi_i} & = 0 .
\end{align}
}
Equations~\eqref{eq:Lambda}, \eqref{eq:RGEVeff2}, and~\eqref{eq:EffectivePotential} then imply
\begin{align}
    \beta_\lambda & = -\frac{\partial^4}{\prod_{i=1}^4 \partial \Phi_i} \frac{\partial V_1(\Phi,\mu_R)}{\partial \log \mu_R} - \sum_{i=1}^4 \gamma_{\Phi_i} \frac{\partial^4 V_0(\Phi)}{\prod_{i=1}^4 \partial \Phi_i} %\nonumber \\
    %& 
    = \frac{1}{32\pi^2} \frac{\partial^4}{\prod_{i=1}^4 \partial \Phi_i} (\text{Tr}\left[ \mathbf{M}^4_S(\Phi)\right] + 3\text{Tr}\left[ \mathbf{M}^4_G(\Phi)\right]) + \beta_{\lambda,FS} ,
    \label{eq:betaraw}
\end{align}
where $\mathbf{M}^4_S(\Phi)$ and $\mathbf{M}^4_G(\Phi)$ are defined in Eqs.~\eqref{eq:field-dependent-MS2} and~\eqref{eq:field-dependent-MG2}, $\beta_\lambda$ denotes the desired one-loop quartic scalar beta function satisfying
\begin{align}
\beta_{\lambda} & = \frac{d \lambda}{d \log \mu_R}, 
\end{align}
and
\begin{align}
\label{eq:AnomalBetaDef}
    \beta_{\lambda,FS} & = -\lambda\sum_{i=1}^4 \gamma_{\Phi_i}
\end{align}
is the field-strength-dependent part.

%======================================================
\subsection{Field-strength-dependent part\label{app:field-dependent}}
%======================================================

The $\beta_{\lambda,FS}$ part of Eq.~\eqref{eq:betaraw} is closely connected to the 
$\Phi$-field anomalous dimension defined as 
\begin{align}
\label{eq:anomalDef}
\gamma_\Phi & = \frac{1}{2}\frac{1}{Z_\Phi} \frac{\partial Z_\Phi }{\partial \log \mu_R},
\end{align}
where
\begin{align}
\label{eq:FieldStr}
Z_\Phi & = 1+\left.\frac{\partial \Sigma^{\overline{MS}}_\Phi(p^2)}{\partial p^2}\right\vert_{p^2 = m^2_\Phi} 
\end{align}
is the field-strength-renormalization factor of $\Phi$ in the $\overline{\text{MS}}$ renormalization scheme, and $\Sigma^{\overline{MS}}_\Phi(p^2)$ is the corresponding self-energy. The momentum-squared-proportional part of $\Sigma^{\overline{MS}}_\Phi(p^2)$ originates from the diagram
\begin{center}
	\includegraphics[width=0.3\textwidth]{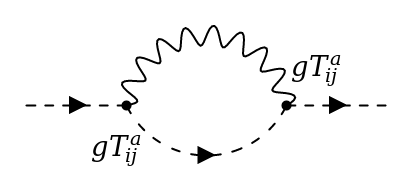}
\end{center}
which yields
\begin{align}
\Sigma^{\overline{MS}}_\Phi(p^2) & = C_2(R)\cdot p^2 \cdot\frac{3g^2}{16\pi^2} \log \mu_R^2+\ldots ,
\end{align}
where $C_2(R)$ is the quadratic Casimir of the irreducible representation $R$ that $\Phi$ belongs to. Hence,
\begin{align}
\label{eq:AnomalFinal}
\gamma_\Phi & = C_2(R) \frac{3g^2}{16\pi^2}.
\end{align}
Combining Eqs.~\eqref{eq:AnomalBetaDef} and~\eqref{eq:AnomalFinal}, the $\beta_{\lambda,FS}$ term in Eq.~\eqref{eq:betaraw} becomes
\begin{align}
\label{eq:AnomalBeta}
\beta_{\lambda,FS} & = -\lambda \frac{3 g^2}{16 \pi^2} \sum_{i=1}^4C_2(R_{\Phi_i}).
\end{align}
%

%======================================================
\subsection{Resulting expressions\label{app:final beta}}
%======================================================

With all this in hand, the one-loop beta functions for the dimensionless scalar-potential couplings are readily obtained: 
\begin{widetext}
\begingroup
\allowdisplaybreaks 
\begin{align}
    \begin{split}
    16 \pi ^2\beta_{\beta_4^\prime} &= \phantom{+} 16 \alpha \beta_4^\prime+16 a_0 \beta_4^\prime+2a_2 \beta_4-4 a_2\beta_4^\prime-\beta_4^2-28 \beta_4 \beta_4^\prime+2 \beta_4 \lambda_2+6 \beta_4 \lambda_4 + \\
    &\quad +80 \beta_4 \lambda_4^\prime-124 \beta_4^{\prime 2}+4\beta_4^\prime \lambda_0-12 \beta_4^\prime\lambda_2+20 \beta_4^\prime \lambda_4-144 \beta_4^\prime \lambda_4^\prime+16 \left|\gamma_2\right|^2 - \\
    &\quad  -3g^4-123\beta_4^\prime g^2,\\
    \end{split}\EQSPACE
%%%%%%%%%%%%
    \begin{split}
    16\pi^2 \beta_\alpha &=\phantom{+} 8 \alpha ^2+508 \alpha  \lambda_0+1220 \alpha  \lambda_2+1340 \alpha  \lambda_4+2480 \alpha  \lambda_4^{\prime}+376 \alpha  a_0+80 a_0 \beta_4 + \\
    &\quad +160 a_0
    \beta_4^{\prime}+76 \alpha a_2+16a_2 \beta_4+32a_2 \beta_4^{\prime}+4 \beta_4^2+16 \beta_4 \beta_4^{\prime}+112 \beta_4 \lambda_0 + \\
    &\quad+272 \beta_4 \lambda_2+288 \beta_4 \lambda_4+512 \beta_4 \lambda_4^{\prime}+144 \beta_4^{\prime 2}+224 \beta_4^{\prime} \lambda_0+544
    \beta_4^{\prime} \lambda_2+576 \beta_4^{\prime} \lambda_4 + \\
    &\quad+1024 \beta_4^{\prime} \lambda_4^{\prime}+64 |\gamma_2|^2+12 g^4-123 \alpha  g^2,\\
    \end{split}\EQSPACE
%%%%%%%%%%%%
    \begin{split}
    16\pi^2\beta_{\lambda_0} &= \phantom{+}90 \alpha ^2+40 \alpha  \beta_4+80 \alpha  \beta_4^{\prime}+10 \beta_4^2+80 \beta_4^{\prime 2}+520 \lambda_0^2+2440\lambda_0 \lambda_2+2680 \lambda_0 \lambda_4 + \\
    &\quad+4960 \lambda_0 \lambda_4^{\prime}+3460 \lambda_2^2+7880 \lambda_2 \lambda_4+12320 \lambda_2
    \lambda_4^{\prime}+4660 \lambda_4^2+13280 \lambda_4 \lambda_4^{\prime} + \\
    &\quad+16960 \lambda_4^{\prime 2}+\frac{135 g^4}{2}-150 g^2 \lambda_0,\\
    \end{split}\EQSPACE
%%%%%%%%%%%%
    \begin{split}
    \label{eq:betaa0}
    16 \pi^2\beta_{a_0} &= \phantom{+}126 \alpha^2+56 \alpha  \beta_4+112 \alpha  \beta_4^{\prime}+424 a_0^2+152 a_0 a_2+12 a_2^2+\frac{33 \beta_4^2}{2}+26 \beta_4 \beta_4^\prime + \\
    &\quad+106 \beta_4^{\prime 2}-56 |\gamma_2|^2+\frac{9 g^4}{2}-96 a_0 g^2,
    \\
    \end{split}\EQSPACE
%%%%%%%%%%%%
    \begin{split}
    \label{eq:betaa2}
    16 \pi^2\beta_{a_2} &= \phantom{+} 96 a_0 a_2+76 a_2^2-5 \beta_4^2+60 \beta_4\beta_4^\prime-100 \beta_4^{\prime 2}+560 |\gamma_2|^2 + 3 g^4-96a_2g^2,
    \\
    \end{split}\EQSPACE
%%%%%%%%%%%%
    \begin{split}
    16 \pi^2\beta_{\lambda_2}&=-4 \beta_4^2-32 \beta_4^{\prime 2}+24 \lambda_0 \lambda_2-180 \lambda_2^2-584 \lambda_2 \lambda_4-160 \lambda_2 \lambda_4^{\prime}-656 \lambda_4^2 - \\
    &\quad-800 \lambda_4 \lambda_4^{\prime}-2560 \lambda_4^{\prime 2} -1264 |\eta_2|^2-24 g^4-150 g^2 \lambda_2,
    \\
    \end{split}\EQSPACE
%%%%%%%%%%%%
    \begin{split}
    16 \pi^2\beta_{\lambda_4}&=\phantom{+}2 \beta_4^2+16 \beta_4^{\prime 2}+24 \lambda_0 \lambda_4+16 \lambda_2^2+112 \lambda_2 \lambda_4+128 \lambda_2 \lambda_4^{\prime}+268 \lambda_4^2 + \\
    &\quad+640 \lambda_4 \lambda_4^{\prime}+1408 \lambda_4^{\prime 2}+1328 |\eta_2|^2+12 g^4-150 g^2 \lambda_4,\\
    \end{split}\EQSPACE
%%%%%%%%%%%%
    \begin{split}
    16 \pi^2 \beta_{\lambda_4^\prime}&=\phantom{+}4 \beta_4 \beta_4^{\prime}-4 \beta_4^{\prime 2}+24 \lambda_0 \lambda_4^{\prime}-4 \lambda_2^2-8 \lambda_2 \lambda_4-16 \lambda_2 \lambda_4^{\prime}+4 \lambda_4^2+112 \lambda_4 \lambda_4^{\prime} - \\
    &\quad-240 \lambda_4^{\prime 2}+32 |\eta_2|^2-3 g^4-150 g^2 \lambda_4^{\prime},\\
    \end{split}\EQSPACE
%%%%%%%%%%%%
    \begin{split}
    16\pi^2 \beta_{\beta_4} &=\phantom{+}16 \alpha  \beta_4+16 a_0 \beta_4+16a_2 \beta_4^{\prime}+48 \beta_4^2+80 \beta_4 \beta_4^{\prime}+4 \beta_4\lambda_0-8 \beta_4 \lambda_2 + \\
    &\quad+32 \beta_4\lambda_4+16 \beta_4 \lambda_4^{\prime}+16 \beta_4^{\prime 2}+16 \beta_4^{\prime}\lambda_2+48 \beta_4^{\prime} \lambda_4+640 \beta_4^{\prime} \lambda_4^{\prime}+64 |\gamma_2|^2 + \\
    &\quad+12 g^4-123 \beta_4 g^2,\\
    \end{split}\EQSPACE
%%%%%%%%%%%%
    \begin{split}
    16\pi^2 \beta_{\gamma_2} &= \phantom{+} 8 \alpha  \gamma_2+14 \beta_4 \gamma_2+28 \beta_4^{\prime} \gamma_2+440 \eta _2 \gamma_2^*-123 \gamma_2 g^2,\label{eq:beta-gamma_2}\\
    \end{split}\EQSPACE
%%%%%%%%%%%%
    \begin{split}
    16 \pi^2 \beta_{\eta_2} &=\phantom{+}\frac{32 \gamma_2^2}{3}+24 \eta_2  \lambda_0+160 \eta_2  \lambda_2+600 \eta_2  \lambda_4+640 \eta_2  \lambda_4^\prime -150 \eta_2  g^2 \label{eq:beta-eta_2}.\\
    \end{split}
\end{align}
\endgroup
\end{widetext}
It is perhaps worth noting that several parts of these results, in particular, those corresponding to Green's functions with four adjoint scalar outer legs [such as the $a_0^2$-, $a_0a_2$-, and $a_2^2$-proportional terms in Eqs.~\eqref{eq:betaa0} and~\eqref{eq:betaa2}], have been independently cross-checked by direct diagrammatic methods.

%%%%%%%%%%%%%%%%%%%%%%%%%%%%%%%%%%%%%%%%%%%%%%%%%%%%%%%
\section{Masses in limits\label{App:Masses}}
%%%%%%%%%%%%%%%%%%%%%%%%%%%%%%%%%%%%%%%%%%%%%%%%%%%%%%%

In this appendix the analytical expressions for the tree-level spectrum (Appendices~\ref{app:gauge masses} and~\ref{app:scalar masses}) and the one-loop masses of pseudo-Goldstone bosons (Appendix~\ref{app:PGB masses}) are collected for both symmetry-breaking patterns of interest ($|\omega_R|\gg|\sigma|\gg|\omega_{BL}|$ and $|\omega_{BL}|\gg|\sigma|\gg|\omega_R|$). 
The presented formulas are merely approximate, since they contain only the leading tree-level contributions, while the radiative corrections are computed exclusively in the analytical $a_2\to 0$, $\gamma_2\to 0$, $\sigma\to 0$ limit. It should be stressed that in the actual scans all the masses have been computed numerically at the one-loop level for an arbitrary choice of couplings and VEVs, i.e.,~not only in the aforementioned limit. Nonetheless, the expressions introduced below can be quite useful in providing some qualitative insight into the structure of non-tachyonicity regions in Figs.~\ref{fig:AnalyticalBL} and~\ref{fig:AnalyticalR}.

%======================================================
\subsection{Representation decompositions\label{app:decompositions}}
%======================================================

The decompositions of all the representations that constitute the scalar and gauge sectors of the model can be found in Table~\ref{tab:decompositions}. They correspond to the two-stage breaking scenarios in which either the value $\omega_R$ or $\omega_{BL}$ spontaneously breaks $\mathrm{SO}(10)$ GUT symmetry to the \BLzero or \LR intermediate symmetry, respectively, while $|\sigma|\ll M_{\text{GUT}}$ breaks it further to the Standard Model gauge group.
\begin{table*}[htb]
	\caption{Decomposition of the (a) $45$-dimensional adjoint and (b) $126$-dimensional self-dual representations of $\mathrm{SO}(10)$ into multiplets of \BLzero (the $\omega_{BL}\to 0$, $\sigma\to 0$ limit) in the left columns and \LR (the $\omega_{R}\to 0$, $\sigma\to 0$ limit) in the right columns and their subsequent breaking into Standard Model representations after engaging non-zero $\sigma$; cf.~Table~\ref{tab:BreakingSchemes}.
	The color scheme indicates whether a representation is \emph{real} (\real{\textbf{blue}}), \emph{complex} (\complex{\textbf{black}}) or a \emph{complex conjugate} (\conjugate{\textbf{red}}) of some other multiplet within the same $\mathrm{SO}(10)$ representation. Since the $\mathbf{126}$ is a complex representation, all the multiplets that belong to it are also complex, while the $\mathbf{45}$ is real and thus consists of real multiplets and complex-conjugate pairs. The multiple copies of the same SM multiplets in the $\mathbf{126}$, e.g.,~the $(1,2,+\tfrac{1}{2})$ and $(1,2,-\tfrac{1}{2})$ weak doublets, therefore represent independent degrees of freedom and not just the complex-conjugate counterparts of each other. 
	\label{tab:decompositions}}
    \begin{subtable}[t]{.49\linewidth}
	\caption{$\mathbf{45}$ of $\mathrm{SO}(10)$\label{tab:decomposition-45}}
	\begin{tabular}[t]{l@{$\quad$}l}
		\hline
		$4_C 2_L 1_R$ & $\begin{array}{l} 3_c 2_L 1_Y \end{array}$\\
		\hline
		\hline
		$\real{(1,3,0)}$ & $\begin{array}{l} \real{(1,3,0)} \end{array}$\\
		\hline
		$\real{(15,1,0)}$ & $\begin{array}{l} \real{(8,1,0)} \\ \complex{(\xbar{3},1,-\tfrac{2}{3})} \\ \conjugate{(3,1,+\tfrac{2}{3})} \\ \real{(1,1,0)} \end{array}$\\
		\hline
		$\complex{(6,2,-\tfrac{1}{2})}$ & $\begin{array}{l} \complex{(3,2,-\tfrac{5}{6})} \\ \conjugate{(\xbar{3},2,-\tfrac{1}{6})} \end{array}$ \\
		\hline
		$\conjugate{(6,2,+\tfrac{1}{2})}$ & $\begin{array}{l} \conjugate{(\xbar{3},2,+\tfrac{5}{6})} \\ \complex{(3,2,+\tfrac{1}{6})} \end{array}$ \\
		\hline
		$\complex{(1,1,+1)}$ & $\begin{array}{l} \complex{(1,1,+1)} \end{array}$ \\
		\hline
		$\conjugate{(1,1,-1)}$ & $\begin{array}{l} \conjugate{(1,1,-1)} \end{array}$ \\
		\hline
		$\real{(1,1,0)}$ & $\begin{array}{l} \real{(1,1,0)} \end{array}$ \\
		\hline
	\end{tabular}
\hspace{.01\linewidth}
	\begin{tabular}[t]{l@{$\quad$}l}
		\hline
		$3_c 2_L 2_R 1_{BL}$ & $\begin{array}{l} 3_c 2_L 1_Y \end{array}$\\
		\hline
		\hline
		$\real{(1,3,1,0)}$ & $\begin{array}{l} \real{(1,3,0)} \end{array}$\\
		\hline
		$\real{(8,1,1,0)}$ & $\begin{array}{l} \real{(8,1,0)} \end{array}$\\
		\hline
		$\complex{(3,2,2,-\tfrac{2}{3})}$ & $\begin{array}{l} \complex{(3,2,-\tfrac{5}{6})} \\ \complex{(3,2,+\tfrac{1}{6})} \end{array}$ \\
		\hline
		$\conjugate{(\xbar{3},2,2,+\tfrac{2}{3})}$ & $\begin{array}{l} \conjugate{(\xbar{3},2,+\tfrac{5}{6})} \\ \conjugate{(\xbar{3},2,-\tfrac{1}{6})} \end{array}$ \\
		\hline
		$\real{(1,1,3,0)}$ & $\begin{array}{l} \complex{(1,1,+1)} \\ \conjugate{(1,1,-1)} \\ \real{(1,1,0)} \end{array}$\\
		\hline
		$\complex{(\xbar{3},1,1,-\tfrac{4}{3})}$ & $\begin{array}{l} \complex{(\xbar{3},1,-\tfrac{2}{3})} \end{array}$ \\
		\hline
		$\conjugate{(3,1,1,+\tfrac{4}{3})}$ & $\begin{array}{l} \conjugate{(3,1,+\tfrac{2}{3})} \end{array}$ \\
		\hline
		$\real{(1,1,1,0)}$ & $\begin{array}{l} \real{(1,1,0)} \end{array}$ \\
		\hline
	\end{tabular}
	\end{subtable}
\hfill
	\begin{subtable}[t]{.49\linewidth}
	\caption{$\mathbf{126}$ of $\mathrm{SO}(10)$\label{tab:decomposition-126}}
	\begin{tabular}[t]{l@{$\quad$}l}
		\hline
		$4_C 2_L 1_R$ & $\begin{array}{l} 3_c 2_L 1_Y \end{array}$\\
		\hline
		\hline
		$(\xbar{10},1,+1)$ & $\begin{array}{l} (1,1,+2) \\ (\xbar{3},1,+\tfrac{4}{3}) \\ (\xbar{6},1,+\tfrac{2}{3}) \end{array}$\\
		\hline
		$(10,3,0)$ & $\begin{array}{l} (1,3,-1) \\ (3,3,-\tfrac{1}{3}) \\ (6,3,+\tfrac{1}{3}) \end{array}$\\
		\hline
		$(\xbar{10},1,-1)$ & $\begin{array}{l} (\xbar{6},1,-\tfrac{4}{3}) \\ (\xbar{3},1,-\tfrac{2}{3}) \\ (1,1,0) \end{array}$ \\
		\hline
		$(\xbar{10},1,0)$ & $\begin{array}{l} (\xbar{6},1,-\tfrac{1}{3}) \\ (\xbar{3},1,+\tfrac{1}{3}) \\ (1,1,+1) \end{array}$ \\
		\hline
		$(15,2,+\tfrac{1}{2})$ & $\begin{array}{l} (1,2,+\tfrac{1}{2}) \\ (3,2,+\tfrac{7}{6}) \\ (8,2,+\tfrac{1}{2}) \\ (\xbar{3},2,-\tfrac{1}{6}) \end{array}$ \\
		\hline
		$(15,2,-\tfrac{1}{2})$ & $\begin{array}{l} (1,2,-\tfrac{1}{2}) \\ (\xbar{3},2,-\tfrac{7}{6}) \\ (8,2,-\tfrac{1}{2}) \\ (3,2,+\tfrac{1}{6}) \end{array}$ \\
		\hline
		$(6,1,0)$ & $\begin{array}{l} (\xbar{3},1,+\tfrac{1}{3}) \\ (3,1,-\tfrac{1}{3}) \end{array}$ \\
		\hline
	\end{tabular}
\hspace{.01\linewidth}
	\begin{tabular}[t]{l@{$\quad$}l}
		\hline
		$3_c 2_L 2_R 1_{BL}$ & $\begin{array}{l} 3_c 2_L 1_Y \end{array}$\\
		\hline
		\hline
		$(1,1,3,+2)$ & $\begin{array}{l} (1,1,+2) \\ (1,1,+1) \\ (1,1,0) \end{array}$\\
		\hline
		$(1,3,1,-2)$ & $\begin{array}{l} (1,3,-1) \end{array}$\\
		\hline
		$(\xbar{3},1,3,+\tfrac{2}{3})$ & $\begin{array}{l} (\xbar{3},1,+\tfrac{4}{3}) \\ (\xbar{3},1,+\tfrac{1}{3}) \\ (\xbar{3},1,-\tfrac{2}{3}) \end{array}$ \\
		\hline
		$(3,3,1,-\tfrac{2}{3})$ & $\begin{array}{l} (3,3,-\tfrac{1}{3}) \end{array}$ \\
		\hline
		$(6,3,1,+\tfrac{2}{3})$ & $\begin{array}{l} (6,3,+\tfrac{1}{3}) \end{array}$ \\
		\hline
		$(\xbar{6},1,3,-\tfrac{2}{3})$ & $\begin{array}{l} (\xbar{6},1,-\tfrac{4}{3}) \\ (\xbar{6},1,-\tfrac{1}{3}) \\ (\xbar{6},1,+\tfrac{2}{3}) \end{array}$ \\
		\hline
		$(1,2,2,0)$ & $\begin{array}{l} (1,2,+\tfrac{1}{2}) \\ (1,2,-\tfrac{1}{2}) \end{array}$ \\
		\hline
		$(3,2,2,+\tfrac{4}{3})$ & $\begin{array}{l} (3,2,+\tfrac{7}{6}) \\ (3,2,+\tfrac{1}{6}) \end{array}$ \\
		\hline
		$(\xbar{3},2,2,-\tfrac{4}{3})$ & $\begin{array}{l} (\xbar{3},2,-\tfrac{7}{6}) \\ (\xbar{3},2,-\tfrac{1}{6}) \end{array}$ \\
		\hline
		$(8,2,2,0)$ & $\begin{array}{l} (8,2,+\tfrac{1}{2}) \\ (8,2,-\tfrac{1}{2}) \end{array}$ \\
		\hline
		$(\xbar{3},1,1,+\tfrac{2}{3})$ & $\begin{array}{l} (\xbar{3},1,+\tfrac{1}{3}) \end{array}$ \\
		\hline
		$(3,1,1,-\tfrac{2}{3})$ & $\begin{array}{l} (3,1,-\tfrac{1}{3}) \end{array}$ \\
		\hline
	\end{tabular}
	\end{subtable}
\end{table*}
%

%======================================================
\subsection{Gauge boson masses\label{app:gauge masses}}
%======================================================

Here the tree-level masses of gauge bosons are gathered. Their leading contributions for both scenarios of interest are also shown explicitly. In addition to $33$ non-zero states listed in Table~\ref{tab:tree-level-gauge-boson-masses}, there exist $12$ massless would-be Goldstone modes corresponding to the $12$ generators of the Standard Model gauge group. 
\begin{table*}[htb]
\begin{center}
\caption{Tree-level spectrum of massive gauge bosons produced in $\mathrm{SO}(10)\to\;$\SM breaking is listed for the most general case and for both perturbative limits of interest. In the last column, the changes of the corresponding one-loop $\beta$ coefficients $\Delta a_i$ for gauge-coupling running are gathered; cf.~Eq.~\eqref{eq:ai}. 
Note that massive gauge contributions implicitly include also those of their longitudinal components treated as adding the WGB scalar counterparts at the same scale.
\label{tab:tree-level-gauge-boson-masses}}
\begin{tabular}{l@{$\quad$}l@{$\quad$}l@{$\quad$}l@{$\quad$}r}
\hline
Mass&General case&$\omega_{BL}\to 0$&$\omega_{R}\to 0$&$(\Delta a_3,\Delta a_2,\Delta a_1)$\\
\hline
\\[-10pt]
$M^2_G(3,2,-\tfrac{5}{6})$ & $\frac{1}{2} g^2 \left(\omega_R - \omega_{BL}\right)^2 $ &$\tfrac{1}{2} g^2 \omega_{R}^2$&$\tfrac{1}{2} g^2 \omega_{BL}^2$&$(-7,-\tfrac{21}{2},-\tfrac{35}{2})$\\[6pt]
$M^2_G(3,2,+\tfrac{1}{6})$ & $ \tfrac{1}{2} g^2(\omega_R + \omega_{BL})^2 + 2 g^2 \vert\sigma \vert^2 $&$\tfrac{1}{2} g^2 \omega_{R}^2$&$\tfrac{1}{2} g^2 \omega_{BL}^2$&$(-7,-\tfrac{21}{2},-\tfrac{7}{10})$\\[6pt]
$M^2_G(3,1,+\tfrac{2}{3})$ & $ 2g^2 \omega_{BL}^2 + 2g^2\vert \sigma \vert^2$&
$2 g^2 \vert \sigma \vert^2$&$2 g^2 \omega_{BL}^2$&$(-\tfrac{7}{2},0,-\tfrac{28}{5})$\\[6pt]
$M^2_G(1,1,+1)$ & $ 2g^2\omega_{R}^2 + 2g^2\vert \sigma \vert^2$&$2 g^2 \omega_{R}^2$&$2 g^2 \vert \sigma \vert^2$&$(0,0,-\tfrac{21}{5})$\\[6pt]
$M^2_G(1,1,0)$ &  $ 10 g^2 \vert \sigma \vert^2$&$ 10 g^2 \vert \sigma \vert^2$&$ 10 g^2 \vert \sigma \vert^2$&$(0,0,0)$\\[4pt]
\hline
\end{tabular}
\end{center}
\end{table*}
%

%======================================================
\subsection{Scalar masses\label{app:scalar masses}}
%======================================================

The tree-level scalar masses are presented in the $\omega_{BL}\to 0$ and $\omega_R\to 0$ limits, which should be properly understood as
\hbox{$|\omega_R| \gg |\sigma| \geq \sqrt{|\omega_R||\omega_{BL}|} \gg |\omega_{BL}|$} and 
\hbox{$|\omega_{BL}| \gg |\sigma| \geq \sqrt{|\omega_{BL}||\omega_R|} \gg |\omega_R|$}, respectively, satisfying the perturbativity condition~\eqref{eq:VEVConstraint}.
For the corresponding mass matrices in the general breaking scenario, the interested reader is referred to~\cite{Graf:2016znk}.

The mass squares are, in general, linearly proportional to the $\tau$ parameter, whose value is determined by
the vacuum stationarity condition~\eqref{eq:tau}
\begin{align}
    \tau & = 2 \beta_4' (3 \omega_{BL} + 2 \omega_{R}) + a_2 \chi (\omega_{BL} + \omega_{R}) 
\end{align}
implying an implicit dependence of the spectrum on the $\chi$ parameter defined in Eq.~\eqref{eq:chi-definition}. Its contribution should not be neglected despite the $|\chi| \leq 1$ constraint from perturbativity (and thus, $|\tau|\lesssim M_{\text{GUT}}$) 
and a small $a_2$ coefficient in front of it (with values $|a_2|\lesssim 0.05$ produced in scans; see Fig.~\ref{fig:Sylvester}). 
It turns out that the value of the $\beta_4'$ coupling is also rather small for valid points
(see Fig.~\ref{fig:ScalarParameters}), which makes both terms in $\tau$ 
comparable. Furthermore, $\tau$ is always found to be negative, but its relative size (compared to the GUT scale), and thus its impact on the masses, 
is typically larger in the $\omega_{BL}\to 0$ case due to larger $|\beta_4'|$.

\begin{figure}
    \centering
    \includegraphics[width=0.5\columnwidth]{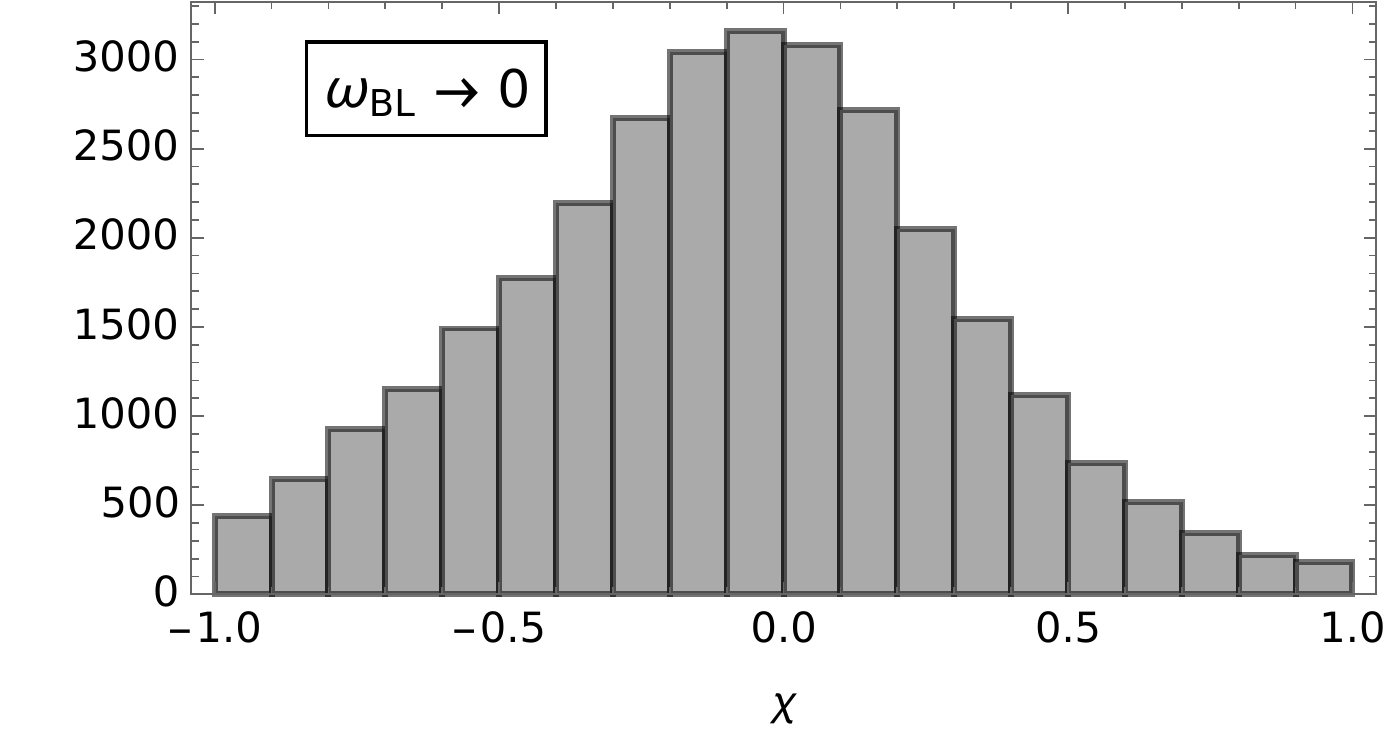}%\\[3pt]
    \includegraphics[width=0.5\columnwidth]{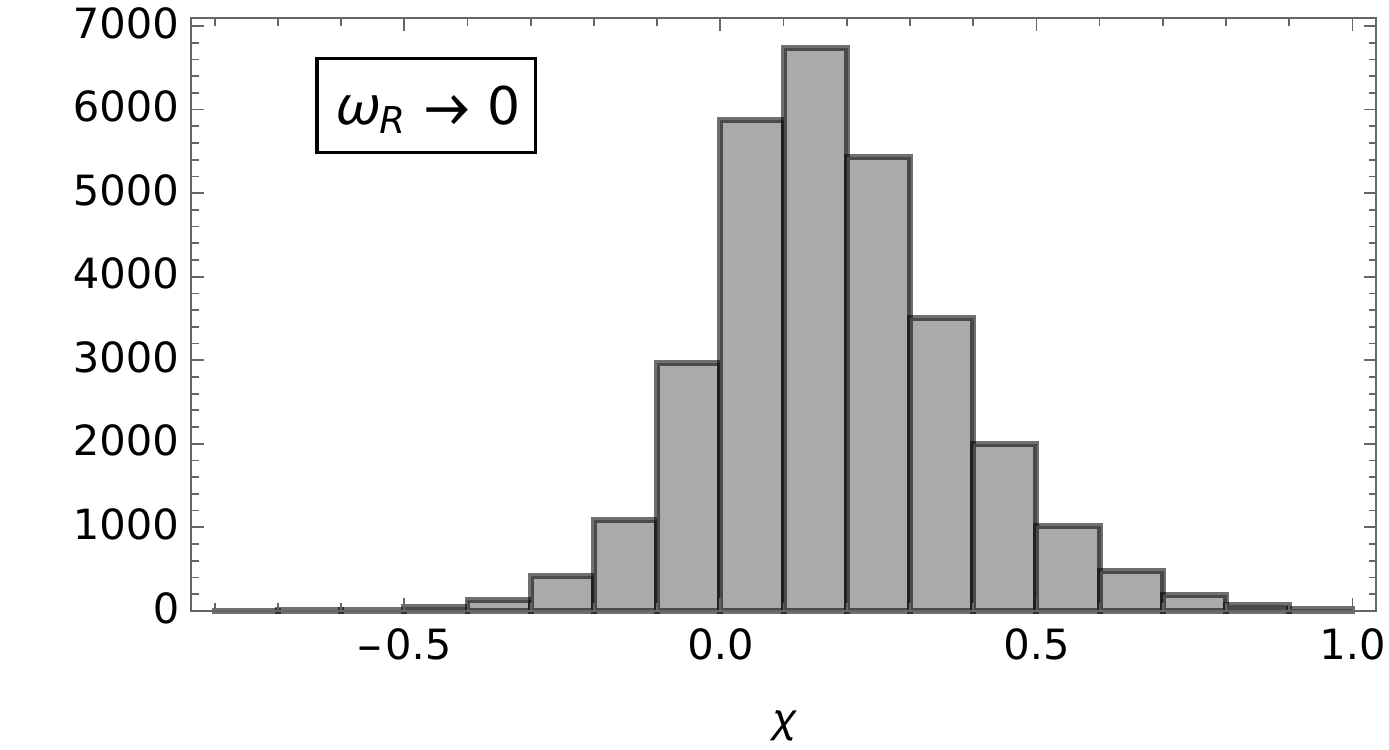}
    \caption{Distribution of values of $\chi$ parameter for all the points in the main unbiased datasets $B_{+}$ ($\omega_{BL} \to 0$) and $R_{+}$ ($\omega_R \to 0$); see Table~\ref{tab:datasets}.}
   \label{fig:chiValues}
\end{figure}

Because of the relation in Eq.~\eqref{eq:chi-definition}, the effect  
of the smallest VEV --- either $\omega_{BL}$ or $\omega_R$ --- on tachyonicity of the scalar spectrum should not be immediately dismissed.
The distribution of $\chi$ presented in Fig.~\ref{fig:chiValues} shows a slight inclination for negative values in the $\omega_{BL} \to 0$ case, while favouring positive values for $\omega_R \to 0$. In other words, the first scenario prefers $\omega_{BL}$ and $\omega_R$ of different sign, while in the other limit, both VEVs are more often of the same sign.

The eigenvalues of tree-level mass matrices for scalars are presented in both limits in Table~\ref{tab:tree-level-scalar-masses}. We show only their leading-order terms, which are proportional either to the $\mathrm{SO}(10)$-breaking VEV ($|\omega_R|$ or $|\omega_{BL}|$) or the intermediate-symmetry-breaking scale $|\sigma|$.
When the dominant contributions to masses are $M_{\text{GUT}}^2$-proportional, the corresponding states belong to approximate \BLzero or \LR intermediate-symmetry multiplets listed in Table~\ref{tab:decompositions}. Their subleading $|\sigma|^2$-proportional tree-level terms can be safely neglected due to the large hierarchy between the scales ($|\sigma|$ is either of order $10^{-4} M_{\text{GUT}}$ or even $10^{-10} M_{\text{GUT}}$; see Fig.~\ref{fig:Scales}). 
These $|\sigma|^2$ terms cannot compete even with the leading one-loop contributions of the order $\sim M_{\text{GUT}}^2/(16\pi^2)$, which become dominant when the tree-level $M_{\text{GUT}}^2$-proportional terms are accidentally small (e.g.,~for PGBs in the $a_2\to 0$ regime). The only exception are those states whose leading-order mass contributions as well as their loop corrections are proportional to $|\sigma|^2$. One can thus distinguish several types of states (cf.~also Sec.~\ref{sec:spectrum}):
\begin{itemize}
    \item The \emph{would-be Goldstone} scalars $(3,2,-\tfrac{5}{6})$, $(1,1,+1)_1$, $(\xbar{3},1,-\tfrac{2}{3})_1$, $(3,2,+\tfrac{1}{6})_1$, and $(1,1,0)_1$ are massless. They get absorbed into the longitudinal components of their massive gauge boson companions listed in Table~\ref{tab:tree-level-gauge-boson-masses} through the Higgs mechanism.
    \item The masses of \emph{generic (heavy) states} can be accurately estimated by their dominant $M_{\text{GUT}}^2$-proportional tree-level contributions. These are $\omega_{R}^{2}$- or $\omega_{BL}^{2}$-proportional in the $\omega_{BL}\to 0$ and $\omega_{R}\to 0$ limits, respectively.
    \item A separate category from the generic states are the \emph{pseudo-Goldstone bosons}, whose tree-level masses are accidentally small, since their leading contributions are proportional to the suppressed value of $a_2$. Therefore, their one-loop corrections can be of comparable size or even dominate. As in the generic case, the subleading $|\sigma|^2$-proportional terms can still be safely neglected, and they are in fact not present at all at tree level for the triplet and octet. The states in this category consist of $(8,1,0)$, $(1,3,0)$, and $(1,1,0)_3$, as well as $(\xbar{3},1,-\tfrac{2}{3})_2$ for $\omega_{BL}\to 0$ and $(1,1,+1)_2$ for $\omega_R\to 0$.
    \item The \emph{intermediate-scale states} (including the intermediate-symmetry-breaking Higgs field) have masses that are $|\sigma|^2$-proportional to all orders in a perturbative expansion, since they are protected by symmetry. Therefore, their tree-level terms suffice. These states are the $(1,1,0)_2$ and either $(\xbar{6},1,-\tfrac{4}{3})$ for $\omega_{BL}\to 0$ or $(1,1,+2)$ for $\omega_R\to 0$. 
\end{itemize}
\begin{table*}[htb]
\begin{center}
\caption{The dominant contributions to tree-level scalar masses in both perturbative scenarios and the associated contributions $\Delta a_i$ to one-loop $\beta$ coefficients for gauge-coupling running. \label{tab:tree-level-scalar-masses}}
	\begin{tabular}{l@{$\quad$}l@{$\quad$}l@{$\quad$}r}
		\hline
		Mass&$\omega_{BL}\to 0$&$\omega_{R}\to 0$&\hspace{-.75cm}$(\Delta a_3,\Delta a_2,\Delta a_1)$\\
		\hline
		\\[-10pt]
		$M^2_S(1,3,0)$ & $+4 a_2 \omega_R^2$ & $-2 a_2 \omega_{BL}^2$ & $(0,\tfrac{1}{3},0)$ \hspace{.5cm} 
		\\[6pt]
		$M^2_S(8,1,0)$ & $-2 a_2 \omega_R^2$ & $+4 a_2 \omega_{BL}^2$ & $(\tfrac{1}{2},0,0)$ \hspace{.5cm}
		\\[6pt]
		$M^2_S(1,1,+2)$ & $-4 (4 \beta_4'+a_2 \chi) \omega_R^2$
		& $+4 (2 \lambda_2+2 \lambda_4+8 \lambda_4'- a_2 \chi^2) |\sigma|^2$ & $(0,0,\tfrac{4}{5})$ \hspace{.5cm}
		\\[6pt]
		$M^2_S(1,3,-1)$ & $-2 (2 \beta_4'+a_2 \chi) \omega_R^2$ & $-6 (6 \beta_4'+a_2 \chi) \omega_{BL}^2$ & $(0,\tfrac{2}{3},\tfrac{3}{5})$ \hspace{.5cm}
		\\[6pt]
		$M^2_S(\xbar{3},1,+\tfrac{4}{3})$ & $-4 (4 \beta_4'+a_2 \chi) \omega_R^2$ & $+2 (\beta_4 - 2 \beta_4'- a_2 \chi) \omega_{BL}^2$ & $(\tfrac{1}{6},0,\tfrac{16}{15})$ \hspace{.5cm}
		\\[6pt]
		$M^2_S(3,3,-\tfrac{1}{3})$ & $-2 (2 \beta_4'+a_2 \chi) \omega_R^2$ & $+2 (\beta_4 - 8 \beta_4'- 2 a_2 \chi) \omega_{BL}^2$ & $(\tfrac{1}{2},2,\tfrac{1}{5})$ \hspace{.5cm}
		\\[6pt]
		$M^2_S(6,3,+\tfrac{1}{3})$ & $-2 (2 \beta_4'+a_2 \chi) \omega_R^2$ & $-2 (2 \beta_4'+ a_2 \chi) \omega_{BL}^2$ & $(\tfrac{5}{2},4,\tfrac{2}{5})$ \hspace{.5cm}
		\\[6pt]
		$M^2_S(\xbar{6},1,-\tfrac{4}{3})$ 
		& $+4 (2 \lambda_2+2 \lambda_4+8 \lambda_4' - a_2 \chi^2) |\sigma|^2$
		& $-4 (4 \beta_4'+ a_2 \chi) \omega_{BL}^2$ & $(\tfrac{5}{6},0,\tfrac{32}{15})$ \hspace{.5cm} 
		\\[6pt]
		$M^2_S(\xbar{6},1,-\tfrac{1}{3})$ & $+2 (\beta_4 - 2 \beta_4' - a_2 \chi) \omega_R^2$ & $-4 (4 \beta_4'+ a_2 \chi) \omega_{BL}^2$ & $(\tfrac{5}{6},0,\tfrac{2}{15})$ \hspace{.5cm}
		\\[6pt]
		$M^2_S(\xbar{6},1,+\tfrac{2}{3})$ & $-4 (4 \beta_4' + a_2 \chi) \omega_R^2$ & $-4 (4 \beta_4'+ a_2 \chi) \omega_{BL}^2$ & $(\tfrac{5}{6},0,\tfrac{8}{15})$ \hspace{.5cm} 
		\\[6pt]
		$M^2_S(1,2,+\tfrac{1}{2})_{1,2}$ & $+\tfrac{1}{2} \left (\beta_4-10 \beta_4' - 4 a_2 \chi \right. \mp$ & $+\tfrac{1}{2} (7 \beta_4-18 \beta_4' - 6 a_2 \chi \mp 4 |\gamma_2|) \omega_{BL}^2$ & $(0,\tfrac{1}{6},\tfrac{1}{10})$ \hspace{.5cm}
		\\[2pt] & $\phantom{+\tfrac{1}{2}.} \mp \left. 2 \sqrt{4 |\gamma_2|^2+(4 \beta_4'+a_2 \chi)^2} \right ) \omega_R^2$ & &
		\\[6pt]
		$M^2_S(3,2,+\tfrac{7}{6})_{1,2}$ & $+\tfrac{1}{2} \left (\beta_4-10 \beta_4' - 4 a_2 \chi \right. \mp$ & $+\tfrac{1}{2} \left (\beta_4-26 \beta_4' - 6 a_2 \chi \right. \mp$ & $(\tfrac{1}{3},\tfrac{1}{2},\tfrac{49}{30})$ \hspace{.5cm}
		\\[2pt] & $\phantom{+\tfrac{1}{2}.} \mp \left. 2 \sqrt{4 |\gamma_2|^2+(4 \beta_4'+a_2 \chi)^2} \right ) \omega_R^2$ & $\phantom{+\tfrac{1}{2}.} \mp \left. 4 \sqrt{|\gamma_2|^2+(6 \beta_4'+a_2 \chi)^2} \right ) \omega_{BL}^2$ &
		\\[6pt]
		$M^2_S(8,2,+\tfrac{1}{2})_{1,2}$ & $+\tfrac{1}{2} \left (\beta_4-10 \beta_4' - 4 a_2 \chi \right. \mp$ & $+\tfrac{1}{2} (\beta_4-18 \beta_4' - 6 a_2 \chi \mp 4 |\gamma_2|) \omega_{BL}^2$ & $(2,\tfrac{4}{3},\tfrac{4}{5})$ \hspace{.5cm}
		\\[2pt] & $\phantom{+\tfrac{1}{2}.} \mp \left. 2 \sqrt{4 |\gamma_2|^2+(4 \beta_4'+a_2 \chi)^2} \right ) \omega_R^2$ & &
		\\[6pt]
		$M^2_S(\xbar{3},1,+\tfrac{1}{3})_{1,3}$ & $+(\beta_4-4 \beta_4' - 2 a_2 \chi \mp 4 |\gamma_2|) \omega_R^2$ & $+\left (\beta_4-10 \beta_4' - 3 a_2 \chi \right. \mp$ & $(\tfrac{1}{6},0,\tfrac{1}{15})$ \hspace{.5cm}
		\\[2pt] & & $\phantom{+.} \mp \left. \sqrt{16 |\gamma_2|^2+(6 \beta_4'+a_2 \chi)^2} \right ) \omega_{BL}^2$ &
		\\[6pt]
		$M^2_S(\xbar{3},1,+\tfrac{1}{3})_2$ & $+2(\beta_4 -2 \beta_4' - a_2 \chi) \omega_R^2$ & $+2(\beta_4 -2 \beta_4' - a_2 \chi) \omega_{BL}^2$ & $(\tfrac{1}{6},0,\tfrac{1}{15})$ \hspace{.5cm}
		\\[6pt]
		$M^2_S(1,1,+1)_2$ & $+2 (\beta_4-2 \beta_4'-a_2 \chi) \omega_R^2$ & $-2 a_2 \omega_{BL}^2$ & $(0,0,\tfrac{1}{5})$ \hspace{.5cm}
		\\[6pt]
		$M^2_S(\xbar{3},1,-\tfrac{2}{3})_2$ & $-2 a_2 \omega_R^2$ & $+2 (\beta_4-2 \beta_4'-a_2 \chi) \omega_{BL}^2$ & $(\tfrac{1}{6},0,\tfrac{4}{15})$ \hspace{.5cm}
		\\[6pt]
		$M^2_S(3,2,+\tfrac{1}{6})_{2,3}$ & $+\tfrac{1}{2} \left (\beta_4-10 \beta_4' - 4 a_2 \chi \right. \mp$ & $+\tfrac{1}{2} \left (\beta_4-26 \beta_4' - 6 a_2 \chi \right. \mp$ & $(\tfrac{1}{3},\tfrac{1}{2},\tfrac{1}{30})$ \hspace{.5cm}
		\\[2pt] & $\phantom{+\tfrac{1}{2}.} \mp \left. 2 \sqrt{4 |\gamma_2|^2+(4 \beta_4'+a_2 \chi)^2} \right ) \omega_R^2$ & $\phantom{+\tfrac{1}{2}.} \mp \left. 4 \sqrt{|\gamma_2|^2+(6 \beta_4'+a_2 \chi)^2} \right ) \omega_{BL}^2$ &
		\\[6pt]
		$M^2_S(1,1,0)_2$ 
		& $+\left (8 \lambda_0 - \frac{8 ((\alpha + \beta_4')^2 + (\alpha + \beta_4') a_2 \chi)}{4 a_0 + a_2} + \right.$ & $+\left (8 \lambda_0 - \frac{12 ((\alpha + \beta_4')^2 + (\alpha + \beta_4') a_2 \chi)}{6 a_0 + a_2} + \right.$ & $(0,0,0)$ \hspace{.5cm} 
		\\[2pt] & $\phantom{+.} \left. + \frac{4 (6 a_0 + a_2) a_2 \chi^2}{4 a_0 + a_2} \right ) |\sigma|^2$ & $\phantom{+.} \left. + \frac{(24 a_0 + a_2) a_2 \chi^2}{6 a_0 + a_2} \right ) |\sigma|^2$ &
		\\[6pt]
		$M^2_S(1,1,0)_3$ & $-2 a_2 \omega_R^2$ & $-2 a_2 \omega_{BL}^2$ & $(0,0,0)$ \hspace{.5cm}
		\\[6pt]
		$M^2_S(1,1,0)_4$ & $+4(4 a_0 + a_2) \omega_R^2$ & $+4(6 a_0 + a_2) \omega_{BL}^2$ & $(0,0,0)$ \hspace{.5cm}
		\\[4pt]
		\hline
	\end{tabular}
\end{center}
\end{table*}

As an interesting aside, note that the actual mass eigenvalues are not necessary to check for either non-tachyonicity of the spectrum or gauge-coupling unification --- in both cases, the quantities of interest can be directly extracted from mass matrices by computing their characteristic polynomial or from knowledge of their principal minors. Positive (semi-)definiteness can be examined by applying Sylvester's criterion on the matrices; cf.~Sec.~\ref{sec:Tachyonicity}. This, for example, leads to conditions in Eqs.~\eqref{eq:nontachyonicity-analytic-BL} and~\eqref{eq:nontachyonicity-analytic-R} for the $a_2\to 0$ limit. Alternatively, alternating signs of coefficients in the characteristic polynomial also imply positive definiteness of the spectrum. For positive semi-definite
matrices with $n$ zero-eigenvalues, the $n$ lowest-order coefficients in the characteristic polynomial vanish. Unification constraints at one loop, on the other hand, require only knowledge of the product of non-zero eigenvalues $\prod_{k=1}^{n_R} m^2_{R_k}$ for every SM representation $R$ with multiplicity $n_{R}$. This number can be extracted as the absolute value of the lowest degree coefficient in the characteristic polynomial that does not vanish. The reason that this product suffices is that all $n_R$ eigenvalues of 
multiplets in the representation $R$ contribute with the same $\beta$ coefficient $\Delta a_i^{(R)}$ to Eq.~\eqref{eq:aSolution2}.

%======================================================
\subsection{One-loop pseudo-Goldstone masses\label{app:PGB masses}}
%======================================================

An accurate assessment of the PGB masses in the small $a_2$ regime requires also the computation of their one-loop corrections. In addition to the recurring octet, triplet, and singlet pseudo-Goldstone fields, the two limits $\omega_{BL}\to 0$ and $\omega_{R}\to 0$ provide another
PGB in the form of the $(\xbar{3},1,-\tfrac{2}{3})_2$ or $(1,1,+1)_2$ state, respectively.

In the $\omega_{BL}\to 0$ scenario with \BLzero intermediate stage, the new PGB state $(\xbar{3},1,-\tfrac{2}{3})_2$ belongs to the $(15,1,0)$ multiplet of that symmetry, together with the octet and singlet PGBs; see Table~\ref{tab:decomposition-45}. This confirms its PGB character, since its mass must be degenerate with the octet and singlet up to subdominant corrections of the intermediate-symmetry-breaking VEV $\sigma$. Similarly, in the $\omega_{R}\to 0$ scenario with the \LR intermediate stage, the new PGB state $(1,1,+1)_2$ is embedded together with the SM singlet into a $(1,1,3,0)$ representation.

An interesting observation is that the weak-triplet and SM-singlet PGBs receive the same one-loop \emph{gauge} contributions to their masses in the $\omega_{R}\to 0$ limit. Note that the two mentioned states belong to different representations of the intermediate \emph{left-right} symmetry, namely, to $(1,3,1,0)$ and $(1,1,3,0)$, respectively. However, these intermediate-stage representations are connected by \emph{D-parity} ($L \leftrightarrow  R$ exchange), which corresponds to a transformation $\omega_{BL} \to - \omega_{BL}$ at the VEV level. In the $\omega_{R}\to 0$ case,  the dominant one-loop gauge contributions to masses must be proportional to $g^4 \omega_{BL}^2$ and are hence the same for pairs of states that exchange under D-parity. The one-loop contributions from scalars, on the other hand, are not parity invariant due to the implicit presence of massive scalar-potential parameters $\mu$, $\nu$, and $\tau$ obtained by solving stationarity conditions, where a change of sign in $\omega_{BL}$ does have an effect.

The one-loop contributions to masses of PGB states in the $a_2\to 0$, $\gamma_2\to 0$ limit\footnote{Note that in this limit, the tree-level PGB masses exactly vanish (see Table~\ref{tab:tree-level-scalar-masses}), so the one-loop corrections actually represent their entire masses at the one-loop level.} are presented in Table~\ref{tab:one-loop-scalar-masses}. 
Although this limit is apparently unphysical since the numerical scans show that $|\gamma_2|\gtrsim 0.1$ for valid points (cf.~Fig.~\ref{fig:Sylvester}), it is well suited for obtaining analytical expressions. For both the $\omega_{BL}\to 0$ and $\omega_{R}\to 0$ scenario, we show only the dominant, $\mathrm{SO}(10)$-breaking terms proportional to the largest VEV, while neglecting those with $\sigma$. The qualitative features of PGBs discussed in the beginning of this section are confirmed by the explicit formulae in Table~\ref{tab:one-loop-scalar-masses}, at least in the given limit. For more general formulae with an arbitrary $\omega_{BL}$ and $\omega_R$ hierarchy, but still in the $a_2\to 0$, $\gamma_2\to 0$, $\sigma\to 0$ limit, the interested reader is referred to~\cite{Graf:2016znk, JarkovskaThesis}.

The one-loop expressions in Table~\ref{tab:one-loop-scalar-masses} are organized as follows: The \emph{polynomial} contributions from gauge bosons and scalars are followed by gauge and scalar \emph{logarithmic} terms. The arguments of the logarithms in the $\omega_{BL}\to 0$ scenario correspond to the mass squares of the \BLzero multiplets ordered as
    \begin{align*}
        &M^2_G(6,2,-\tfrac{1}{2}),
        &&M^2_G(1,1,+1),
        &&M^2_S(\xbar{10},1,0), \\[3pt]
        &M^2_S(15,2,+\tfrac{1}{2})_1,
        &&M^2_S(15,2,+\tfrac{1}{2})_2,
        &&M^2_S(\xbar{10},1,+1), \\[3pt]
        &M^2_S(10,3,0),
        &&M^2_S(6,1,0),
    \end{align*}
and for $\omega_{R}\to 0$ to mass squares of the \LR multiplets arranged as
    \begin{align*}
        &M^2_G(3,2,2,-\tfrac{2}{3}), 
        &&M^2_G(3,1,1,+\tfrac{4}{3}), 
        &&M^2_S(1,3,1,-2),  \\[3pt]
        &M^2_S(3,2,2,+\tfrac{4}{3})_1, 
        &&M^2_S(3,2,2,+\tfrac{4}{3})_2,
        &&M^2_S(\xbar{6},1,3,-\tfrac{2}{3}), \\[3pt] 
        &M^2_S(6,3,1,+\tfrac{2}{3}), 
        &&M^2_S(3,1,1,-\tfrac{2}{3})_1, 
        &&M^2_S(3,1,1,-\tfrac{2}{3})_2, \\[3pt] 
        &M^2_S(3,3,1,-\tfrac{2}{3}),
        &&M^2_S(\xbar{3},1,3,+\tfrac{2}{3}), 
        &&M^2_S(1,2,2,0), \\[3pt]
        &M^2_S(8,2,2,0).
    \end{align*}
The tree-level masses in the log arguments are taken in the $a_2\to 0$, $\gamma_2\to 0$ limit, as confirmed by cross-referencing Tables~\ref{tab:decompositions}, \ref{tab:tree-level-gauge-boson-masses}, and~\ref{tab:tree-level-scalar-masses}. Note that for all terms where the logarithm diverges, e.g.,~when the argument of the log contains the mass of a PGB in the $a_2\to 0$ limit or a would-be Goldstone boson, the coefficient in front of the log vanishes, so these terms do not contribute and are omitted.

\begin{table*}[htb]
	\begin{center}
			\caption{The dominant one-loop contributions to masses of PGB scalars in the $a_2\to 0$, $\gamma_2\to 0$ limit for both perturbative scenarios and containing polynomial as well as logarithmic terms with gauge bosons and scalar fields running in the loop. \label{tab:one-loop-scalar-masses}}
			\begin{tabular}{l@{$\quad$}l@{$\quad$}l}	
		    \hline
		    Mass&$\omega_{BL}\to 0$&$\omega_{R}\to 0$\\
		    \hline \\[-10pt]      			%
			$M_S^2(1,3,0)$ & $\frac{\omega_R^2}{16\pi^2} \Big ( 16 g^4 + 40 (\beta_4^2 + 36 \beta_4'^2)$ & $\frac{\omega_{BL}^2}{16\pi^2} \Big ( 19 g^4 + (31 \beta_4^2 + 60 \beta_4 \beta_4' +2900 \beta_4'^2)$
			\\[2pt] & $+18 g^4 \log \left[\frac{g^2  \omega_R^2}{2 \mu_R^2} \right] - 12 g^4 \log \left[\frac{2 g^2  \omega_R^2}{\mu_R^2} \right]$ & $+21 g^4 \log \left[\frac{g^2  \omega_{BL}^2}{2 \mu_R^2} \right] - 24 g^4 \log \left[\frac{2 g^2 \omega_{BL}^2}{\mu_R^2}\right]$
			\\[2pt] & $+40(-\beta_4^2 +2 \beta_4 \beta_4') \log \left[\frac{2(\beta_4 - 2\beta_4') \omega_R^2}{\mu_R^2}\right]$ & $-72 (\beta_4 \beta_4'+ 6 \beta_4'^2)\log \left[-\frac{36\beta_4' \omega_{BL}^2}{\mu_R^2}\right]$
			\\[2pt] & $+15 (\beta_4^2 +14 \beta_4 \beta_4')\log \left[\frac{(\beta_4 -18 \beta_4')\omega_R^2}{2\mu_R^2}\right]$ & $+(7\beta_4^2-136\beta_4\beta_4' -1100 \beta_4'^2) \log \left[\frac{(\beta_4 - 50\beta_4')\omega_{BL}^2}{2\mu_R^2}\right]$
			\\[2pt] & $+15 (\beta_4^2 - 6\beta_4 \beta_4' + 8 \beta_4'^2) \log \left[\frac{(\beta_4 - 2\beta_4')\omega_R^2}{2\mu_R^2}\right]$ & $+(7\beta_4^2-40\beta_4\beta_4' +52 \beta_4'^2)\log \left[\frac{(\beta_4 - 2\beta_4')\omega_{BL}^2}{2\mu_R^2}\right]$
			\\[2pt] & $-1280 \beta_4'^2\log \left[-\frac{16 \beta_4' \omega_R^2}{\mu_R^2}\right]$ & $-768 \beta_4'^2 \log \left[-\frac{16\beta_4' \omega_{BL}^2}{\mu_R^2}\right]$
			\\[2pt] & $+80 (-\beta_4 \beta_4' + 12 \beta_4'^2) \log\left[-\frac{4 \beta_4' \omega_R^2}{\mu_R^2}\right] \Big )$ & $+48 (-\beta_4 \beta_4' +14\beta_4'^2) \log \left[-\frac{4\beta_4' \omega_{BL}^2}{\mu_R^2}\right]$
			\\[2pt] & & $-(23 \beta_4^2-376 \beta_4 \beta_4' + 128 \beta_4'^2) \log \left[\frac{(\beta_4 - 16\beta_4')\omega_{BL}^2}{\mu_R^2}\right]$
			\\[2pt] & & $+(\beta_4^2 -8\beta_4 \beta_4' +16 \beta_4'^2) \log \left[\frac{(\beta_4 - 4\beta_4')\omega_{BL}^2}{\mu_R^2}\right]$
			\\[2pt] & & $+12 (3\beta_4^2-24 \beta_4 \beta_4' +64 \beta_4'^2) \log \left[\frac{2(\beta_4 - 8\beta_4')\omega_{BL}^2}{\mu_R^2}\right]$
		\\[2pt] & & $+24 (-\beta_4^2+\beta_4 \beta_4' + 2\beta_4'^2)\log \left[\frac{2(\beta_4 - 2\beta_4')\omega_{BL}^2}{\mu_R^2}\right]$
			\\[2pt] & & $+(-\tfrac{53}{3} \beta_4^2 + 4 \beta_4 \beta_4' +108 \beta_4'^2)\log \left[\frac{(7\beta_4 - 18\beta_4')\omega_{BL}^2}{2\mu_R^2}\right]$
			\\[2pt] & & $+(\tfrac{56}{3} \beta_4^2 + 128 \beta_4 \beta_4' +864 \beta_4'^2) \log \left[\frac{(\beta_4 - 18\beta_4')\omega_{BL}^2}{2\mu_R^2}\right] \Big )$
			\\[16pt]
			$M_S^2(8,1,0)$ & $\frac{\omega_R^2}{16\pi^2} \Big ( 13 g^4 + (19 \beta_4^2 + 60 \beta_4 \beta_4' + 1340 \beta_4'^2)$ & $\frac{\omega_{BL}^2}{16\pi^2} \Big ( 22 g^4 + (\tfrac{172}{3} \beta_4^2 + 3000 \beta_4'^2)$
			\\[2pt] & $+9 g^4 \log \left[\frac{g^2  \omega_R^2}{2 \mu_R^2} \right] - 12 g^4 \log \left[\frac{2 g^2  \omega_R^2}{\mu_R^2} \right]$ & $+12 g^4 \log \left[\frac{g^2  \omega_{BL}^2}{2 \mu_R^2} \right] - 6 g^4 \log \left[\frac{2 g^2 \omega_{BL}^2}{\mu_R^2}\right]$
			\\[2pt] & $-24 (\beta_4 \beta_4' -6 \beta_4'^2)  \log \left[\frac{2(\beta_4 - 2\beta_4') \omega_R^2}{\mu_R^2}\right]$ & $-1296 \beta_4'^2\log \left[-\frac{36\beta_4' \omega_{BL}^2}{\mu_R^2}\right]$
			\\[2pt] & $+(\tfrac{23}{2}\beta_4^2 +38 \beta_4 \beta_4'+54 \beta_4'^2)\log \left[\frac{(\beta_4 -18 \beta_4')\omega_R^2}{2\mu_R^2}\right]$ & $+4 (\beta_4^2+29\beta_4\beta_4' -350 \beta_4'^2) \log \left[\frac{(\beta_4 - 50\beta_4')\omega_{BL}^2}{2\mu_R^2}\right]$
			\\[2pt] & $+(\tfrac{23}{2}\beta_4^2 - 30\beta_4 \beta_4' + 126 \beta_4'^2) \log \left[\frac{(\beta_4 - 2\beta_4')\omega_R^2}{2\mu_R^2}\right]$ & $+4 (\beta_4^2-7\beta_4\beta_4' +10 \beta_4'^2)\log \left[\frac{(\beta_4 - 2\beta_4')\omega_{BL}^2}{2\mu_R^2}\right]$
			\\[2pt] & $-64 (\beta_4 \beta_4' + 11 \beta_4'^2) \log \left[-\frac{16 \beta_4' \omega_R^2}{\mu_R^2}\right]$ & $+48 (-3 \beta_4\beta_4' + 14\beta_4'^2) \log \left[-\frac{16\beta_4' \omega_{BL}^2}{\mu_R^2}\right]$
			\\[2pt] & $-48 (\beta_4 \beta_4' - 9 \beta_4'^2) \log\left[-\frac{4 \beta_4' \omega_R^2}{\mu_R^2}\right]$ & $-12 (3\beta_4 \beta_4' -38\beta_4'^2) \log \left[-\frac{4\beta_4' \omega_{BL}^2}{\mu_R^2}\right]$
			\\[2pt] & $-2 (9 \beta_4^2 - 34 \beta_4 \beta_4' - 24 \beta_4'^2) \log\left[\frac{(\beta_4 - 4 \beta_4') \omega_R^2}{\mu_R^2}\right] \Big )$ & $+(\beta_4^2+22 \beta_4 \beta_4' - 32 \beta_4'^2) \log \left[\frac{(\beta_4 - 16\beta_4')\omega_{BL}^2}{\mu_R^2}\right]$
			\\[2pt] & & $+(\beta_4^2 -14\beta_4 \beta_4' +40 \beta_4'^2) \log \left[\frac{(\beta_4 - 4\beta_4')\omega_{BL}^2}{\mu_R^2}\right]$
			\\[2pt] & & $-6 (\beta_4^2-46 \beta_4 \beta_4' +16 \beta_4'^2) \log \left[\frac{2(\beta_4 - 8\beta_4')\omega_{BL}^2}{\mu_R^2}\right]$
			\\[2pt] & & $-6 (\beta_4^2+8\beta_4 \beta_4' - 20\beta_4'^2)\log \left[\frac{2(\beta_4 - 2\beta_4')\omega_{BL}^2}{\mu_R^2}\right]$
			\\[2pt] & & $-(\tfrac{224}{9}\beta_4^2 - 64 \beta_4 \beta_4')\log \left[\frac{(7\beta_4 - 18\beta_4')\omega_{BL}^2}{2\mu_R^2}\right]$
			\\[2pt] & & $+(\tfrac{152}{9} \beta_4^2 - 88 \beta_4 \beta_4' +1296 \beta_4'^2) \log \left[\frac{(\beta_4 - 18\beta_4')\omega_{BL}^2}{2\mu_R^2}\right] \Big )$
            \\[10pt]
			\hline
			\multicolumn{3}{|c|}{Continued on next page} \\
			\hline
			\end{tabular}
	\end{center}
\end{table*}
\begin{table*}[htb]
	\begin{center}
			\begin{tabular}{l@{$\quad$}l@{$\quad$}l}	
		    \multicolumn{3}{l}{
		    \tablename\ \thetable{} --- continued from previous page} \\[4pt]
			\hline
		    Mass&$\omega_{BL}\to 0$&$\omega_{R}\to 0$\\
		    \hline \\[-10pt]
			$M_S^2(1,1,0)_3$ & \multicolumn{1}{c}{\text{same as} $M_S^2(8,1,0)$} & $\frac{\omega_{BL}^2}{16\pi^2} \Big ( 19 g^4 + (31 \beta_4^2 + 60\beta_4 \beta_4' + 2900 \beta_4'^2)$
			\\[2pt] & \phantom{$(9 \beta_4^2 - 34 \beta_4 \beta_4' - 24 \beta_4'^2) \log\left[\frac{(\beta_4 - 4 \beta_4') \omega_R^2}{\mu_R^2}\right] \Big )$} & $+21 g^4 \log \left[\frac{g^2  \omega_{BL}^2}{2 \mu_R^2} \right] - 24 g^4 \log \left[\frac{2 g^2 \omega_{BL}^2}{\mu_R^2}\right]$
			\\[2pt] & & $-1296 \beta_4'^2\log \left[-\frac{36\beta_4' \omega_{BL}^2}{\mu_R^2}\right]$
			\\[2pt] & & $+(7 \beta_4^2+104\beta_4\beta_4' -1100 \beta_4'^2) \log \left[\frac{(\beta_4 - 50\beta_4')\omega_{BL}^2}{2\mu_R^2}\right]$
			\\[2pt] & & $+(7 \beta_4^2+8\beta_4\beta_4' +52 \beta_4'^2)\log \left[\frac{(\beta_4 - 2\beta_4')\omega_{BL}^2}{2\mu_R^2}\right]$
			\\[2pt] & & $+192 (-\beta_4\beta_4' + 8\beta_4'^2) \log \left[-\frac{16\beta_4' \omega_{BL}^2}{\mu_R^2}\right]$
			\\[2pt] & & $+96 \beta_4'^2 \log \left[-\frac{4\beta_4' \omega_{BL}^2}{\mu_R^2}\right]$
			\\[2pt] & & $+(\beta_4^2-8 \beta_4 \beta_4' - 128 \beta_4'^2) \log \left[\frac{(\beta_4 - 16\beta_4')\omega_{BL}^2}{\mu_R^2}\right]$
			\\[2pt] & & $-(23\beta_4^2 -88\beta_4 \beta_4' -16 \beta_4'^2) \log \left[\frac{(\beta_4 - 4\beta_4')\omega_{BL}^2}{\mu_R^2}\right]$
			\\[2pt] & & $-24 (\beta_4^2-10 \beta_4 \beta_4' +16 \beta_4'^2) \log \left[\frac{2(\beta_4 - 8\beta_4')\omega_{BL}^2}{\mu_R^2}\right]$
			\\[2pt] & & $+12 (3 \beta_4^2-12\beta_4 \beta_4' + 28\beta_4'^2)\log \left[\frac{2(\beta_4 - 2\beta_4')\omega_{BL}^2}{\mu_R^2}\right]$
			\\[2pt] & & $+(-\tfrac{53}{3}\beta_4^2 + 100 \beta_4 \beta_4'+108\beta_4'^2)\log \left[\frac{(7\beta_4 - 18\beta_4')\omega_{BL}^2}{2\mu_R^2}\right]$
			\\[2pt] & & $+(\tfrac{56}{3} \beta_4^2 - 256 \beta_4 \beta_4' +864 \beta_4'^2) \log \left[\frac{(\beta_4 - 18\beta_4')\omega_{BL}^2}{2\mu_R^2}\right] \Big )$
			\\[16pt]
			$M_S^2(\xbar{3},1,-\tfrac{2}{3})_2$ & \multicolumn{1}{c}{\text{same as} $M_S^2(8,1,0)$} & \multicolumn{1}{c}{\text{not a PGB}}
			\\[16pt]
			$M_S^2(1,1,+1)_2$ & \multicolumn{1}{c}{\text{not a PGB}} & \multicolumn{1}{c}{\text{same as} $M_S^2(1,1,0)_3$}
			\\[4pt]
			\hline
			\end{tabular}
	\end{center}
\end{table*}
%

%%%%%%%%%%%%%%%%%%%%%%%%%%%%%%%%%%%%%%%%%%%%%%%%%%%%%%%%%%%%%%%%%%%%%%%%%%%%%%%%
% % %   BIBLIOGRAPHY
%\bibliographystyle{h-physrev5}
%\bibliography{bibliography1}

%%%%%%%%%%%%%%%%%%%%%%%%%%%%%%%%%%%%%%%%%%%%%%%%%%%%%%%%%%%%%%%%%%%%%%%%%%%%%%%%%%
%%%%%%%%%%%%%%%%%%%%%%%%%%%%%%%%%%%%%%%%%%%%%%%%%%%%%%%%%%%%%%%%%%%%%%%%%%%%%%%%%%

\end{document}